\documentclass[12pt]{article}
\usepackage{amsfonts}
\usepackage{amssymb}
\usepackage{amsbsy}

\input epsf.sty

\newcommand{\BEQ}{\begin{equation}}     % Gleichungen Anfang ..
\newcommand{\BEA}{\begin{eqnarray}}
\newcommand{\BD}{\begin{displaymath}}
\newcommand{\EEQ}{\end{equation}}       % .. und Ende
\newcommand{\EEA}{\end{eqnarray}}
\newcommand{\ED}{\end{displaymath}}
\newcommand{\al}{\alpha}                % griechische Buchstaben

\newcommand{\del}{\delta}

\newcommand{\eps}{\varepsilon}          % epsilon

\newcommand{\g}{{\mathfrak{g}}}
\newcommand{\h}{{\mathfrak{h}}}
\newcommand{\s}{{\mathfrak{s}}}

\newcommand{\sch}{{\mathfrak{sch}}}
\newcommand{\sv}{{\mathfrak{sv}}}
\newcommand{\se}{{\mathfrak{se}}}
\newcommand{\ns}{{\mathfrak{ns}}}
\newcommand{\kk}{{\mathfrak{k}}}
\newcommand{\sns}{{\mathfrak{sns}}}
\newcommand{\symp}{{\mathfrak{sp}}}
\newcommand{\slin}{{\mathfrak{sl}}}
\newcommand{\osp}{{\mathfrak{osp}}}
\newcommand{\spin}{{\mathfrak{spin}}}
\newcommand{\sh}{{\mathfrak{sh}}}

\newcommand{\gl}{{\mathfrak{gl}}}
\newcommand{\GL}{{\mathrm{Gl}}}
\newcommand{\conf}{{\mathfrak{conf}}}
\newcommand{\sgal}{{\mathfrak{sgal}}}
\newcommand{\Vir}{{\mathfrak{vir}}}
\newcommand{\R}{\mathbb{R}}
\newcommand{\C}{\mathbb{C}}
\newcommand{\Z}{\mathbb{Z}}

\newcommand{\eop}{\hfill $\Box$}        % quod erat demonstrandum ... 

\newcommand{\D}{{\rm d}}                % gerades d fuer Ableitungen
\newcommand{\II}{{\rm i}}               % gerades i fuer komplexe Einheit
\newcommand{\half}{{1\over 2}}          % 1/2 als Bruch
\newcommand{\ad}{{\mathrm{ad\,}}}       % ad aus den Liealgebren
\renewcommand{\deg}{{\rm deg\,}}        % Degree deg
\newcommand{\gra}{{\rm gra\,}}          % Grad gra
\newcommand{\cdim}{{\rm cdim\,}}        % konforme Dimension cdim
\newcommand{\wit}[1]{\widetilde{#1}}    % weite Schlange
\newcommand{\zeile}[1]{\vskip #1 \baselineskip} % N Zeilen ueberschlagen
\newcommand{\vekz}[2]
     {\mbox{${\begin{array}{c} #1  \\ #2 \end{array}}$}}
\newcommand{\appsection}[2]{\setcounter{equation}{0} \section*{Appendix #1. #2}
\renewcommand{\theequation}{#1\arabic{equation}}
              \renewcommand{\thesection}{#1} }
\catcode`\@=11
\def\numberbysection{\@addtoreset{equation}{section}
        \def\theequation{\thesection.\arabic{equation}}}
\numberbysection

\begin{document}

\begin{titlepage}

%{\hfill \tt \today; pr\'eliminaire !}

\vskip 1.5 cm
\begin{center}
{\Large \bf Supersymmetric extensions of Schr\"odinger-invariance}
\end{center}

\vskip 2.0 cm
   
\centerline{ {\bf Malte Henkel}$^a$ and {\bf J\'er\'emie Unterberger}$^b$}
\vskip 0.5 cm
\centerline {$^a$Laboratoire de Physique des 
Mat\'eriaux,\footnote{Laboratoire associ\'e au CNRS UMR 7556} 
Universit\'e Henri Poincar\'e Nancy I,} 
\centerline{ B.P. 239, 
F -- 54506 Vand{\oe}uvre-l\`es-Nancy Cedex, France}
\vskip 0.5 cm
\centerline {$^b$Institut Elie Cartan,\footnote{Laboratoire 
associ\'e au CNRS UMR 7502} Universit\'e Henri Poincar\'e Nancy I,} 
\centerline{ B.P. 239, 
F -- 54506 Vand{\oe}uvre-l\`es-Nancy Cedex, France}

\begin{abstract}

The set of dynamic symmetries of the scalar free Schr\"odinger equation in 
$d$ space dimensions gives a realization of the Schr\"odinger algebra that 
may be extended into a representation of the conformal algebra in $d+2$ 
dimensions, which yields the set of dynamic 
symmetries of the same equation where the mass is not viewed as a constant,
but as an additional coordinate. An analogous construction also holds for the 
spin-$\half$ L\'evy-Leblond equation.  An $N=2$ supersymmetric extension of 
these equations leads, respectively, to a `super-Schr\"odinger' model 
and to the $(3|2)$-supersymmetric model. Their dynamic supersymmetries
form the Lie superalgebras $\osp(2|2)\ltimes\sh(2|2)$ and $\osp(2|4)$, 
respectively. The Schr\"odinger algebra and its supersymmetric counterparts 
are found to be the largest finite-dimensional Lie subalgebras
of a family of infinite-dimensional Lie superalgebras that are systematically 
constructed in a Poisson algebra setting, including the 
Schr\"odinger-Neveu-Schwarz algebra $\sns^{(N)}$ with $N$ supercharges.

Covariant two-point functions of quasiprimary superfields are calculated
for several subalgebras of $\osp(2|4)$. If one includes both
 $N=2$ supercharges and  time-inversions, then the sum
of the scaling dimensions is  restricted to a finite set of possible values. 
\end{abstract}

\zeile{2} \noindent 
\underline{PACS:} 02.20Tw, 05.70Fh, 11.25Hf, 11.30.Pb\\
\underline{Keywords:} conformal field-theory, correlation functions, 
algebraic structure of integrable models, Schr\"odinger-invariance, 
supersymmetry  
\end{titlepage}

%%%%%%%%%%%%%%%%%%%%%%%%%%%%%%%%%%%%%%%%%%%%%%%%%%%%%%%%%%%%%%%%%%%%%%%%%%%%%%%%
\section{Introduction}
%%%%%%%%%%%%%%%%%%%%%%%%%%%%%%%%%%%%%%%%%%%%%%%%%%%%%%%%%%%%%%%%%%%%%%%%%%%%%%%%

Symmetries have always played a central r\^ole in mathematics and physics. 
For example, it is well-known since the work of Lie (1881) that the
free diffusion equation in one spatial dimension has a non-trivial symmetry
group. It was recognized much later that the same group also appears to be  the 
maximal dynamic invariance group of the free Schr\"odinger equation in $d$ 
space dimensions, and it is therefore referred to as the {\em Schr\"odinger 
group} \cite{Nied72}. Its Lie algebra is denoted by $\mathfrak{sch}_d$. 
In the case $d=1$, one may realize $\sch_1$ by the following differential 
operators 
\BEA 
X_{-1} = -\partial_t \;\; , \;\; Y_{-1/2} = -\partial_r & &
\mbox{\rm time and space translations} \nonumber \\
Y_{1/2} = - t\partial_r - {\cal M} r & &
\mbox{\rm Galilei transformation} \nonumber \\
X_{0} = -t\partial_t - \frac{1}{2} r\partial_r - \frac{x}{2} & &
\mbox{\rm dilatation} \nonumber \\
X_{1} = -t^2\partial_t - tr\partial_r - \frac{\cal M}{2} r^2 - 2 x t & &
\mbox{\rm special transformation} \nonumber \\
M_{0} = - {\cal M} & &
\mbox{\rm phase shift}
\label{gl:1:sch1}
\EEA
Here, ${\cal M}$ is a (real or complex) number and $x$ is the scaling 
dimension of the wave function $\phi$ on which the generators of 
$\mathfrak{sch}_1$ act. The Lie algebra $\sch_1$ realizes dynamical symmetries 
of the $1D$ free Schr\"odinger/diffusion equation 
${\cal S}\phi=(2{\cal M}\partial_t -\partial_r^2)\phi=0$ only if $x=1/2$.

In particular, $\mathfrak{sch}_1$ is isomorphic to the semi-direct
product $\mathfrak{sl}_2(\R)\ltimes \mathfrak{h}_1$, where
$\mathfrak{sl}_2(\R)$ is spanned by the three $X$-generators whereas the
 Heisenberg algebra in one space-dimension $\mathfrak{h}_1$ is spanned by 
$Y_{\pm 1/2}$ and $M_0$.

Schr\"odinger-invariance has been found in physically very different systems
such as non-relativistic field-theory \cite{Kast68,Hage72,Jack90b}, 
celestial mechanics \cite{Duva91}, 
the Eulerian equations of motion of a viscous fluid \cite{Hass00,ORai01} 
or the slow dynamics of statistical systems far from 
equilibrium \cite{Henk03b,Pico04,Stoi05}, just to mention a few. 
In this paper, we investigate the following two 
important features of Schr\"odinger-invariance  in a supersymmetric setting. 
The consideration of supersymmetries
in relation with Schr\"odinger-invariance may be motivated from the 
long-standing topic of supersymmetric quantum mechanics \cite{Crom83} 
and from the application of Schr\"odinger-invariance to 
the long-time behaviour of systems undergoing ageing, e.g. in the context 
of phase-ordering kinetics.
Equations such as the Fokker-Planck or Kramers
equations, which are habitually used to describe non-equilibrium statistical 
systems, are naturally supersymmetric, see \cite{Junk96,Tail05} and
references therein.

1. First, there is a certain analogy between Schr\"odinger- and
conformal invariance. This is less surprising than it might appear at first
sight since there is an embedding of the (complexified) Schr\"odinger
Lie algebra in $d$ space dimensions  into the conformal algebra in $(d+2)$ 
space dimensions,  
$\mathfrak{sch}_d \subset (\mathfrak{conf}_{d+2})_{\mathbb{C}}$ 
\cite{Burd73,Henk03a}.\footnote{In the literature, the invariance under 
the  generator of special transformations  $X_1$ is sometimes referred to as
 `conformal invariance', 
but we stress that the embedding 
$\mathfrak{sch}_d \subset (\mathfrak{conf}_{d+2})_{\mathbb{C}}$ is
considerably more general. In this paper, conformal invariance always 
means invariance under the whole conformal algebra $\conf_{d+2}$.} 
This embedding comes out naturally when one thinks of the mass parameter 
$\cal M$ in the Schr\"odinger equation as an additional {\em coordinate}. 
Then a Laplace-transform of the Schr\"odinger equation with respect to $\cal M$ 
yields a Laplace-like equation which is known to be invariant under the 
conformal group.

2. Second, we recall the fact, observed by one of us long ago \cite{Henk94}, 
that the six-dimensional Lie algebra 
$\mathfrak{sch}_1$ can be embedded into the following infinite-dimensional 
Lie algebra with the non-vanishing commutators
\newpage \typeout{ *** Seitenvorschub vor Gl. (1.2) ***}
\BEA
\left[ X_n, X_{n'} \right] &=& (n-n') X_{n+n'} 
+ \frac{c}{12}\left( n^3 - n\right) \delta_{n,n'} \;\; , \;\;
\left[ X_n, Y_m \right] \:=\: \left(\frac{n}{2}-m\right) Y_{n+m} \;\; , \;\;
\nonumber \\
\left[ X_n, M_{n'} \right] &=& -n' M_{n+n'} 
\hspace{4.15truecm} \;\; , \;\; 
\left[ Y_m, Y_{m'}\right] \:=\: (m-m') M_{m+m'}. 
\label{gl:1:sv}
\EEA
where $n,n'\in \mathbb{Z}$, $m\in \mathbb{Z}+\frac{1}{2}$, and $c$ is 
the central charge. It can be shown that no further non-trivial central
extension of this algebra is possible \cite{Henk94}. We shall call the
algebra (\ref{gl:1:sv}) the {\em Schr\"odinger-Virasoro algebra} and denote
it by $\mathfrak{sv}$. For $c=0$,
a differential-operator representation of the algebra $\mathfrak{sv}$ is
\BEA
X_n &=& -t^{n+1}\partial_t - \frac{n+1}{2}\ t^n r\partial_r 
- (n+1)\frac{x}{2}  t^n
-\frac{n(n+1)}{4}{\cal M} t^{n-1} r^2 \nonumber \\
Y_m &=& -t^{m+1/2} \partial_r - \left( m+\frac{1}{2}\right) t^{m-1/2} r{\cal M}
\nonumber \\
M_n &=& - {\cal M} t^n
\label{gl:1:Schr}
\EEA
where $\cal M$ is a parameter and $x$ is again a scaling dimension. 
Extensions to higher space-dimensions are straightforward, see \cite{Henk02}. 
Remarkably, there is a `no-go'-theorem forbidding any reasonable double 
extension of the Schr\"odinger algebra both by the conformal algebra 
{\it and} by the Schr\"odinger-Virasoro algebra \cite{RogUnt}. The differential
equations which are conditionally invariant under the representation (\ref{gl:1:Schr}) of $\sv$ with ${\cal M}=0$ are given in \cite{Cher04}. 

The algebra $\mathfrak{sv}$ can be further extended by considering the
generators \cite{Henk02}
\BEA
X_n &=& -t^{n+1}\partial_t - \frac{n+1}{2}\ t^n r\partial_r 
-(n+1)\frac{x}{2} t^n -\frac{n(n+1)}{4}{\cal M} t^{n-1} r^2 
-\frac{n^3-n}{8}{\cal M}'\, t^{n-2} r^4 \nonumber \\
Y_m &=& -t^{m+1/2} \partial_r - \left( m+\frac{1}{2}\right){\cal M} t^{m-1/2} r
-\left(m^2-\frac{1}{4}\right) {\cal M}'\, t^{m-3/2} r^3
\nonumber \\
Z_n^{(2)} &=& -n t^{n-1} r^2 
\label{gl:1:schext} \\
Z_m^{(1)} &=& -2 t^{m-1/2} r \nonumber \\
Z_n^{(0)} &=& -2 t^n \nonumber 
\EEA
where $\cal M$ and ${\cal M}'$ are (real or complex) 
parameters\footnote{In \cite{Henk02}, the notation $B_{10}={\cal M}/2$ 
and $B_{20}=12{\cal M}'$ was used.} and $x$ is a scaling dimension. The 
non-vanishing commutators of the generators (\ref{gl:1:schext}) read 
\BEA
\left[ X_n, X_{n'} \right] &=& (n-n') X_{n+n'}  
\hspace{0.8truecm} \;\; , \;\;
\left[ X_n, Y_m \right] \:=\: \left(\frac{n}{2}-m\right) Y_{n+m} \;\; , \;\;
\nonumber \\
\left[ Y_m, Y_{m'} \right] &=& \left( m - m'\right) 
\left( 48 {\cal M}'\, Z_{m+m'}^{(2)} + \frac{\cal M}{2} Z_{m+m'}^{(0)} \right) 
\nonumber \\
\left[ X_n, Z_{n'}^{(2)} \right] &=& -n' Z_{n+n'}^{(2)} 
\hspace{1.5truecm} \;\; , \;\;
\left[ Y_m, Z_n^{(2)} \right] \:=\: -n Z_{n+m}^{(1)} 
\label{gl:1:ext} \\
\left[ X_n, Z_m^{(1)} \right] &=& -\left( \frac{n}{2} +m\right) Z_{n+m}^{(1)} 
\;\; , \;\;
\left[ Y_m, Z_{m'}^{(1)} \right] \:=\: - Z_{m+m'}^{(0)} 
\nonumber \\
\left[ X_n , Z_{n'}^{(0)} \right] &=& - n' Z_{n+n'}^{(0)} 
\nonumber 
\EEA
and it can be shown that for $c=0$, this is the maximal extension of 
$\mathfrak{sv}$ through first-order differential operators such that 
the time- and space- translations $X_{-1}, Y_{-1/2}$ and the dilatation
$X_0$ are unmodified compared to (\ref{gl:1:sch1}) \cite{Henk02}. 
For ${\cal M}'=0$, the algebra $\mathfrak{sv}$ is recovered as a subalgebra. 

Rather than proceeding from example to example, it would be valuable to have 
a systematic approach for the construction of infinite-dimensional 
(supersymmetric) extensions of $\mathfrak{sch}_1$.

Generalizing the correspondence between Schr\"odinger- and conformal 
invariance, we shall in this paper introduce a supersymmetric extension of 
the free Schr\"odinger equation in $d=1$ space dimension with two 
super-coordinates (which we call the {\em super-Schr\"odinger model} 
below), whose Lie symmetries form a supersymmetric
extension of the Schr\"odinger algebra that is isomorphic to a semi-direct 
product $\osp(2|2)\ltimes\sh(2|2)$ of an orthosymplectic Lie algebra by a 
super-Heisenberg Lie algebra. We  relate this model to a classical $N=2$
supersymmetric model in $(3+2)$ dimensions, giving  at the same time 
an explicit  embedding of our `super-Schr\"odinger algebra' into $\osp(2|4)$ .  
Note that supersymmetric extensions of the 
Schr\"odinger algebra have been discussed several times in the past 
\cite{Beck86,Beck91,Beck92,Gaun90,Duva94,Ghos01,Ghos04}, some of
them in the context of supersymmetric quantum mechanics.
Here, we consider the problem from a field-theoretical perspective.

We shall also present a systematic 
construction of a family of infinite-dimensional supersymmetric extensions of 
the Schr\"odinger algebra. Our main examples will be  the 
{\em Schr\"odinger-Neveu-Schwarz algebras} 
$\sns^{(N)}$ with $N$ supercharges. The $N=1$ Neveu-Schwarz superalgebra 
\cite{Neve71,Kac98} is recovered as a subalgebra of $\sns^{(1)}$, while 
$\sns^{(0)}$ is the Schr\"odinger-Virasoro algebra $\sv$.

The link between the two parts is given by a realization of the 
infinite-dimensional Lie algebra $\sns^{(2)}$
providing an extension of the realization of $\osp(2|2)\ltimes\sh(2|2)$ as Lie 
symmetries of the super-Schr\"odinger model (see Proposition 4.3).

We begin in section~2 by recalling some useful facts about the 
Schr\"odinger-invariance of the scalar free Schr\"odinger equation and then
give a generalization to its spin-$\half$ analogue, the L\'evy-Leblond 
equation. By considering  the `mass' as an additional variable, we
extend the spinor representation of the Schr\"odinger
algebra $\sch_1$ into a  representation of $\conf_3$. As an application,
we derive the Schr\"odinger-covariant two-point spinorial correlation 
functions. In section~3, we combine the free Schr\"odinger and L\'evy-Leblond
equations (together with a scalar auxiliary field) into a 
{\it  super-Schr\"odinger model}, and show, by using a superfield formalism in 
$3+2$ dimensions, that this model has a kinematic supersymmetry algebra  with 
$N=2$ supercharges. Including then time-inversions, we compute the full 
dynamical symmetry algebra and prove that it
is isomorphic to the Lie algebra of symmetries $\osp(2|2)\ltimes\sh(2|2)$ found 
in several mechanical systems with a finite number of particles.
By treating the `mass' as a coordinate, we  obtain a well-known supersymmetric 
model (see \cite{Fre99}) that we call the  
{\em $(3|2)$-supersymmetric model}. Its dynamical symmetries form the Lie 
superalgebra $\osp(2|4)$.  The derivation of these results is greatly 
simplified through the correspondence with Poisson structures and the 
introduction of several gradings which will be described in detail. 
In section~4,  we use a Poisson algebra formalism to construct for every $N$  
an infinite-dimensional supersymmetric
extension with $N$ supercharges of the Schr\"odinger algebra that we call 
{\em Schr\"odinger-Neveu-Schwarz algebra} and denote by  $\sns^{(N)}$. 
At the same time, we give  an extension of the differential-operator 
representation
of $\osp(2|4)$ into a differential-operator representation of $\sns^{(2)}$.
We compute in section~5 the two-point correlation functions that are covariant 
under $\osp(2|4)$ or under some of its subalgebras. Remarkably, in 
many instances, the requirement of supersymmetric covariance is enough 
to allow only a finite number of possible quasiprimary superfields. 
Our conclusions are given in section~6. 
In appendix~A we present the details for the 
calculation of the supersymmetric two-point functions, whereas in appendix~B, 
we collect for easy reference the numerous Lie superalgebras introduced 
in the paper and their differential-operator realization 
as Lie symmetries of the $(3|2)$-supersymmetric model. 

%%%%%%%%%%%%%%%%%%%%%%%%%%%%%%%%%%%%%%%%%%%%%%%%%%%%%%%%%%%%%%%%%%%%%%%%%%%%%%%%
\section{On the Dirac-L\'evy-Leblond equation}
%%%%%%%%%%%%%%%%%%%%%%%%%%%%%%%%%%%%%%%%%%%%%%%%%%%%%%%%%%%%%%%%%%%%%%%%%%%%%%%%
Throughout this paper we shall use the following notation:  
$[A,B]_{\mp} := AB \mp BA$ stand for the commutator and anticommutator, 
respectively. We shall often simply write $[A,B]$ if it is clear which one
should be understood. Furthermore $\{A,B\}:=
{\partial A\over \partial q}\ {\partial B\over \partial p}-
{\partial A\over \partial p}\ {\partial B\over \partial q}
$ denotes the Poisson bracket or supersymmetric 
extensions thereof which will be introduced below. We shall use the Einstein
summation convention unless explicitly stated otherwise. 

In this section we first recall some properties of the one-dimensional free
Schr\"odinger equation before considering a reduction to a system of 
first-order equations introduced by L\'evy-Leblond \cite{Levy67}.
 
%%==============================================================================
\begin{figure}
\epsfxsize=90mm
\centerline{\epsffile{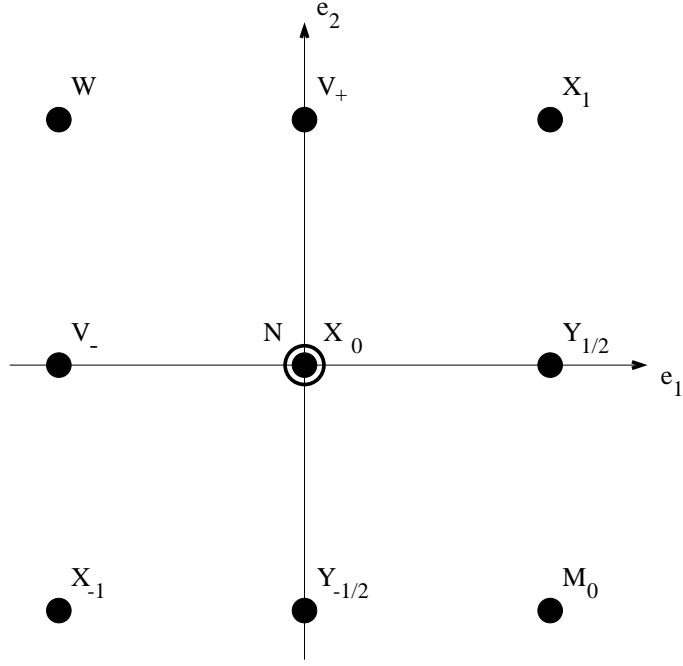}}
\caption[Abbildung 1]{\small Roots of the three-dimensional complexified 
conformal Lie algebra $(\mathfrak{conf}_3)_{\mathbb{C}}\cong B_2$ and 
their relation with the generators of the
Schr\"odinger algebra $\mathfrak{sch}_1$. The double circle in the centre
denotes the Cartan subalgebra $\mathfrak{h}$. 
\label{Abb1}}
\end{figure}
%%==============================================================================

Consider the free  Schr\"odinger or diffusion equation 
\BEQ
{\cal S}\tilde{\phi}= (2{\cal M}\partial_t-\partial_r^2)\tilde{\phi}=0
\EEQ
in one space-dimension, where the
Schr\"odinger operator may be expressed in terms of the generators of
$\mathfrak{sch}_1$ as ${\cal S} := 2 M_0 X_{-1} - {Y}_{-1/2}^2$.  
That the Schr\"odinger algebra realized by (\ref{gl:1:sch1}) 
is indeed a dynamical symmetry of the Schr\"odinger equation if $x=1/2$ can 
be seen from the commutators
\BEQ
{}\left[ {\cal S}, X_1\right] = -2t {\cal S} - 2\left( x-\frac{1}{2}\right) M_0 
\;\; , \;\;
{} \left[ {\cal S}, X_{-1} \right] = \left[ {\cal S}, Y_{-\half}\right] 
%%= \left[ {\cal S}, M_0\right] 
= 0
\EEQ
while the symmetry with respect to the other generators follows from the
Jacobi identities. In many situations, it is useful to go over from the
representation eq.~(\ref{gl:1:sch1}) to another one obtained by
formally considering the `mass' ${\cal M}$ as an additional variable such
that $\tilde{\phi}=\tilde{\phi}_{\cal M}(t,r)$. As a general rule, 
we shall denote in this article by $\zeta$ 
the variable conjugate to $\cal M$ via a
Fourier-Laplace transformation, and the corresponding wave function by the same
letter but without the tilde, here 
$\phi=\phi(\zeta,t,r)$. In this way, one may show that there
is a complex embedding of the Schr\"odinger algebra  into the conformal 
algebra, viz.  
$(\mathfrak{sch}_1)_{\mathbb{C}}\subset(\mathfrak{conf}_{3})_{\mathbb{C}}$ 
\cite{Henk03a}, whereas the representation (\ref{gl:1:sch1}) 
(after a Fourier-Laplace transform) may be extended into the usual 
representation of $\conf_3$ as infinitesimal conformal transformations of
$\R^3$ for a certain choice of coordinates.\footnote{This apparently 
abstract extension becomes important for the explicit calculation of the
two-time correlation function in phase-ordering kinetics \cite{Henk04}.}

We illustrate this for the one-dimensional case $d=1$
in figure~\ref{Abb1}, where the root diagram for 
$(\mathfrak{conf}_3)_{\mathbb{C}}\cong B_2$ is shown. The identification 
with the generators of $\mathfrak{sch}_1$ is clear, see eq.~(\ref{gl:1:sch1}), 
and we also give a name to  the extra conformal generators. In particular, 
$\langle N,X_0\rangle$ form a Cartan subalgebra and the eigenvalue of 
$\ad N$ on any root vector is given by the
coordinate along $-e_1$.\footnote{For example, 
$\ad N(Y_{\half})=[N,Y_{\half}]=-Y_{\half}$ or  
$\ad N(Y_{-\half})=[N,Y_{-\half}]=0$.}  Furthermore, the conformal
invariance of the Schr\"odinger equation follows from $[{\cal S},V_{-}]=0$
\cite{Henk03a}. 

One of the main applications of the (super-)symmetries studied in this article
will be the calculation of covariant correlation functions and we now
define this important concept precisely, generalizing a basic concept
of conformal field-theory \cite{Bela84}. 

\noindent{\bf Definition 1.}
{\it  Let ${\cal L}({\cal H})$ be the set of linear 
operators on a Hilbert space $\cal H$, let 
$R: \mathfrak{g}\to{\cal L}({\cal H})$ be a 
representation of a (super) Lie algebra $\mathfrak{g}$ and  
$R^{(n)}: \mathfrak{g}\to {\cal L}({\cal H})^{\otimes n}$ be
the tensor representation for $n$-particle operators. 
If $\phi_1,\ldots,\phi_n\in {\cal H}$ are fields, then their $n$-point function
$\langle\phi_1\ldots\phi_n\rangle$ may be defined by an averaging function 
$\mathrm{Av}: {\cal H}^{\otimes n} \to\C$ such that 
$\mathrm{Av\,}(\phi_1\otimes\ldots\otimes\phi_n)
=\langle\phi_1\ldots\phi_n\rangle$. 
Then one says that the $n$-point function 
$\langle\phi_1\cdots\phi_n\rangle$ is {\em $\mathfrak{g}$-covariant under the
representation $R$}, if for any generator ${\cal X}\in\mathfrak{g}$
\BEQ
{\mathrm{Av}}\left( R^{(n)}({\cal X}) (\phi_1\otimes\ldots\phi_n)\right)=
 \langle ({\cal R}({\cal X})\phi_1)\phi_2\ldots\phi_n\rangle
+\ldots+\langle\phi_1\ldots\phi_{n-1}({\cal R}({\cal X})\phi_n)\rangle = 0
\EEQ
In this case, the fields $\phi_i$ are called {\em $\mathfrak{g}$-quasiprimary
with respect to $R$}, or simply quasiprimary. }
 
As a specific example, let us consider here  $n$-point functions that are 
covariant under $(\mathfrak{conf}_3)_{\mathbb{C}}$ or one of its Lie 
subalgebras, which for our purposes will be either $\mathfrak{sch}_1$ or the 
{\it parabolic} subalgebra 
\BEQ
\wit{\sch}_1:=\C N\ltimes\sch_1
\EEQ 
(see \cite{Knap86} for the definition of parabolic subalgebras). 
In the extension of the Fourier-Laplace transform of the representation
(\ref{gl:1:sch1})  to $\conf_3$, the generator $N$ is given by 
\BEQ
N=-t\partial_t+\zeta \partial_{\zeta}+\nu
\EEQ
with $\nu=0$ \cite{Henk03a}. But any choice for the value of $\nu$  also gives 
a representation of $\widetilde{\sch}_1$,
although it does not extend to the whole conformal Lie algebra. So one may 
consider more generally $\widetilde{\sch}_1$-quasiprimary fields characterized 
by a scaling exponent $x$ and a {\it $N$-exponent} $\nu$.

It will turn out later to be useful to work with the variable $\zeta$ 
conjugate to $\cal M$.
If we arrange for ${\cal M}=-\half\partial_{\zeta}$ through a Laplace 
transform, see eq.~(\ref{gl:LaplaceTr}) below, it is easy to see
that the $\sch_1$-covariant two-point function under the representation 
(\ref{gl:1:sch1}) is given by \cite{Henk03a}
\BEQ \label{gl:2:cov}
\left\langle \phi_1\phi_2\right\rangle := 
\left\langle \phi_1(\zeta_1,t_1,r_1)\phi_2(\zeta_2,t_2,r_2)\right\rangle =
\psi_0 \delta_{x_1,x_2} (t_1-t_2)^{-x_1} 
f\left( \zeta_1-\zeta_2 +\frac{1}{4}\frac{(r_1-r_2)^2}{t_1-t_2}\right)
\EEQ
where $x_{1,2}$ are the scaling dimensions of the $\sch_1$-quasiprimary fields 
$\phi_{1,2}$, $f$ is an undetermined scaling function and $\psi_0$ 
a normalization constant. If $\phi_{1,2}$ are $\widetilde{\sch}_1$-quasiprimary 
fields with $N$-exponents $\nu_{1,2}$, then $f(u)=u^{-x_1-\nu_1-\nu_2}$. 
If in addition $\psi_{1,2}$ are $\mathfrak{conf}_3$-quasiprimary, 
then $f(u)=u^{-x_1}$ and,
after an inverse Laplace transform, this two-point 
function becomes the well-known heat kernel
$\langle\tilde{\phi}_1 \tilde{\phi}_2\rangle 
= \phi_0 \delta_{x_1,x_2} t^{-x_1} \exp(-r^2/(2{\cal M}t))$,
together with the causality condition $t>0$ \cite{Henk03a}. The same form for 
$\langle\tilde{\phi}_1\tilde{\phi}_2\rangle$ also holds true for 
$\sch_1$-quasiprimary fields, 
since the function $f$ in (\ref{gl:2:cov}) simply gives, after the Laplace 
transform, a mass-dependent normalization constant $\phi_0$.

We now turn to the Dirac-L\'evy-Leblond equations. They were constructed by 
L\'evy-Leblond \cite{Levy67} by adapting to a non-relativistic setting 
Dirac's square-root method for finding a relativistically covariant partial 
differential equation of first order.
Consider in $d$ space dimensions a first-order vector wave equation of the form 
\BEQ
{\cal R} \tilde{\Phi}:= 
\left(A\frac{\partial}{\partial t}
+\sum_{i=1}^d B_i \frac{\partial}{\partial r^i} + {\cal M} C\right)
\tilde{\Phi}=0
\EEQ
where $A,B_i$ and $C$ are matrices to be determined 
such that the square of the operator $\cal R$ is equal to the free 
Schr\"odinger operator ${\cal R}^2 \stackrel{!}{=}{\cal S}=2{\cal M}
\partial_t-\sum_{i=1}^d ({\partial_{r^i}})^2$. It is easy to see that
the matrices $A,B_i,C$ give a representation of a Clifford algebra (with 
an unconventional metric) in $d+2$ dimensions. Namely, if one sets
\BEA
\mathfrak{B}_j &:=& \II \sqrt{2\,} B_j \;\; ; \;\; j=1,\ldots,d
\nonumber \\
\mathfrak{B}_{d+1} &:=& A + \frac{1}{2} C \;\; , \;\;
\mathfrak{B}_{d+2} \: :=\: \II \left( A - \frac{1}{2} C\right)
\EEA
then the condition on ${\cal R}$ is equivalent to 
$[\mathfrak{B}_j,\mathfrak{B}_k]_+ = 2\delta_{j,k}$ for 
$j,k=1,\ldots,d+2$. 

We are interested in the case $d=1$. Then the Clifford algebra generated by 
$\mathfrak{B}_j$, $j=1,2,3$, has exactly one irreducible representation up to 
equivalence, which is given for instance by 
\BEQ
\mathfrak{B}_1=\sigma^3=\left(\begin{array}{cc} 1&0\\0&-1 \end{array}\right)
\;\; , \;\; 
\mathfrak{B}_2=\sigma^2=\left(\begin{array}{cc} 0&\II\\ -\II&0 
\end{array}\right) 
\;\; , \;\;  
\mathfrak{B}_3=\sigma^1=\left(\begin{array}{cc} 0&1\\1&0\end{array}\right).
\EEQ
Then the wave equation 
${\cal R}\tilde{\Phi}={\cal R}
\left(\begin{array}{cc} \tilde{\psi}  \\ \tilde{\phi}\end{array}\right)=0$
becomes explicitly, after a rescaling $r\mapsto \sqrt{2\,} r$
\BEQ \label{gl:LL}
\partial_t \tilde{\psi}=\partial_r \tilde{\phi} \;\; , \;\; 
2{\cal M}\tilde{\phi}= \partial_r \tilde{\psi} 
\EEQ
These are the {\em Dirac-L\'evy-Leblond equations} in one space dimension.

Since the masses $\cal M$ are by physical convention real and positive, it
is convenient to define their conjugate $\zeta$ through a Laplace transform
\BEQ \label{gl:LaplaceTr}
{\psi}(\zeta,t,r) = \int_{0}^{\infty} \!\D{\cal M}\: 
e^{-2{\cal M}\zeta} \tilde{\psi}({\cal M};t,r)
\EEQ
and similarly for $\phi$. Then eqs.~(\ref{gl:LL}) become 
\BEQ \label{gl:LLzeta}
\partial_t {\psi}=\partial_r {\phi}
\;\; , \;\;
\partial_{\zeta}{\phi}=-\partial_r {\psi}.
\EEQ
Actually, it is easy to see that these eqs.~(\ref{gl:LLzeta}) are equivalent 
to the three-dimensional massless free
Dirac equation $\gamma^{\mu}\partial_{\mu} \Phi=0$, where 
$\partial_{\mu}=\partial/\partial \xi^{\mu}$ and $\xi^{\mu}$ with $\mu=1,2,3$
are the coordinates. If we set $t=\half(\xi^1+\II\xi^2)$, 
$\zeta=\half(\xi^1-\II\xi^2)$ 
and finally $r=\xi^3$, and choose the representation 
$\gamma^{\mu}=\sigma^{\mu}$, then we recover indeed eq.~(\ref{gl:LLzeta}). 

The invariance of the free massless Dirac equation under the conformal group 
is well-known.\footnote{The Schr\"odinger-invariance of a free 
non-relativistic particle of spin $S$ is proven in \cite{Hage72}.} 
The generators of $\mathfrak{conf}_3$ act as follows on the spinor field 
$\Phi=\left(\begin{array}{c} \psi\\ \phi\end{array}\right)$, 
(see again figure~\ref{Abb1}) 
\BEA
X_{-1}      &=&  -\partial_t \;\; , \;\; 
Y_{-\half} \:=\: -\partial_r\;\; , \;\;   
M_0        \:=\: \half\partial_{\zeta} 
\nonumber \\
V_- &=& - r\partial_t + 2 \zeta\partial_r
        -\left(\begin{array}{cc} 0&1\\ 0&0 \end{array}\right)
\nonumber \\ 
N &=& -t\partial_t+\zeta\partial_{\zeta}
      +\half\left(\begin{array}{cc} 1 & 0 \\ 0 & -1 \end{array}\right)
\nonumber \\  
Y_{\half} &=& -t\partial_r + \half r\partial_{\zeta}
              -\half\left(\begin{array}{cc} 0&0\\ 1&0 \end{array}\right) 
\nonumber \\
X_0 &=& -t\partial_t - \half r\partial_r -\half\left( \begin{array}{cc}
x & 0 \\ 0 & x+1 \end{array} \right) 
\label{gl:2:sch2}\\
%D &=& -t\partial_t-\zeta\partial_{\zeta}-r\partial_r
%-\half\left(\begin{array}{cc} 2x+1 & 0 \\ 0 & 2x+1\end{array}\right) \\
W &=& -\half r^2\partial_t+2\zeta^2\partial_{\zeta}
      +2\zeta r\partial_r
      +\left(\begin{array}{cc} 2(x+1)\zeta & -r\\ 0 &  2x \zeta 
      \end{array}\right)
\nonumber \\
V_+&=& -tr\partial_t -\zeta r\partial_{\zeta}-\half(r^2-4\zeta t)\partial_r
       -\half\left(\begin{array}{cc} (2x+1)r & 2t\\ -2\zeta & (2x+1)r 
       \end{array}\right) 
\nonumber \\
X_1 &=& -t^2\partial_t-tr\partial_r + \frac{1}{4}r^2\partial_{\zeta}
        -\left(\begin{array}{cc} x t & 0 \\ r/2 & (x+1) t 
	\end{array}\right)
\nonumber
\EEA
For solutions of (\ref{gl:LLzeta}), one has $x=\half$. 
As in the case of the scalar representation (\ref{gl:1:sch1}), 
arbitrary values of the scaling exponent $x$ also give a
representation of the conformal algebra.

There are three `translations' ($X_{-1}, Y_{-\half}, M_{0}$), three 
`rotations' ($V_{-}, N, Y_{\half}$), one `dilatation' ($X_0$) and three
`inversions' or special transformations ($W, V_+, X_1$). It is sometimes
useful to work with the generator $D:= 2X_0 -N$ (whose differential part is 
the Euler operator
$-t\partial_t-r\partial_r-\zeta\partial_{\zeta}$) instead of either $X_0$ or
$N$. We also see that the individual components $\psi,\phi$ of the spinor $\Phi$
have scaling dimensions $x_{\psi}=x$ and $x_{\phi}=x+1$, respectively. 
If we write the Dirac operator as
\BEQ
{\cal D}=\frac{1}{\II} {\cal R} = \left( \begin{array}{cc}
\partial_r  & \partial_{\zeta} \\
\partial_t & -\partial_r \end{array} \right)
\EEQ
then the Schr\"odinger- and also the full conformal invariance of the 
Dirac-L\'evy-Leblond equation ${\cal D}\Phi=0$ follows from the commutators
\BEA
{}\left[ {\cal D}, X_1\right] &=& - t{\cal D} -\left(x-\half\right) 
\left( \begin{array}{cc} 0 & 0 \\ 0 & 1\end{array} \right) 
\nonumber \\
{}\left[ {\cal D}, X_{-1} \right] &=& \left[ {\cal D}, Y_{-\half}\right] \:=\: 
%%\left[ {\cal D}, M_0\right] \:=\: 
\left[ {\cal D}, V_{-} \right] \:= \: 0
\EEA
It is clear that  dynamical symmetries of the Dirac-L\'evy-Leblond equation 
are obtained only if $x=\half$. Since %%$X_1,X_{-1},Y_{-\half},M_0,V_-$ 
$X_1,X_{-1},Y_{-\half},V_-$ generate $(\mathfrak{conf}_3)_{\mathbb{C}}$, 
as can be seen from the root 
structure represented in figure~\ref{Abb1}, the symmetry under the remaining 
generators of $(\mathfrak{conf}_3)_{\mathbb{C}}$ follows from the Jacobi 
identities.

Let $\Phi_i =\left(\vekz{\psi_i}{\phi_i}\right)$ , $i=1,2$ be two quasiprimary 
spinors under the representation (\ref{gl:2:sch2}) of either 
$\sch_1,\widetilde{\sch}_1$ or $\conf_3$, with  scaling
dimensions $\left(\vekz{x_i}{x_i+1}\right)$ of the component fields. 
We now consider the covariant 
two-point functions; from translation-invariance it is
clear that these will only depend on $\zeta=\zeta_1-\zeta_2$, 
$t=t_1-t_2$ and $r=r_1-r_2$. 

\noindent {\bf Proposition 2.1.} {\it Suppose $\Phi_1$, $\Phi_2$ are
quasiprimary spinors under the representation (\ref{gl:2:sch2}) of 
$\widetilde{\sch}_1$. Then their two-point
functions vanish unless $x_1=x_2$ or $x_1=x_2\pm 1$, in which case 
they read ($\phi_0,\psi_0$ are normalization constants)}

\noindent {\it (i) if $x_1=x_2$, then}
\BEA
\langle \psi_1 \psi_2\rangle &=& \psi_0 t \left(4\zeta t+r^2\right)^{-x_1-1}
\nonumber \\
\langle \psi_1 \phi_2\rangle &=& \langle \phi_1 \psi_2\rangle 
\: = \: -\half \psi_0 r \left(4\zeta t+r^2\right)^{-x_1-1}
\label{gl:Fall1}\\
\langle \phi_1 \phi_2 \rangle &=& \frac{\psi_0}{4} \frac{r^2}{t} 
\left(4\zeta t+r^2\right)^{-x_1-1} + 
\phi_0 \frac{1}{t} \left(4\zeta t+r^2\right)^{-x_1}
\nonumber
\EEA
\noindent {\it (ii) if $x_1=x_2+1$, then} 
\BEA
\langle \psi_1 \psi_2 \rangle &=& \langle \phi_1 \psi_2\rangle \:=\: 0
\nonumber \\
\langle \psi_1 \phi_2\rangle &=& \psi_0 \left(4\zeta t+r^2\right)^{-x_1}
\label{gl:Fall2}\\ 
\langle \phi_1 \phi_2\rangle &=& -\frac{\psi_0}{2} \frac{r}{t} 
\left(4\zeta t+r^2\right)^{-x_1}
\nonumber
\EEA
{\it The case $x_1=x_2-1$ is obtained by exchanging $\Phi_1$ with $\Phi_2$.} 

For brevity, the arguments of the two-point functions in 
eqs.~(\ref{gl:Fall1},\ref{gl:Fall2}) were suppressed. Let us emphasize that
the scaling dimensions of the component fields  with a standard Schr\"odinger 
form eq.~(\ref{gl:2:cov}) 
($\langle\psi_1\psi_2\rangle$ and $\langle\phi_1\phi_2\rangle$ 
in eq.~(\ref{gl:Fall1}), and $\langle\psi_1\phi_2\rangle$ in 
eq.~(\ref{gl:Fall2})) must agree, which is not the case for the other 
two-point functions 
which are obtained from them by applying derivative operators. 

On the other hand, the covariance under the whole conformal group implies the 
supplementary constraint $x_1=x_2$ (equality of the scaling exponents), and we 
have  

\noindent {\bf Proposition 2.2.} {\it The non-vanishing two-point functions, 
$(\mathfrak{conf}_3)_{\mathbb{C}}$-covariant under the representation 
(\ref{gl:2:sch2}), of the fields
$\psi$ and $\phi$ are obtained from eq.~(\ref{gl:Fall1}) with $x_1=x_2$ and 
the extra condition $\phi_0=-\psi_0/4$, which gives}
\BEQ
\langle \phi_1\phi_2\rangle = -\psi_0 \zeta \left(4\zeta t+r^2\right)^{-x_1-1}
\EEQ

\noindent {\bf Proof:} In proving these two propositions, we merely outline the 
main ideas since the calculations are straightforward. We begin with 
Proposition 2.1. Given the obvious
invariance under the translations, we first consider the invariance under the
special transformation $X_1$ and use the form (\ref{gl:2:sch2}). 
With the help of dilatation invariance ($X_0$) and
Galilei-invariance ($Y_{\half}$) this simplifies to
\BEA
(x_1-x_2)t\langle \psi_1\psi_2\rangle &=& 0 \;\; , \;\;
(x_1-x_2-1)t\langle\psi_1\phi_2\rangle-\half r\langle\psi_1\psi_2\rangle\:=\: 0 
\;\; ,\;\;
\nonumber \\
(x_1-x_2+1)t\langle\phi_1\psi_2\rangle+\half r\langle\psi_1\psi_2\rangle&=& 0 
\;\; ,\;\;
(x_1-x_2)t\langle\phi_1\phi_2\rangle
+\half r\langle\psi_1\phi_2\rangle-\half r\langle\phi_1\psi_2\rangle\:=\:0
\nonumber
\EEA
Considering  the first of these equations leads us to distinguish two cases: 
either (i) $x_1=x_2$ or (ii) $\langle\psi_1\psi_2\rangle=0$. 

In the first case, we get from the remaining three equations
\BD
\langle\psi_1\phi_2\rangle=\langle\phi_1\psi_2\rangle=
-\frac{r}{2t}\langle\psi_1\psi_2\rangle
\ED
and the covariance under $Y_{\half}$, $N$ and $X_0$, respectively, leads to
the following system of equations
\BEA
\left( -t\frac{\partial}{\partial r}+\frac{r}{2}\frac{\partial}{\partial\zeta}
\right)\langle\psi_1\psi_2\rangle &=& 0 
\;\; , \;\;
\left( -t\frac{\partial}{\partial t}+\zeta\frac{\partial}{\partial \zeta}
+1 \right) \langle\psi_1\psi_2\rangle \:=\: 0 
\;\; , \;\;
\nonumber \\
\left( -t\frac{\partial}{\partial t}-\frac{r}{2}\frac{\partial}{\partial r}
-x_1\right)\langle\psi_1\psi_2\rangle &=& 0 
\nonumber
\EEA
with a  unique solution (up to a multiplicative factor) given by the first  
line of eq.~(\ref{gl:Fall1}).
Similarly, covariance under the same three generators leads to a
system of three linear inhomogeneous equations for $\langle\phi_1\phi_2\rangle$
whose general solution is also given in eq.~(\ref{gl:Fall1}). 

In the second case, the remaining conditions coming from $X_1$ are
\BEA
(x_1-x_2-1)t \langle \psi_1\phi_2\rangle \:=\: 0 \;\; , \;\;
(x_1-x_2+1)t \langle \phi_1\psi_2\rangle &=& 0 \;\; , \;\;
\nonumber \\ 
(x_1-x_2)t \langle \phi_1\phi_2\rangle 
+\frac{r}{2}\left(\langle \psi_1\phi_2\rangle 
- \langle \phi_1\psi_2\rangle\right) &=& 0
\nonumber
\EEA
and one of the conditions $x_1=x_2\pm 1$ must hold true. Supposing that 
$x_1=x_2+1$, we get 
$\langle\phi_1\phi_2\rangle=-\half (r/t)\langle \psi_1\phi_2\rangle$ and 
an analogous relation holds (with the first and second
field exchanged) in the other case. Again, covariance under $Y_{\half},N,X_0$
leads to a system of three linear equations for $\langle\psi_1\phi_2\rangle$ 
whose general solution in given in eq.~(\ref{gl:Fall2}). 

To prove Proposition 2.2, it is now sufficient to verify covariance under the
generator $V_-$. Direct calculation shows that eq.~(\ref{gl:Fall1}) is 
compatible with this condition only if $\phi_0=-\psi_0/4$.  On the other
hand, compatibility with the second case eq.~(\ref{gl:Fall2}) requires
that $\psi_0=0$. \eop

\noindent {\bf Remark:} If we come back to the original fields 
$\tilde{\psi},\tilde{\phi}$ by inverting the Laplace transform 
(\ref{gl:LaplaceTr}), the $\wit{\sch}_1$-covariant two-point 
functions of eq.~(\ref{gl:Fall1}) take the form
\BEA
\langle \tilde{\psi}_1\tilde{\psi}_2\rangle 
&=& \psi_0' \left(\frac{\cal M}{t}\right)^{x_1}
\exp\left(-\frac{{\cal M}}{2}\frac{r^2}{t}\right) 
\nonumber \\
\langle \tilde{\psi}_1\tilde{\phi}_2\rangle &=& 
\langle \tilde{\phi}_1\tilde{\psi}_2\rangle \:=\: 
- \psi_0' \frac{r}{2t} \left(\frac{\cal M}{t}\right)^{x_1}
\exp\left(-\frac{{\cal M}}{2}\frac{r^2}{t}\right) 
\\
\langle \tilde{\phi}_1\tilde{\phi}_2\rangle &=& \frac{\psi_0'}{4}\frac{r^2}{t} 
\left(\frac{\cal M}{t}\right)^{x_1}
\exp\left(-\frac{{\cal M}}{2}\frac{r^2}{t}\right) 
+ \phi_0' \left(\frac{\cal M}{t}\right)^{x_1 -1}
\exp\left(-\frac{{\cal M}}{2}\frac{r^2}{t}\right)
\nonumber
\EEA
where $\psi_0'=\psi_0/(\Gamma(x_1+1)2^{x_1+1})$, 
$\phi_0'=\phi_0/(\Gamma(x)2^{x_1})$, and $\Gamma(x)$ is the Gamma function.  

\noindent {\bf Proposition 2.3.} {\it 
(i) Let $f$ be a solution of the Laplace-transformed Schr\"odinger equation 
$(\partial_{\zeta}\partial_t +\partial_r^2)f=0$.
Then $\Phi=\left(\vekz{\psi}{\phi}\right):=
\left(\begin{array}{c} -\partial_{\zeta}f\\ \partial_r f\end{array}\right)$
satisfies the Dirac-L\'evy-Leblond equations (\ref{gl:LLzeta}). \\
(ii) Suppose that $f_1,f_2$ are $\widetilde{\sch}_1$-quasiprimary fields 
with scaling exponents
$x=x_1=x_2$ and $N$-exponents $\nu_1=\nu_2=-\half$, and let
$\Phi_i:=\left(\begin{array}{c} -\partial_{\zeta}f_i \\ \partial_r f_i 
\end{array}\right)$. Then the covariant two-point function 
$$\langle f_1 f_2\rangle=t^{-x} (\zeta+r^2/4t)^{1-x}$$
implies a particular case of eq.~(\ref{gl:Fall1}), given by 
$\psi_0=-x(x-1)2^{2x+2}$ and $\phi_0=(x-1) 2^{2x-1}$.}

Both assertions are easily checked by straightforward calculations. 

\noindent {\bf Remark:} In the case (i) of Proposition 2.3, the correspondence
$\Phi_i =\left(\vekz{\psi_i}{\phi_i}\right)
=\left(\vekz{-\partial_{\zeta}f_i}{\partial_r f_i}\right)$ induces from
(\ref{gl:1:sch1}) a representation of the Schr\"odinger group on the fields
$\phi_i,\psi_i$ in terms of integro-differential operators. 
It is at first sight
not obvious that the two-point function of spinors that are quasiprimary
under (\ref{gl:2:sch2}) should be derived from $\langle f_1 f_2\rangle$ in 
such  a simple way.

%%%%%%%%%%%%%%%%%%%%%%%%%%%%%%%%%%%%%%%%%%%%%%%%%%%%%%%%%%%%%%%%%%%%%%%%%%%%%%%%
\section{Supersymmetry in three dimensions and supersymmetric 
Schr\"odinger-invariance}
%%%%%%%%%%%%%%%%%%%%%%%%%%%%%%%%%%%%%%%%%%%%%%%%%%%%%%%%%%%%%%%%%%%%%%%%%%%%%%%%

\subsection{From $N=2$ supersymmetry to the super-Schr\"odinger equation}

We begin by recalling the construction of super 
space-time \cite[Lectures 3 \& 4]{Fre99}. 
Take as $n$-dimensional space-time $\R^n$ 
(or, more generally, any $n$-dimensional Lorentzian manifold). 
One has a quite general construction
of (non-supercommutative) superspace-time $\R^{n|s}$, with
$s$ odd coordinates, as the exponential of the Lie superalgebra $V\oplus S$, 
where the even part $V$ is an $n$-dimensional
vector space, and the $s$-dimensional odd part $S$ is
a spin representation of dimension $s$ of Spin$(n-1,1)$, provided with 
non-trivial Lie super-brackets $(f_1,f_2)\in S\times S
\mapsto [f_1,f_2]_+\in V$ which define a Spin$(n-1,1)$-equivariant  pairing
$\Gamma : {\rm Sym}^2(S)\to V$ from symmetric two-tensors on $S$ into $V$ 
(see \cite{Fre99}, Lecture 3).  Super-spacetime
$\R^{n|s}$ can then be extended in a natural way into the exponential of the 
super-Poincar\'e algebra $(\spin(n-1,1)\ltimes V)\oplus S$, with
the canonical action of $\spin(n-1,1)$ on $V$ and on $S$.

Let us make this construction explicit in space-time dimension $n=3$,  which
is the only case that we shall study in this paper. Then the minimal spin 
representation is two-dimensional, so we consider super-spacetime $\R^{3|2}$ 
with two odd coordinates $\theta=(\theta^1,\theta^2)$. 
We shall denote by $D_{\theta^a}$, $a=1,2$, the associated
left-invariant derivatives, namely, the left-invariant super-vector 
fields that coincide
with $\partial_{\theta^1},\partial_{\theta^2}$ when $\theta^1,\theta^2\equiv 0$.
Consider $\R^2$ with the coordinate vector fields 
$\partial_{y^1}, \partial_{y^2}$ and the associated symmetric two-tensors with 
components $\partial_{y^{ij}}$, $i,j=1,2$. These form a three-dimensional 
vector space with natural coordinates $y=(y^{11},y^{12},y^{22})$ defined by
\BEQ \label{gl:3:3.1}
\left[ \partial_{y^{cd}}, y^{ab} \right]_{-} := \delta_{ca}\delta_{db} +
\delta_{cb}\delta_{da}
\EEQ
Then define the map $\Gamma$ introduced above to be
\BEQ
\Gamma(\partial_{\theta^a},\partial_{\theta^b}) :=\partial_{y^{ab}}
\EEQ 
Hence, one has the simple relation 
$[D_{\theta^a},D_{\theta^b}]_+=\partial/\partial y^{ab}$ for the odd generators 
of $\R^{3|2}$. So, by the Campbell-Hausdorff formula,  
\BEQ 
D_{\theta^a}=\partial_{\theta^a}+\theta^b\partial_{y^{ab}}.
\EEQ
In this particular case, $\spin(2,1)\cong\slin(2,\R)$. The usual 
action of $\gl(2,\R)\supset\spin(2,1)$ on $\R^2$ is given by the two-by-two 
matrices $E_{ab}$ such that $E_{ab}\partial_{y^c}=-\delta_{ac} \partial_{y^b}$ 
and extends naturally to the following action on symmetric 2-tensors 
\BEQ
E_{ab}\partial_{y^{cd}}=-\del_{ac}\partial_{y^{bd}}-\del_{ad}\partial_{y^{cb}}
\:=\: \left[y^{a\bar{a}} \partial_{y^{\bar{a}b}},\partial_{y^{cd}}\right]_{-},
\EEQ
so $E_{ab}$ is represented by the vector field on $V\oplus S$
\BEQ
E_{ab}= y^{a\bar{a}} \partial_{y^{\bar{a}b}}+\theta^a \partial_{\theta^b}
\EEQ
One may verify that the adjoint action of $E_{ab}$ on the left-covariant 
derivatives is given by the usual matrix action, namely, 
$[E_{ab},D_{\theta^c}]=-\del_{ac}D_{\theta^b}$.

Consider now a superfield $\Phi(y^{11},y^{12},y^{22};\theta^1,\theta^2)$: 
we introduce the Lagrangian density
\BEQ
{\cal L}(\Phi)=\half \eps^{ab} (D_{\theta^a}\Phi)^* (D_{\theta^b} \Phi)
\EEQ
where $\eps^{ab}$ is the totally antisymmetric two-tensor 
defined by $\eps^{12}=-\eps^{21}=1,\ \eps^{11}
=\eps^{22}=0.$ 
It yields the equations of motion
\BEQ \label{gl:eq_mouvement}
\eps^{ab} D_{\theta^a} D_{\theta^b} \Phi=(D_{\theta^1} D_{\theta^2}
-D_{\theta^2} D_{\theta^1})\Phi=0.
\EEQ
This equation is invariant under even translations $\partial_{y^{ab}}$, and 
under right-invariant super-derivatives
\BEQ
\bar{D}_{\theta^a}=\partial_{\theta^a}-\theta^b\partial_{y^{ab}}
\EEQ
since these anticommute with the $D_{\theta^a}$. Furthermore,
the Lagrangian density is multiplied by $\det(g)$ under the 
action of $g\in \GL(2,\R)$, hence all elements in $\gl(2,\R)$  
leave equation (\ref{gl:eq_mouvement}) invariant. 

Note that the equations of motion are also invariant under the left-invariant 
super-derivatives $D_{\theta^a}$ since
these commute with the coordinate vector fields $\partial_{y^{bc}}$
(this is true for flat space-time manifolds only).

All these translational and rotational symmetries
form by linear combinations a Lie superalgebra that we shall call (in the 
absence of any better name)
the {\it `super-Euclidean Lie algebra of $\R^{3|2}$'}, 
and denote by $\se(3|2)$, viz.
\BEQ
\se(3|2) = \left\langle \partial_{y^{ab}}, D_{\theta^a}, \bar{D}_{\theta^a}, 
E_{ab}; a,b\in\{1,2\} \right\rangle
\EEQ 
We shall show later that it can be included in a larger Lie super-algebra
which is more interesting for our purposes.

Let us look at this more closely by using proper coordinates. 
The vector fields $\partial/\partial y^{ij}$ 
are related to the physical-coordinate vector fields by
\BEQ \label{gl:3:trzeta}
\frac{\partial}{\partial t} = \frac{\partial}{\partial y^{11}} \;\; , \;\;
\frac{\partial}{\partial r} = \frac{\partial}{\partial y^{12}} \;\; , \;\;
\frac{\partial}{\partial \zeta} = \frac{\partial}{\partial y^{22}}
\EEQ
hence by eq.~(\ref{gl:3:3.1}) we have 
$t=2y^{11}, r=y^{12}, \zeta=2y^{22}$. We set 
\BEQ \label{gl:3:superc}
\Phi(\zeta,t,r;\theta^1,\theta^2)=f(\zeta,t,r)+\theta^1\phi(\zeta,t,r)
+\theta^2 \psi(\zeta,t,r) + \theta^1\theta^2 g(\zeta,t,r).
\EEQ
Then the left-invariant superderivatives read 
\BEQ
D_{\theta^1}=\partial_{\theta^1}+\theta^1 \partial_t+\theta^2 \partial_r, \
D_{\theta^2}=\partial_{\theta^2}+\theta^1 \partial_r+\theta^2 \partial_{\zeta}.
\EEQ
The equations of motion (\ref{gl:eq_mouvement}) become 
\BEQ
\left(\partial_{\theta^1}\partial_{\theta^2}
+\theta^1\theta^2(\partial_{\zeta}\partial_t-\partial_r^2)+\theta^1
(\partial_{\theta^2}\partial_t-\partial_{\theta^1}\partial_r)
+\theta^2(\partial_{\theta^2}\partial_r-\partial_{\theta^1}
\partial_{\zeta}) \right)\Phi =0 
\EEQ
which yields the following system of equations in the coordinate fields:
\begin{eqnarray}
\label{gl:SSzeta}
g &=& 0  \nonumber \\
\partial_r \phi &=& \partial_t \psi
\;\; , \;\; 
\partial_r \psi = \partial_{\zeta}\phi \nonumber\\
(\partial_r^2-\partial_{\zeta}\partial_t)f &=&0.
\end{eqnarray}
We shall call this system the {\it $(3|2)$-supersymmetric model}.
{}From the two equations in the second line of (\ref{gl:SSzeta}) we recover 
the Dirac-L\'evy-Leblond equations (\ref{gl:LLzeta}) after the 
change of variables $\zeta\mapsto -\zeta$.

The equations (\ref{gl:SSzeta}) may be obtained in turn from the action
\BEQ
S=\int \!\D\zeta\ \D t\ \D r\ \D\theta^2\ \D\theta^1\ {\cal L}(\Phi)
=\int \!\D\zeta\ \D t\ \D r\ L(f,\phi,\psi,g)
\EEQ
where 
\BEQ
L(f,\phi,\psi,g)=f^* (\partial_{\zeta}\partial_t-\partial_r^2)f
                +\phi^* (\partial_t \psi-\partial_r \phi)
		+\psi^* (\partial_{\zeta}\phi-\partial_r \psi)
		+g^* g.
\EEQ
Now consider the field $\Phi=(f,\psi,\phi,g)$ as the Laplace transform 
${\Phi}=\int\!\D{\cal M}\,e^{2{\cal M}\zeta}\,\tilde{\Phi}_{\cal M}$ 
of the field $\tilde{\Phi}_{\cal M}$ with respect to $\zeta$, 
so that the derivative operator 
$\partial_{\zeta}$ corresponds to the multiplication by twice the 
mass coordinate 
$2{\cal M}$. The equations of motion (\ref{gl:SSzeta}) then read as follows:
\begin{eqnarray}
\label{gl:SS}
\tilde{g} &=& 0 \nonumber \\
\partial_r \tilde{\phi} &=& \partial_t \tilde{\psi}
\;\; , \;\;  \partial_r \tilde{\psi}=2{\cal M}\tilde{\phi}
\nonumber \\
(\partial_r^2-2{\cal M} \partial_t)\tilde{f} &=&0
\end{eqnarray}
We shall refer to equations (\ref{gl:SS}) as 
the {\it super-Schr\"odinger model}.

In this context, $g$ or $\tilde{g}$ can be interpreted as an auxiliary field, 
while $(\psi,\phi)$ is a spinor field satisfying the Dirac equation
in (2+1) dimensions (\ref{gl:LLzeta}) and its inverse Laplace transform 
$(\tilde{\psi},\tilde{\phi})$ satisfies the Dirac-L\'evy-Leblond
equation in one space dimension, see (\ref{gl:LL}), and $\tilde{f}$ 
is a solution of the free  Schr\"odinger equation in one space dimension.

%%++++++++++++++++++++++++++++++++++++++++++++++++++++++++++++++++++++++++++++++
\begin{table}[t]
\caption{\small Defining equations of motion of the supersymmetric models. 
The kinematic and dynamic symmetry algebras (see the text for the definitions) 
are also listed. \label{tab0}}
\begin{center}
\begin{tabular}{|l|cc|} \hline
model & $(3|2)$-supersymmetric & super-Schr\"odinger \\ \hline
      & $g=0$                  & $\tilde{g}=0$       \\
      & $\partial_r\phi=\partial_t\psi$ & 
        $\partial_r\tilde{\phi}=\partial_t\tilde{\psi}$ \\
      & $\partial_r\psi=\partial_{\zeta}\phi$ & 
        $\partial_r\tilde{\psi}=2{\cal M}\tilde{\phi}$ \\
      & $(\partial_r^2-\partial_{\zeta}\partial_t)f=0$ & 
        $(\partial_r^2-2{\cal M}\partial_t)\tilde{f}=0$ \\ \hline
kinematic algebra & $\se(3|2)$ & $\mathfrak{sgal}$ \\ \hline
dynamic algebra   & $\s^{(2)}\cong\osp(2|4)$ & 
                    $\tilde{\s}^{(2)}\cong \osp(2|2)\ltimes\mathfrak{sh}(2|2)$
                    \\ \hline 
\end{tabular} \end{center}
\end{table}
%%++++++++++++++++++++++++++++++++++++++++++++++++++++++++++++++++++++++++++++++

Let us now study the kinematic Lie symmetries of the $(3|2)$-supersymmetric 
model (\ref{gl:SSzeta}) and of the
super-Schr\"odinger model (\ref{gl:SS}). For convenience, we collect their 
definitions in table~\ref{tab0}. By definition, {\em kinematic} symmetries are
(super)-translations and (super-)rotations, and also scale transformations, 
that leave invariant
the equations of motion. 
Generally speaking, the kinematic Lie 
symmetries of the super-Schr\"odinger model contained in $\mathfrak{sgal}$ 
correspond to those symmetries of the $(3|2)$ supersymmetric model such that  
the associated vector 
fields do not depend on the coordinate $\zeta$, in other words which leave the 
`mass' invariant. Below, we shall also consider the so-called {\em dynamic}
symmetries of the two free-field models which  arise when also
inversions $t\mapsto -1/t$ are included, and form a strictly 
larger Lie algebra. We anticipate on  later results and already
include the dynamic algebras in table~\ref{tab0}. 

Let us summarize the results obtained so far
on the kinematic symmetries of the two supersymmetric models.
\newpage \typeout{ *** Seitenvorschub vor Prop. 3.1 ***}
\noindent {\bf Proposition 3.1} {\it 
\begin{enumerate}
\item {\it The Lie algebra of kinematic Lie symmetries of the
$(3|2)$-supersymmetric model (\ref{gl:SSzeta}) contains a subalgebra which is
isomorphic to $\se(3|2)$. The Lie algebra
$\se(3|2)$ has dimension $11$, and a basis of $\se(3|2)$ in its realization as 
Lie symmetries is given by the following generators. 
There are the three even translations}
\BD
X_{-1}\;\; , \;\; Y_{-\half}\;\; , \;\; M_0 
\ED
{\it the four odd translations}
\BD 
G_{-\half}^1=-\half(D_{\theta^1}+\bar{D}_{\theta^1})\;\; , \;\;
G_{-\half}^2=-\half(D_{\theta^1}-\bar{D}_{\theta^1})\;\; , \;\;
\bar{Y}_0^1=-\half(D_{\theta^2}+\bar{D}_{\theta^2})\;\; , \;\;
\bar{Y}_0^2=-\half(D_{\theta^2}-\bar{D}_{\theta^2})
\ED
{\it and the four  generators in  $\gl(2,\R)$}
\BD
Y_{\half}=-\half E_{12} \;\; , \;\; 
X_0=-\half E_{11}-\frac{x}{2} \;\; , \;\; 
D=-\half(E_{11}+E_{22})-x \;\; ,\;\; 
V_-=-\half E_{21}
\ED
{\it An explicit realization in terms of differential operators is}
\BEA
X_{-1} &=& -\partial_t \;\; , \;\;
Y_{-\half}\:=\:-\partial_r\;\; ,\;\; 
M_0\:=\:-\half\partial_{\zeta}
\nonumber \\
G_{-\half}^1 &=& -\partial_{\theta^1} \;\; ,\;\;
G_{-\half}^2 \:=\: -\theta^1 \partial_t - \theta^2\partial_r
\nonumber \\
\bar{Y}_0^1 &=& -\partial_{\theta^2} \;\; ,\;\; 
\bar{Y}_0^2 \:=\: -\theta^1 \partial_r-\theta^2\partial_{\zeta}
\nonumber \\
Y_{\half} &=&-t\partial_r-\half r \partial_{\zeta}
             -\half \theta^1\partial_{\theta^2}
\label{gl:3:se32}\\
X_0 &=&-t\partial_t-\half(r\partial_r+\theta^1\partial_{\theta^1})-\frac{x}{2}
\nonumber \\
D&=&-t\partial_t-r\partial_r-\zeta\partial_{\zeta}
-\half(\theta^1\partial_{\theta^1}+\theta^2\partial_{\theta^2})-x
\nonumber \\
V_- &=&-\zeta\partial_r-\half r\partial_t-\half\theta^2\partial_{\theta^1}.
\nonumber
\EEA
{\it Here a scaling dimension $x$ of the superfield $\Phi$ has been added such 
that for $x=1/2$ the generators  $X_0$ and $D$ (which correspond to the action 
of non trace-free elements of $\gl(2,\R)$) leave invariant the Lagrangian 
density. By changing the value of $x$ one finds another realization 
of $\se(3|2)$.}

\item {\it The `super-galilean' Lie subalgebra $\sgal\subset\se(3|2)$  of
symmetries of the super-Schr\"odinger model
(\ref{gl:SS}) is 9-dimensional. Explicitly}
\BEQ
\mathfrak{sgal} = \left\langle X_{-1,0}, Y_{\pm \half}, M_0, G_{-\half}^{1,2},
\bar{Y}_0^{1,2}\right\rangle
\EEQ
\end{enumerate}
}
We stress the strong asymmetry between the two odd coordinates $\theta^{1,2}$ 
as they appear
in the dilatation generator $X_0$. This is a consequence of our identification 
$X_0=-\half E_{11}-\half x$, which is dictated by the requirement that the 
system exhibit 
a non-relativistic behaviour with a dynamic exponent $z=2$.
As we shall show in section~5, this choice will have important consequences
for the calculation of covariant two-point functions. In comparison, in 
relativistic systems with an extended ($N=2$) supersymmetry (see e.g. 
\cite{Dola02,Nagi05b,Park00}), one needs a dynamic exponent $z=1$. In our
notation, the generator $D$ would then be identified as the generator of 
dilatations, leading to a complete symmetry between $\theta^1$ and $\theta^2$.  

The supersymmetries of the free non-relativistic particle with a fixed mass 
have been discussed by Beckers {\it et al.} long ago \cite{Beck86,Beck92} and, 
as we shall recall in subsection~3.3, $\mathfrak{sgal}$ is a subalgebra 
of their dynamical algebra $\osp(2|2)\ltimes \mathfrak{sh}(2|2)$. 

Let us give the Lie brackets of these generators for convenience, and also 
for later use. The three generators $(X_{-1},Y_{-\half},M_0)$ commute with 
all translations, even or odd. The commutators
of the odd translations yield four non-trivial relations:
\BEA
[G_{-\half}^1,G_{-\half}^2]_+ &=& -X_{-1} \;\; , \;\; 
[\bar{Y}_0^1,\bar{Y}_0^2]_+\:=\:-2M_0
\nonumber \\
{}[G_{-\half}^1,\bar{Y}_0^2]_+&=&[G_{-\half}^2, \bar{Y}_0^1]_+\:=\: 
-Y_{-\half}.
\EEA
The rotations act on left- or right-covariant odd derivatives by the 
same formula
\BEQ
[E_{ab},D_{\theta^c}]=-\del_{ac}D_{\theta^b} \;\; , \;\;
[E_{ab},\bar{D}_{\theta^c}]=-\del_{ac}\bar{D}_{\theta^b}, 
\EEQ 
which gives in our basis
\BEA
[X_0,G_{-\half}^{1,2}] &=& \half G_{-\half}^{1,2}
\;\; , \;\;  [X_0,\bar{Y}_0^{1,2}]\:=\:0 
\nonumber \\
{}[Y_{\half},G_{-\half}^{1,2}]&=&\half \bar{Y}_0^{1,2} \;\; , \;\; 
[Y_{\half},\bar{Y}_0^{1,2}]\:=\:0
\nonumber \\
{}[V_-,G_{-\half}^{1,2}]&=& 0 \;\; , \;\; \hspace{0.8truecm}
[V_-,\bar{Y}_0^{1,2}]\:=\: \half G_{-\half}^{1,2}.
\EEA
Finally,  the commutators of elements in $\gl(2,\R)$ may be computed
by using the usual bracket of matrices,
and brackets between elements in $\gl(2,\R)$ and even translations are obvious.

\subsection{Dynamic symmetries of the super-Schr\"odinger model}

Let us consider the symmetries of the super-Schr\"odinger model, starting 
{}from the $9$-dimensional Lie algebra of symmetries $\mathfrak{sgal}$ 
that was introduced in Proposition~3.1.
This Lie algebra may be enlarged by adding the generator
\BEQ \label{gl:3:N0}
N_0=-\theta^1\partial_{\theta^1}-\theta^2\partial_{\theta^2}+x
\EEQ
(Euler operator on odd coordinates), together with three special 
transformations $X_1,G_{\half}^{1,2}$
that will be defined shortly. First notice
that the operators 
\BEA {\cal S}&:=& (2{\cal M}\partial_t-\partial_r^2) \;\; , \;\; 
\hspace{0.3truecm}
{\cal S}'' :=\partial_{\theta^1}\partial_{\theta^2}
\nonumber \\
{\cal S}' &:=& 2{\cal M}\partial_{\theta^1}-\partial_{\theta^2}\partial_r
\;\; ,\;\; 
\bar{\cal S}':=\partial_{\theta^1}\partial_r-\partial_{\theta^2}\partial_t
\label{gl:3:Sops}
\EEA
cancel on solutions of the equations of motion. So
\BEQ \label{gl:3:X1}
X_1:=-{1\over 2{\cal M}}(Y^2_{\half}+t^2{\cal S}+t\theta^1 {\cal S}')
=-t^2\partial_t-t(r\partial_r+\theta^1\partial_{\theta^1})-
xt-{{\cal M}\over 2}r^2-\half r\theta^1\partial_{\theta^2} 
\EEQ
is also a symmetry of (\ref{gl:SS}), extending the special Schr\"odinger
transformation introduced in (\ref{gl:1:sch1}). One obtains two more generators 
by straightforward computations, namely
\BEA
G_{\half}^1\: :=\: [X_1,G_{-\half}^1] &=& -t\partial_{\theta^1}
                   -\half r\partial_{\theta^2}
\nonumber \\
G_{\half}^2\: :=\: [X_1,G_{-\half}^2] &=& -t(\theta^1\partial_t +\theta^2
\partial_r)-\half \theta^1 r\partial_r-x\theta^1-{\cal M}r\theta_2
           +\half \theta^1\theta^2 \partial_{\theta^2}.
\label{gl:3:Gh}
\EEA

\noindent {\bf Proposition 3.2.} {\it The vector space generated by 
$\mathfrak{sgal}$ introduced in Proposition 3.1, together with 
$N_0$ and the three special transformations $X_1,G_{\half}^{1,2}$,
closes into a 13-dimensional Lie superalgebra. We shall call this Lie 
algebra the {\em $(N=2)$-super-Schr\"odinger algebra} and denote 
it by $\tilde{\s}^{(2)}$. Explicitly, 
\BEQ
\tilde{\s}^{(2)} = \left\langle X_{\pm 1,0}, G_{\pm 1/2}^{1,2}, Y_{\pm 1/2},
\bar{Y}_0^{1,2}, M_0, N_0 \right\rangle
\EEQ
and the generators are listed in 
eqs.~(\ref{gl:3:se32},\ref{gl:3:N0},\ref{gl:3:X1},\ref{gl:3:Gh}). See also
appendix~B.}

\noindent {\bf Proof.} One may check very easily the following formulas 
(note that the correcting terms of the type
function times ${\cal D}$, where  ${\cal D}={\cal S},{\cal S}',\bar{\cal S}'$ 
or ${\cal S}''$,  are here for definiteness but yield  $0$ modulo the 
equations of motion when commuted against elements of $\se(3|2)$, so
they can be dismissed altogether when computing brackets)
\BEA
M_0 &=& [Y_{\half},Y_{-\half}]
\nonumber \\
X_{-1} &=& -{1\over 2{\cal M}} Y_{-\half}^2- {1\over 2{\cal M}} {\cal S}
\nonumber \\
X_0&=&-{1\over 4{\cal M}}(Y_{-\half}Y_{\half}+Y_{\half}Y_{-\half})
      -{t\over 2{\cal M}}{\cal S} -{\theta^1\over 4{\cal M}}{\cal S}' 
\nonumber \\
G_{-\half}^{1} &=& {1\over 2{\cal M}} \bar{Y}_0^{1}Y_{-\half}
                   -{1\over 2{\cal M}}{\cal S}' 
\nonumber \\ 
G_{-\half}^{2} &=& {1\over 2{\cal M}} \bar{Y}_0^{2}Y_{-\half}
                   -{\theta^1\over 2{\cal M}}{\cal S}
\nonumber \\
G_{\half}^{1} &=& {1\over 2{\cal M}} \bar{Y}_0^{1}Y_{\half}
                 -{t\over 2{\cal M}}{\cal S}'
\nonumber \\
G_{\half}^{2} &=& {1\over 2{\cal M}} \bar{Y}_0^{2}Y_{\half}
                 -{t\theta^1\over 2{\cal M}}{\cal S}
\nonumber \\
N_0&=&-{1\over 4{\cal M}}( \bar{Y}_0^2\bar{Y}_0^1-\bar{Y}_0^1\bar{Y}_0^2)
      -{\theta^1\over 2{\cal M}}{\cal S}'.
\nonumber 
\EEA
So it takes only a short time to compute the adjoint action 
of $G_{\half}^{1,2}$ on $\se(3|2)$. On the even translations we have
\BD
[G_{\half}^{1,2},X_{-1}]=G_{-\half}^{1,2} \;\; ,\;\; 
[G_{\half}^{1,2},Y_{-\half}]=\half \bar{Y}_0^{1,2}\;\; , \;\; 
[G_{\half}^{1,2},M_0]=0. 
\ED
By commuting the $G$-generators we find
\BD
[G_{\half}^{1,2},G_{-\half}^{1,2}]_+=0 \;\; , \;\; 
[G_{\half}^{1},G_{-\half}^{2}]_+=-\half N_0- X_0\;\; ,\;\;
[G_{\half}^{2},G_{-\half}^{1}]_+= \half N_0- X_0.
\ED
The action on the odd translations is given by
\BD
[G_{\half}^{1,2},\bar{Y}_0^{1,2}]_+=0 \;\; , \;\; 
[G_{\half}^{1,2},\bar{Y}_0^{2,1}]_+=- Y_{\half}.
\ED
Finally,
\BD
[G_{\half}^{1,2},Y_{\half}]=0 \;\; , \;\;
[G_{\half}^{1,2},X_0]=\half G_{\half}^{1,2}.
\ED
The generator $N_0$ acts diagonally on the generators of $\se(3|2)$ : the 
eigenvalue of ad $N_0$ on a generator without upper index is $0$, while it 
is $+1$ (resp. $-1$) on generators with upper index $1$ (resp. $2$). Note
that this is also true for the action of $N_0$ on $G_{\half}^{1,2}$.

The proof may now be finished by verifying that 
$[G_{\half}^i,G_{\half}^i]_+=0$ (for both $i=1,2$),
$X_1=-[G_{\half}^1,G_{\half}^2]_+$ and $[X_1,G_{\half}^{1,2}]=0$.   \eop

\noindent {\bf Remark:}
In order to prove the invariance of the equations of motion under 
$\tilde{\s}^{(2)}$ it is actually enough to prove the
invariance under  $Y_{\pm\half}$ and $\bar{Y}_0^{1,2}$ since all other 
generators are given ({\it modulo} the equations of motion) as
quadratic expressions in these four generators.

\subsection{Some physical applications} 

We now briefly recall some earlier results on supersymmetric non-relativistic
systems with a dynamic supersymmetry algebra which contains $\osp(2|2)$. 

Beckers {\it et al.} \cite{Beck86,Beck91,Beck92} studied the supersymmetric
non-relativistic quantum mechanics in one spatial dimension and derived the 
dynamical Lie superalgebras for any given superpotential $W$. The largest
superalgebras are found for the free particle, the free fall or the harmonic
oscillator, where the dynamic algebra is \cite{Beck92}
\BEQ
\tilde{\s}^{(2)} \cong \osp(2|2) \ltimes \mathfrak{sh}(2|2)
\EEQ
where $\mathfrak{sh}(2|2)$ is the Heisenberg super-algebra. We explicitly
list the correspondence for the harmonic oscillator with total
Hamiltonian, see \cite{Beck86} 
\BEQ 
H=H_B+H_F = \half \left( p^2 + \frac{1}{4} x^2 + \half \sigma_3\right)
\EEQ
The
$\osp(2|2)$-subalgebras of symmetries of our $(3|2)$-supersymmetric model and 
of the harmonic oscillator in the notation of \cite{Beck86} may be identified 
by setting  
\BEQ
H_B = X_0 \;\; , \;\; H_F = \half N_0 \;\; , \;\;
C_{\pm} = \pm \II X_{\mp 1} \;\; , \;\;
Q_+ = G_{\half}^1 \;\; , \;\; Q_- = -G_{-\half}^{2} \;\; , \;\;
S_+ = G_{-\half}^1 \;\; , \;\; S_- = - G_{\half}^2
\EEQ
while the identification of the symmetries in  $\sh(2|2)$ of both models
is given by
\BEQ
P_{\pm} = Y_{\mp\half} \;\; , \;\; 
T_{\pm} = \frac{\II}{\sqrt{2}}\bar{Y}_0^{1,2} \;\; , \;\;
I = - M_0
\EEQ
We remark that the total Hamiltonian corresponds to $H=X_0+\half N_0$ in 
our notation. 

Duval and Horvathy \cite{Duva94} systematically constructed supersymmetric
extensions with $N$ supercharges of the Schr\"odinger algebra $\sch_d$ as 
subalgebras of the extended affine orthosymplectic superalgebras. In general,
there is only one `standard' possible type of such extensions, but in two
space-dimensions, there is a further `exotic' superalgebra with a different
structure. Relationships with Poisson algebras (see below) are also discussed. 
While the kind of supersymmetries discussed above \cite{Beck86,Beck92} belong 
to the first type, the `exotic' type arises for example in Chern-Simons matter
systems, whose  $N=2$ supersymmetry was first described by Leblanc
{\it et al.} \cite{Lebl92}.\footnote{In {\em non-}commutative space-time, 
extended supersymmetries still persist, but scale- and Galilei-invariance
are broken \cite{Loza05}.} In \cite{Duva94}, the supersymmetries of a scalar
particle in a Dirac monopole and of a magnetic vortex are also discussed.  

The uniqueness of $\osp(2|2)$-supersymmetry constructions has been addressed
by Ghosh \cite{Ghos04}. Indeed, the generators of the $\osp(2|2)$ algebra
can be represented in two distinct ways in terms of the coordinates of the
super-Calogero model. This leads to two distinct types of superhamiltonians,
which in the simplest case of $N$ free superoscillators read \cite{Ghos04}
\BEA
H_{\pm} &=& \frac{1}{4} \sum_{i=1}^{N}\left[  \left( p_i^2 + x_i^2 \right) 
\pm  \left( \psi_i^{\dag}\psi_i -\psi_i \psi_i^{\dag} \right) \right]
\label{gl:3:H1} \\
\hat{H}_{\pm} &=&\frac{1}{4} \sum_{i=1}^{N} \left( p_i^2 + x_i^2 \right) 
\pm \frac{\gamma_5}{4} \left[ N - \II \sum_{i,j=1}^{N} 
\left(\psi_i^{\dag}\psi_j^{\dag} +\psi_i \psi_j+ 
\psi_i^{\dag}\psi_j- \psi_j^{\dag}\psi_i\right) L_{ij} \right]
\label{gl:3:H2}
\EEA
where $x_i$ and $p_i$ are bosonic coordinates and momenta,
$L_{ij}=x_ip_j-x_jp_i$ are angular momenta, the $\psi_i$
are fermionic variables satisfying $[\psi_i, \psi_{j}^{\dag}]_+ = \delta_{ij}$
and the operator $\gamma_5$ anticommutes with the $\psi_i$. The Hamiltonian
$H_{\pm}$ in eq.~(\ref{gl:3:H1}) is identical to the one discussed in 
\cite{Beck86,Beck92,Duva94}. 
Further examples discussed in \cite{Ghos04} include superconformal quantum
mechanics and Calogero models but will not be detailed here. 
Dynamical $\osp(2|2)$-supersymmetries also occur in the $d$-dimensional
Calogero-Marchioro model \cite{Ghos01}. 

Finally, we mention that the $SU(2)_0$ Wess-Zumino-Witten model has a hidden
$\osp(2|2)_{-2}$ symmetry, with a relationship to logarithmic conformal
field-theories \cite{Koga02}. 
  
\subsection{Dynamic symmetries of the $(3|2)$-supersymmetric model}

So far, we have considered the mass $\cal M$ as  fixed. Following what
has been done for the simple Schr\"odinger equation, 
we now relax this condition
and ask what happens if $\cal M$ is treated as a variable \cite{Henk03b}.
We then add the generators $D$ and $V_{-}$ to 
$\tilde{\s}^{(2)}$ which generates, through commutation with $X_1$ and 
$G_{\half}^{1,2}$, the following new generators 
\BEA
V_+ &=& 4[X_1,V_-]=
       -2tr\partial_t-2\zeta r\partial_{\zeta}-(r^2+4\zeta t)\partial_r
       -r(\theta^1\partial_{\theta^1}+\theta^2\partial_{\theta^2})
       -2t\theta^2\partial_{\theta^1}-2\zeta\theta^1\partial_{\theta^2}-2xr
\nonumber \\ 
W &=& [V_+,V_-]= -2\zeta^2\partial_{\zeta}
                 -2\zeta(r\partial_r+\theta^2\partial_{\theta^2})
		 -{r^2\over 2}\partial_t-r \theta^2\partial_{\theta^1}-2x\zeta
\nonumber \\
\bar{Z}_0^{1} &=& [G_{\half}^{1},V_-]= 
                  -\half\left(\zeta\partial_{\theta^2}
		   + \half r\partial_{\theta^1}\right)
\\
\bar{Z}_0^2 &=& [G_{\half}^2,V_-]=
-\half\left(\zeta (\theta^2\partial_{\zeta}+\theta^1\partial_r)
+\half \theta^2 r\partial_r + \half r\theta^1 \partial_t
+\half \theta^1\theta^2\partial_{\theta^1} + x\theta^2\right).
\nonumber 
\EEA

\noindent
{\bf Proposition 3.3.} {\it The 19-dimensional vector space}
\BEQ \label{gl:3:osp}
\s^{(2)} = \left\langle X_{\pm 1,0}, Y_{\pm\half}, M_0, D, N_0, 
G_{\pm\half}^{1,2}, \bar{Y}_0^{1,2}, V_{\pm}, W, \bar{Z}_0^{1,2} \right\rangle
\EEQ
{\it closes as a Lie superalgebra and leaves invariant the equations of motion
(\ref{gl:SSzeta}) of the $(3|2)$-supersymmetric model.}

\begin{figure}
\epsfxsize=110mm
\centerline{\epsffile{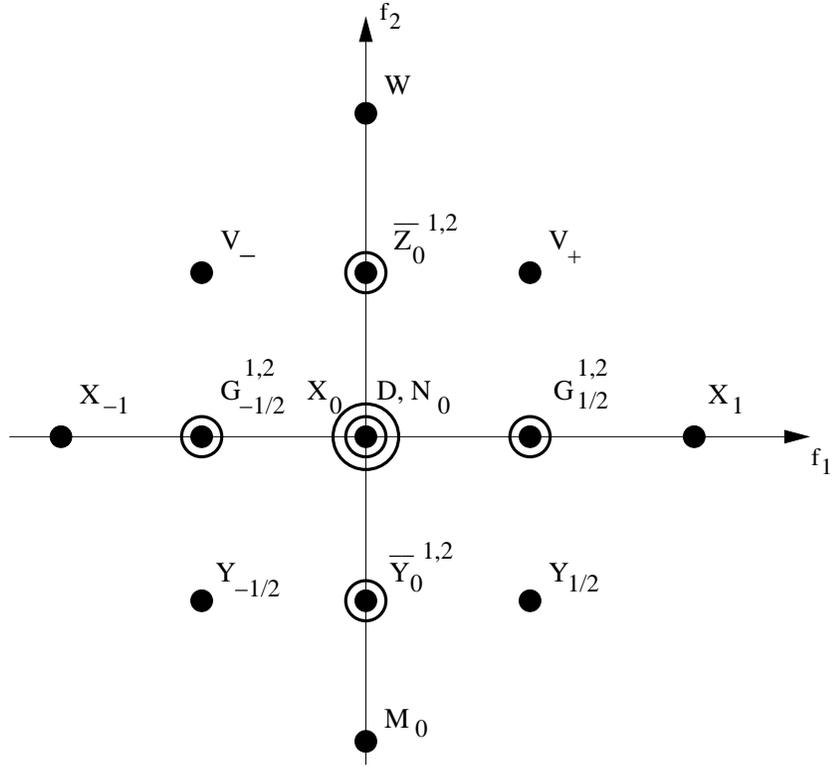}}
\caption[Abbildung 2]{\small Root vectors of the complexified  Lie 
superalgebra $\s^{(2)}\cong\mathfrak{osp}(2|4)$. 
The double circles indicate the presence
of two generators corresponding to opposite values of the root projection 
along $\al$, while the triple circle in the centre
corresponds to the Cartan subalgebra $\mathfrak{h}$ (see proposition 3.5). 
\label{Abb2}}
\end{figure}

We shall prove this in a simple way in subsection 3.6, by establishing a 
correspondence between $\s^{(2)}$ and  a Lie subalgebra of a Poisson algebra. 
This will also show that $\s^{(2)}$ is isomorphic to the 
Lie superalgebra $\osp(2|4)$ - hence one may in the end abandon the notation
$\s^{(2)}$ altogether. The root diagramme of $\osp(2|4)$ is
shown in figure~\ref{Abb2} and the correspondence of the roots 
with the generators of $\s^{(2)}$ is made explicit. 

%%==============================================================================
%%==============================================================================

\subsection[First]{First correspondence with Poisson structures: 
the case of \\ 
$\protect{\tilde{\mathfrak{s}}^{(2)}}$ $\protect{\cong}$
$\protect{\osp(2|2)\ltimes\sh(2|2)}$ or the super-Schr\"odinger model}
%% ici, il fallait bricoler, parce que le titre est trop long.

\begin{figure}
\centerline{\epsfxsize=65mm\epsffile{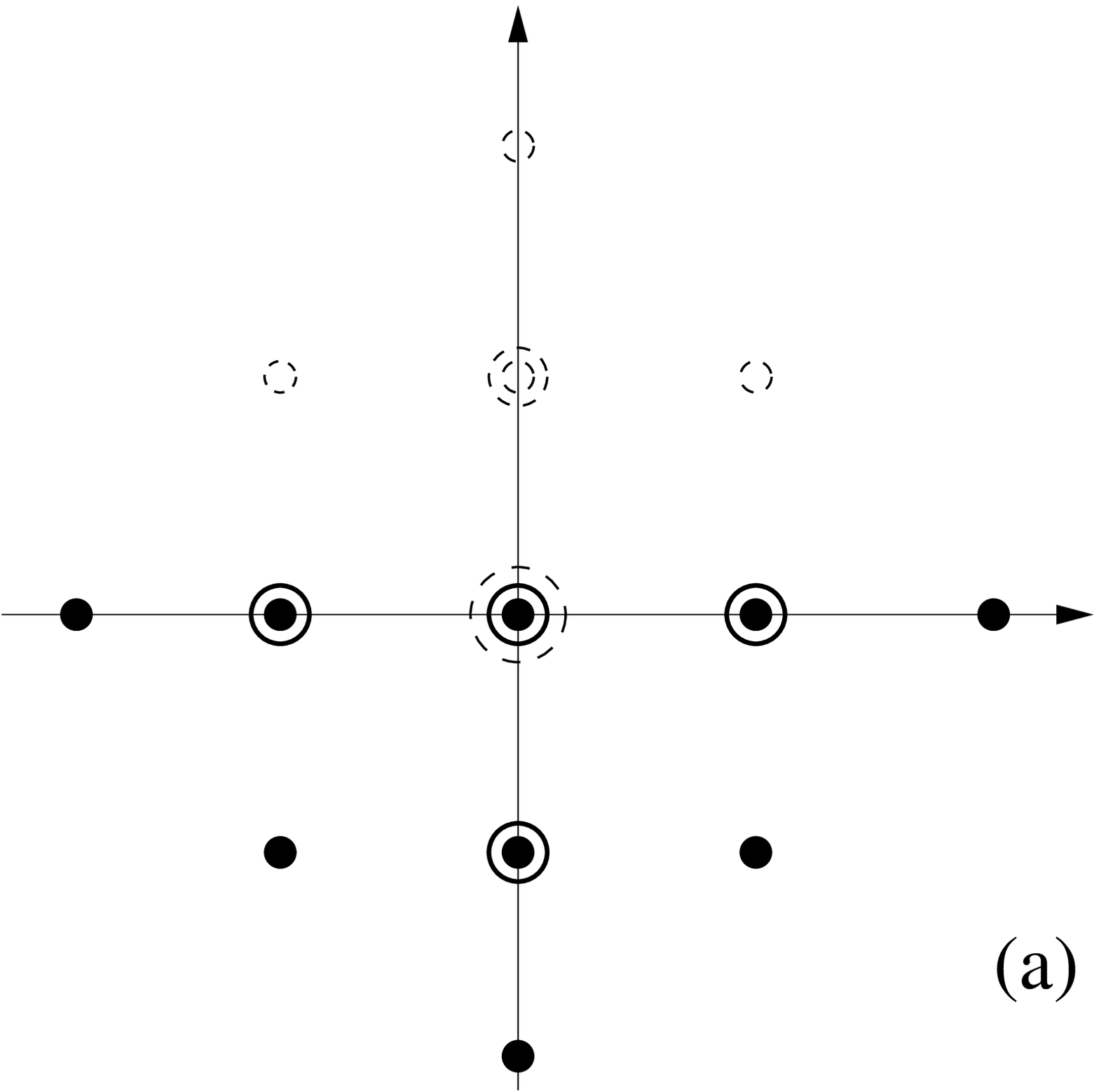} ~\epsfxsize=65mm
\epsffile{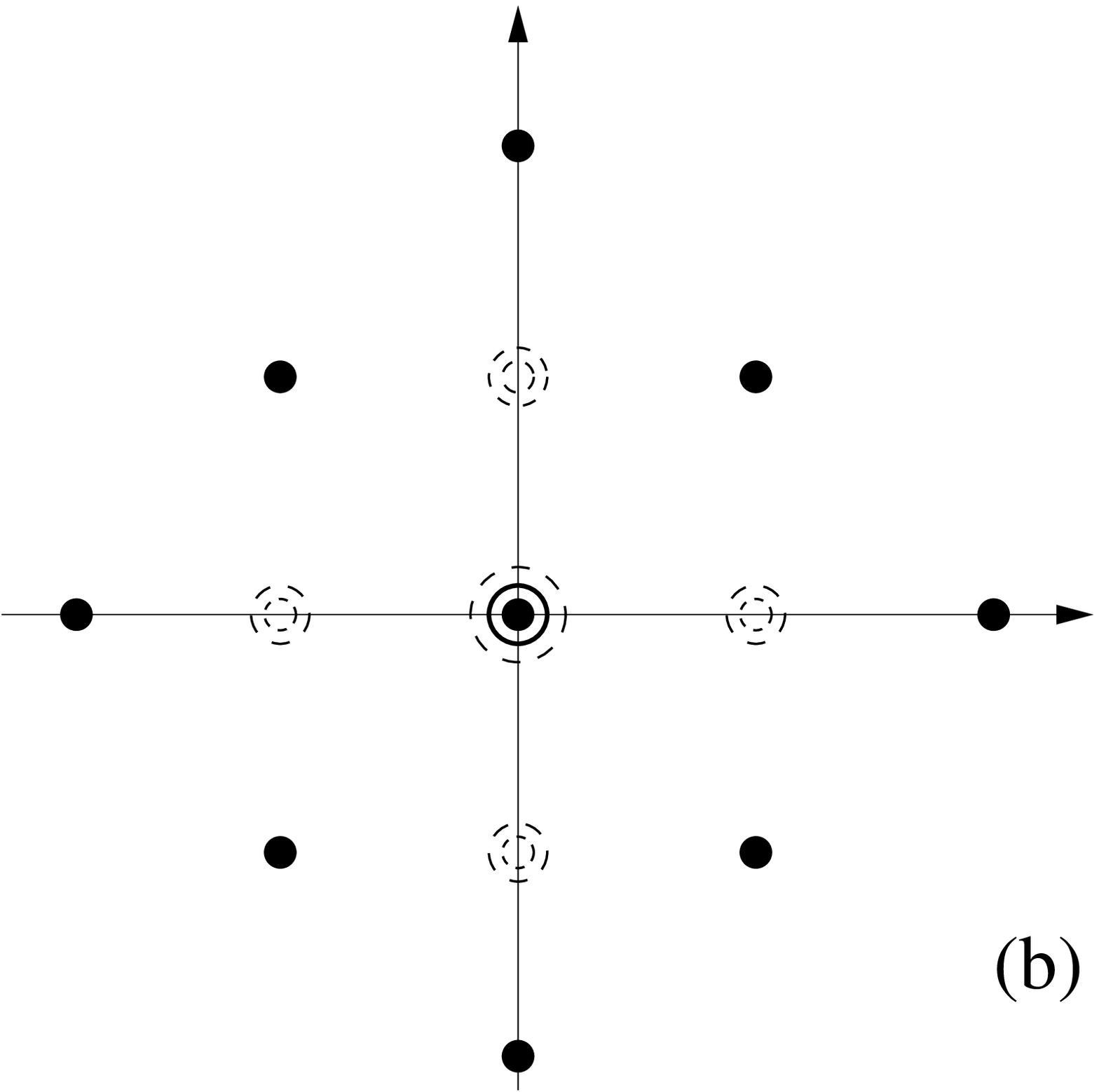} ~\epsfxsize=65mm\epsffile{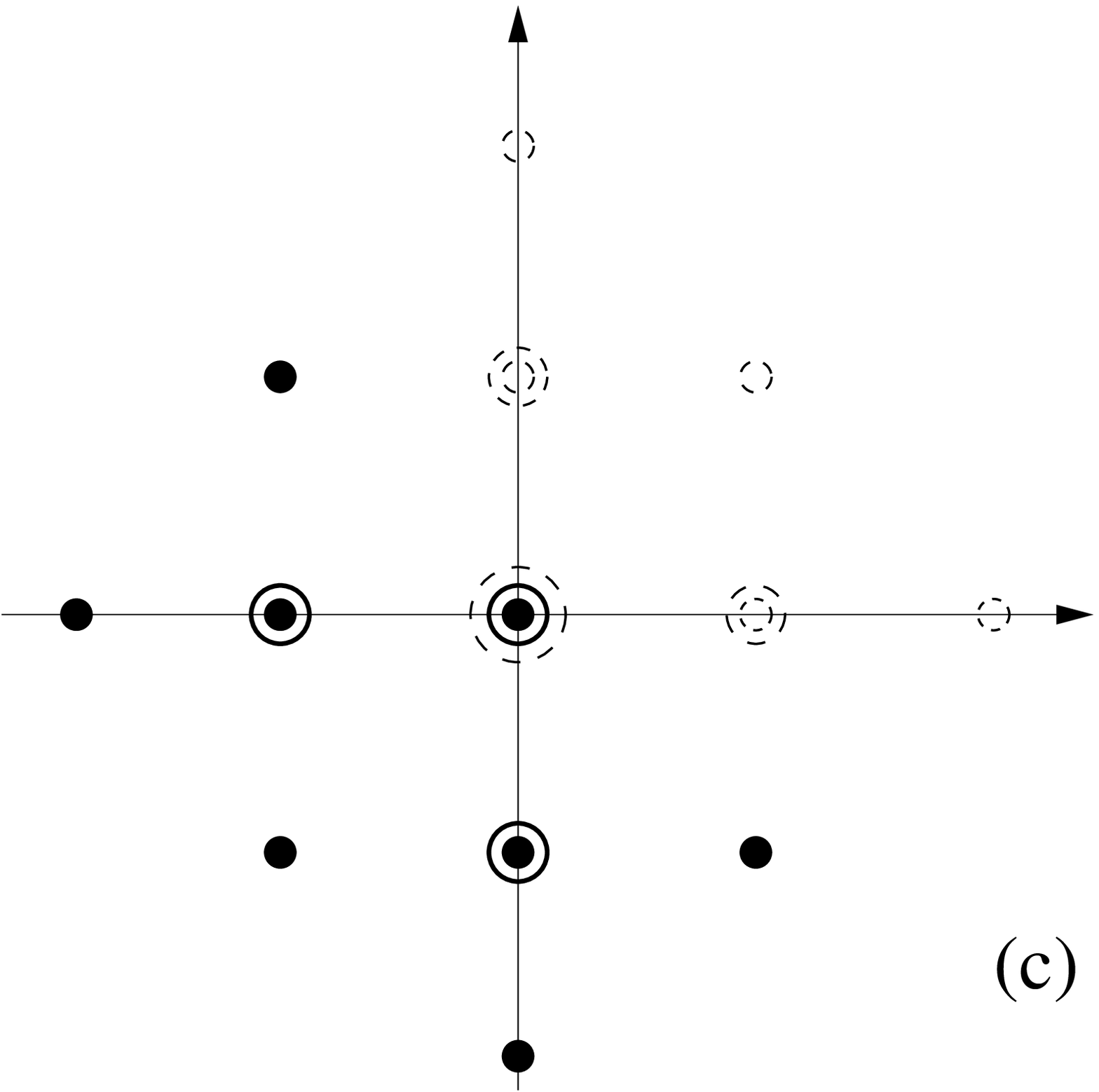}}
\caption[Abbildung 3]{\small Root vectors of several Lie subalgebras of 
$\mathfrak{osp}(2|4)$, arranged in the same way as in figure~\ref{Abb2}. The
full dots and circles give the generators in the respective subalgebra
whereas the open dots and broken circles merely stand for the remaining 
generators of $\mathfrak{osp}(2|4)$. Note that in each case only
two generators of the Cartan subalgebra $\mathfrak{h}$ are retained. The
subalgebras are (a) $\tilde{\mathfrak{s}}^{(2)}$, (b) 
$(\mathfrak{conf}_3)_{\mathbb{C}}$ and
(c) $\mathfrak{se}(3|2)$. 
\label{Abb3}}
\end{figure}

We shall give in this subsection a much simpler-looking presentation
of  $\tilde{\s}^{(2)}$ by embedding it into the Poisson 
algebra of superfunctions on a supermanifold, the Lie bracket of 
$\tilde{\s}^{(2)}$ corresponding to the Poisson
bracket of the superfunctions. In figure~\ref{Abb3}a, we show how 
$\tilde{\s}^{(2)}$ sits inside $\s^{(2)}\cong \osp(2|4)$. 
For comparison, we display in figure~\ref{Abb3}b the even subalgebra
$(\conf_3)_{\mathbb{C}}$ and in figure~\ref{Abb3}c the superalgebra 
$\se(3|2)$. We see that both
$\C D\oplus\tilde{\s}^{(2)}$ and $\C N_0\oplus\se(3|2)$ are maximal
Lie subalgebras of $\osp(2|4)$.  

%%==============================================================================
%%==============================================================================

We  first need some definitions 
(see \cite{Dufo05,GuieuRoger,Leit03} for details).

\noindent 
{\bf Definition 2.} {\em A commutative associative algebra $\cal A$ is a 
{\em Poisson algebra} if there exists
a Lie bracket $\{ \  ,\ \}~:~{\cal A}\times{\cal A}\to{\cal A}$ (called
Poisson bracket) which is {\em compatible} with the associative product 
$f,g\mapsto fg$, that is to say,  such that
the so-called {\em Leibniz identity} holds true}
\BEQ 
\{ fg,h\}=f\{g,h\}+g\{f,h\},\quad \forall f,g,h\in {\cal A}.
\EEQ

This definition is naturally superizable and leads to the notion of a
{\em super-Poisson algebra}. Standard examples are Poisson or 
super-Poisson algebras of smooth functions on supermanifolds.

In this and the following subparagraphs, we shall consider a Poisson algebra, 
denoted ${\cal P}^{(2m|N)}$, of functions on
the $(2m|N)$-supertorus, where $m=1,2$ and $N=0,1,2$. 
As an associative algebra, it may be written as  the tensor product 
\BEQ {\cal P}^{(2m|N)}={\cal P}^{(2m|0)}\otimes\Lambda(\R^N)
\EEQ 
where $\Lambda(\R^N)$ is the Grassmann algebra in the anticommuting variables 
$\theta^1,\ldots,\theta^N$, and ${\cal P}^{(2m|0)}$ is the associative algebra 
generated by the functions $(q_i,q_i^{-1},p_i,p_i^{-1})$, $i=1,\ldots,m$, 
corresponding to finite Fourier series. 
(Note that the Poisson algebra of {\it smooth} functions 
on the $(2|N)$-supertorus is a kind of completion of ${\cal P}^{(2|N)}$.)

\noindent 
{\bf Definition 3.} {\em We denote by $\delta$ the graduation on 
${\cal P}^{(2m|N)}$ defined by setting
 $\del(f)=k$, with $k=0,1,\ldots,N$, for the monomials 
$f(q,p,\theta)=f_0(q,p)\theta^{i_1}\ldots\theta^{i_k}$.}

The Poisson bracket is defined to be 
\BEQ \label{gl:3:leites}
\{f,g\}=\sum_{i=1}^m  \frac{\partial f}{\partial q_i}
\frac{\partial g}{\partial p_i}
-\frac{\partial f}{\partial p_i}\frac{\partial g}{\partial q_i} 
-(-1)^{\del(f)} \sum_{i,j=1}^N  \eta^{i\,j}
\frac{\partial f}{\partial{\theta^i}} 
\frac{\partial g}{\partial {\theta^j}}\ \ ,
\EEQ
where $\eta^{ij}$ is a non-degenerate {\it symmetric} two-tensor. 
Equivalently (by the Leibniz identity), 
it may be defined through the relations
\BEQ
\{q_i,p_j\}=\del_{i,j},\ \{q_i,\theta^j\}=0,\ \{p_i,\theta^j\}=0,
\ \{\theta^i,\theta^j\}=\eta^{ij}.
\EEQ

We warn the reader who is not familiar with  Poisson structures
in a supersymmetric setting against the familiar idea that the Poisson bracket
of two functions should be obtained in a more or less straightforward way from
their products. One has for instance in the case $N=1$
\BD
(\theta^1)^2=0,\quad \{\theta^1,\theta^1\}=\eta^{11}\not=0
\ED
which might look a little confusing at first. 

It is a well-known fact (see \cite{Guil84} for instance) 
that the Schr\"odinger Lie algebra $\sch_1$ generated 
by $X_{\pm 1,0}, Y_{\pm \half},M_0$ is isomorphic to the Lie algebra 
 of polynomials in $(q,p)=(q_1,p_1)$ 
with degree $\le 2$: an explicit isomorphism is given by
\BEQ
X_{-1}\to \half q^2 \;\; , \;\;
X_0\to -\half qp \;\; , \;\;
X_1\to \half  p^2 \;\; , \;\;
Y_{-\half}\to q \;\; , \;\;
Y_{\half}\to -p \;\; , \;\; 
M_0\to 1.
\EEQ
In particular, the Lie subalgebra $\langle X_{-1},X_0,X_1\rangle$ of 
quadratic polynomials in ${\cal P}^{(2|0)}$ is 
isomorphic to the Lie algebra $\symp(2,\R)$ 
of linear infinitesimal canonical transformations of $\R^2$, which is a mere 
reformulation of the canonical
isomorphism $\slin(2,\R)\cong \symp(2,\R)$ (see subsection 3.6 for an 
extension of this result).

We now give a natural extension of  this isomorphism  to a 
supersymmetric setting. In what follows we take $m=1,N=2$ and 
$\{\theta^1,\theta^1\}=\{\theta^2,\theta^2\}=0,\{\theta^1,\theta^2\}=2.$ 

\noindent {\bf Definition 4}
{\it We denote by ${\cal P}^{(2|2)}_{\le 2}\subset {\cal P}^{(2|2)}$
the Lie algebra  of superfunctions that are polynomials in 
$(p,q,\theta^1,\theta^2)$ of degree $\le 2$.}

\noindent {\bf Proposition 3.4.} {\it  One has an isomorphism
from $\tilde{\s}^{(2)}$ to  the Lie algebra ${\cal P}^{(2|2)}_{\le 2}$ 
given explicitly as
\BEA
& & X_{-1}\to \half q^2 \;\; , \;\; X_0\to -\half qp \;\; , \;\;
X_1\to \half p^2 \;\; , \;\;  Y_{-\half}\to q \;\; , \;\; 
Y_{\half}\to -p \;\; , \;\; M_0\to 1 
\nonumber \\
& & N_0\to -\half\theta^1\theta^2
\nonumber \\
& & \bar{Y}_0^1\to -\theta^1 \;\; , \;\;  \quad \bar{Y}_0^2\to \theta^2
\label{gl:3:s2Pois} \\
& & G_{-\half}^1\to -\half q\theta^1 \;\; , \;\;  
G_{-\half}^2\to \half q\theta^2 
\nonumber \\
& & G_{\half}^1\to \half p\theta^1 \;\; , \;\; 
G_{\half}^2\to -\half p\theta^2.
\nonumber
\EEA
}
{\bf Remark : } An equivalent statement of this result and its extension to 
higher spatial
dimensions was given in \cite{Duva94}, eq.~(4.10). This Lie isomorphism allows 
a rapid computation of Lie  brackets in $\tilde{\s}^{(2)}$.

\noindent {\bf Proof.} The subalgebra of ${\cal P}_{\le 2}^{(2|2)}$ 
made up of the 
monomials of degree $0$ in $p$ decomposes as a four-dimensional
commutative algebra $\langle q^2,q,1,\theta^1\theta^2\rangle$ (for the even 
part), plus four odd generators $\theta^i,q\theta^i$, $i=1,2$.
One may easily check the identification with the 3 even translations and the 
`super-Euler operator' $N_0$, plus the 4 odd translations of $\se(3|2)$, see 
figure~\ref{Abb2}.  

Then the two allowed rotations, $X_0$ and $Y_{\half}$, form together with the 
translations a nine-dimensional algebra that
is also easily checked to be isomorphic to its image in 
${\cal P}^{(2|2)}_{\le 2}$.

Finally, one sees immediately that the quadratic expressions 
(appearing just before
Proposition 3.2 and inside its proof)  that give 
$X_1,G_{\half}^{1,2}$ in terms of $Y,\bar{Y}$ also
hold in the associative algebra ${\cal P}^{(2|2)}_{\le 2}$ with the suggested 
identification (actually, this is also true
for the generators $N_0,X_{-1},X_0,G_{-\half}^{1,2}$, so one may still reduce 
the number of verifications.) \eop

Let us finish this paragraph  by coming back to the original 
$N=2$ supersymmetry algebra (see subsection 3.1). 
Suppose we want to consider only left-invariant 
odd translations $D_{\theta^1}=-G_{-\half}^1-G_{-\half}^2$ and 
$D_{\theta^2}=-\bar{Y}_0^1-\bar{Y}_0^2$. It is then natural to 
consider the vector space 
\BEQ
\tilde{\s}^{(1)}:=\langle 
X_{-1},X_0,X_1,Y_{-\half},Y_{\half},M_0,
G_{-\half}^1+G_{-\half}^2, G_{\half}^1+G_{\half}^2, 
\bar{Y}_0^1+\bar{Y}_0^2\rangle
\EEQ
and to ask whether this is a Lie subalgebra of $\tilde{\s}^{(2)}$. 
The answer is yes\footnote{An isomorphic Lie superalgebra was first constructed
by Gauntlett {\it et al.} \cite{Gaun90}.} and
this is best proved by using  the Poisson algebra formulation. 
Since restricting to this subalgebra amounts to 
considering functions that depend only
on $p,q$ and $\tilde{\theta}:=(\theta^1-\theta^2)/(2\II)$, with
$(\tilde{\theta})^2=0$ and 
$\{\tilde{\theta},\tilde{\theta}\}=1$, 
$\tilde{\s}^{(1)}$ can be seen as the Lie algebra of
polynomials of degree $\le 2$  in ${\cal P}^{(2|1)}$: 
it sits inside ${\cal P}^{(2|1)}$ just 
in the same way as $\tilde{\s}^{(2)}$ sits inside
${\cal P}^{(2|2)}$. Of course, the
conjugate algebra obtained by taking the same linear combinations, but with a 
minus sign instead (that is, 
generated by  $X_{\pm 1,0},Y_{\pm\half}, G_{\half}^1-G_{\half}^2$ and  the 
right-invariant odd translations  
$\bar{D}_{\theta^1}=G_{-\half}^2-G_{-\half}^1$ and 
$\bar{D}_{\theta^2}=\bar{Y}_0^2-\bar{Y}_0^1$) is isomorphic to 
$\tilde{\s}^{(1)}$. The commutation relations of $\tilde{\s}^{(1)}$ are again
illustrated in figure~\ref{Abb3}a, where the four double circles of a pair of
generators should be replaced by a single generator. 
We shall consider this algebra once again in section~5.

\subsection{Second correspondence with Poisson structures: the case of 
$\mathfrak{osp}(2|4)$, or the $(3|2)$-supersymmetric model}

We shall prove in this subsection Proposition~3.3 by giving an embedding 
of the vector space  $\s^{(2)}$  into a Poisson algebra, from which the fact 
that $\s^{(2)}$ closes as a Lie algebra becomes self-evident.

Let us first recall the definition of the orthosymplectic superalgebras.

\noindent {\bf Definition 5.}
{\it Let $n,m=1,2,\ldots$ The {\em Lie superalgebra $\osp(n|2m)$} 
is the set of linear vector fields in the coordinates 
$x^1,\ldots,x^{2m},\theta^1,\ldots,\theta^n$ preserving the 2-form 
$\sum_{i=1}^m \D x^i\wedge \D x^{m+i}+\sum_{j=1}^n (\D \theta^i)^2.$ }

In the following proposition, we  recall the folklore result which states
that the  Lie superalgebra $\osp(2|2m)$ may be embedded into a 
super-Poisson algebra of functions on the
$(2m|2)$-supertorus, and detail the root structure in this 
very convenient embedding.

\noindent {\bf Definition 6.}
{\it We denote by ${\cal P}^{(2m|2)}_{(2)}$ the Lie subalgebra of quadratic 
polynomials in the super-Poisson algebra 
${\cal P}^{(2m|2)}$ on the $(2m|2)$-supertorus.}

\noindent {\bf Proposition 3.5.} {\it Equip the super-Poisson
algebra ${\cal P}^{(2m|2)}$ with the super-Poisson bracket
$\{q_i,p_j\}=\del_{i,j},\{\theta^1,\theta^2\}=2,$ and consider its
Lie subalgebra
${\cal P}^{(2m|2)}_{(2)} \subset {\cal P}^{(2m|2)}$. Then
\begin{enumerate}
\item
The  Lie algebra ${\cal P}^{(2m|2)}_{(2)}$ is isomorphic to $\osp(2|2m)$.
\item
Using this isomorphism, a  Cartan subalgebra of $\osp(2|2m)$ is given by 
$p_i q_i$ ($i=1,\ldots,2m$) and $\theta^1\theta^2$.
Let $(f_1,\ldots,f_{2m},\alpha)$ be the dual basis. Then the root-space 
decomposition is given by
\BEA
\osp(2|2m)=\osp(2|2m)_0 & \oplus & \osp(2|2m)_{f_i-f_j}\oplus  
                                   \osp(2|2m)_{\pm(f_i+f_j)} \\
                        & \oplus &  \osp(2|2m)_{\pm 2f_i} \oplus
\osp(2|2m)_{\pm f_i+2\alpha} \oplus \osp(2|2m)_{\pm f_i-2\alpha}\; (i\not=j)
\nonumber
\EEA
Except $\osp(2|2m)_0$ which is equal to the Cartan subalgebra, all other 
root-spaces are one-dimensional, and
\BEA
 \osp(2|2m)_{f_i-f_j}\:=\:\langle p_i q_j\rangle \;\; ,\;\; 
 \osp(2|2m)_{f_i+f_j}&=&\langle p_i p_j\rangle \;\; ,\;\;  
 \osp(2|2m)_{-(f_i+f_j)}\:=\:\langle q_i q_j\rangle 
\nonumber \\
 \osp(2|2m)_{2f_i}&=&\langle p_i^2\rangle\;\;\: \;\; ,\;\; 
 \osp(2|2m)_{-2f_i}\:=\:\langle q_i^2\rangle
\nonumber \\
 \osp(2|2m)_{f_i+2\alpha}  &=&\langle p_i \theta^1\rangle \;\; , \;\;  
 \osp(2|2m)_{-f_i+2\alpha}\:=\:\langle q_i \theta^1\rangle
\nonumber \\
 \osp(2|2m)_{f_i-2\alpha}  &=& \langle p_i \theta^2\rangle \;\; ,\;\;
 \osp(2|2m)_{-f_i-2\alpha}\:=\:\langle q_i \theta^2\rangle.
\EEA
\end{enumerate}  }

\noindent{\bf Proof.} Straightforward.\eop

The root structure is illustrated in figure~\ref{Abb2} in the case $m=2$.
We may now finally state the last ingredient for proving Proposition~3.3.

%\newpage\typeout{ *** Seitenvorschub vor proposition 3.6 ***}
\noindent {\bf Proposition 3.6.} {\it 
\begin{enumerate}
\item
The linear application $\tilde{\s}^{(2)}\to\osp(2|4)$ defined on generators by
\BEA
X_{-1} &\to& \half q_1^2 \;\; , \;\;
X_0\to -\half q_1 p_1 \;\; , \;\;
X_1\to \half p_1^2\;\; , \;\; 
Y_{-\half}\to q_1 q_2 \;\; , \;\; 
Y_{\half}\to- p_1 q_2 \;\; , \;\; M_0\to q_2^2
\nonumber \\
N_0 &\to& \half \theta^1\theta^2
\nonumber \\
\bar{Y}_0^1 &\to& -q_2\theta^1 \;\; , \;\; 
\bar{Y}_0^2\to q_2\theta^2 
\label{gl:3:osp1} \\
G_{-\half}^1 &\to& -\half q_1\theta^1 \;\; , \;\; 
G_{\half}^1\to \half p_1\theta^1  \;\; , \;\; 
G_{-\half}^2\to \half q_1\theta^2 \;\; , \;\; 
G_{\half}^2\to -\half p_1\theta^2,
\nonumber
\EEA
where $i=1,2$, is a Lie algebra morphism and gives an embedding of 
$\tilde{\s}^{(2)}$ into $\osp(2|4)$. 
\item
This application can be extended into a Lie algebra isomorphism from 
$\s^{(2)}$ onto $\osp(2|4)$ by putting
\BEQ \label{gl:3:osp2}
D\to -\half(q_1 p_1+q_2 p_2) \; \;, \; \;
V_+\to p_1 p_2 \;\; , \;\;
W\to \frac{1}{4} p_2^2 \;\; , \;\;
V_-\to -\frac{1}{4} q_1 p_2 \;\; , \;\;  
\bar{Z}_0^{1}\to \frac{1}{8} p_2\theta^{1} \;\; , \;\; 
\bar{Z}_0^2\to -\frac{1}{8} p_2\theta^2.
\EEQ
\end{enumerate}
}

\noindent {\bf Proof.}
The first part is an immediate consequence of 
proposition 3.4. One merely needs to replace $q,p$ by $q_1,p_1$ and then make
all generators quadratic in the variables $p_1,p_2,q_1,q_2,\theta^1,\theta^2$
by multiplying with the appropriate power of $q_2$. 

We now turn to the second part. The root diagram of $\osp(2|4)$ 
in figure~\ref{Abb2} helps to 
understand.  First $\langle \bar{Z}_0^{1,2},G_{\half}^{1,2};W,V_+,X_1\rangle$ 
form a Lie algebra of dimension 7 that
is isomorphic to 
$\langle p_2\theta^{1,2},p_1\theta^{1,2},p_2^2,p_1 p_2,p_1^2\rangle$: 
in particular, the even part $\langle W,V_+,X_1\rangle$
is commutative and commutes with the 4 other generators;  brackets of the odd 
generators $\bar{Z}_0^{1,2},G_{\half}^{1,2}$ yield the whole vector space 
$\langle W,V_+,X_1\rangle$. Note that part of these computations
(commutators of  $X_1,G^{1,2}_{\half}$) come from the preceding subsection, 
the rest must be checked explicitly. So
all there remains to be done is to check for the adjoint action of 
$\bar{Z}^0_{1,2},G_{\half}^{1,2}$ on
$\se(3|2)$. We already computed the action of $G_{\half}^{1,2}$ on 
(even or odd) translations; in particular, $G_{\half}^{1,2}$ preserves 
this subspace. On the other hand, commutators of $G_{\half}^{1,2}$ with 
rotations $V_-,X_0,N_0,Y_{\half}$ yield linear combinations of 
$\bar{Z}_0^{1,2}$ and $G_{\half}^{1,2}$: by definition,
\BD
[G_{\half}^{1,2},V_-]=\bar{Z}_0^{1,2}
\ED 
while other commutators $[G_{\half}^{1,2},X_0]=\half G_{\half}^{1,2}, 
[G_{\half}^1,N_0]=-G_{\half}^1, [G_{\half}^2,N_0]
=G_{\half}^2, [G_{\half}^{1,2},Y_{\half}]=0$ are already known. 
Now the symmetry $t\leftrightarrow \zeta, \theta^1\leftrightarrow \theta^2$ 
preserves $\se(3|2)$ and sends $G_{\half}^{1,2}$
into $2\bar{Z}_0^{1,2}$, and corresponds to the symmetry $p\leftrightarrow q$ 
on $\osp(2|4)\cong {\cal P}^{(4|2)}_{(2)}$, so the action of $\bar{Z}_0^{1,2}$ 
on the rotation-translation symmetry algebra is the right one. Finally, since
$W,V_+$ and $X_1$ are given by commutators of $G_{\half}^{1,2}$ 
and $\bar{Z}_0^2$, and the commutators of $D$ with the other generators
are easily checked to be correct,  we are done. \eop 

In section~5 we shall  
consider two-point functions that are covariant
under the vector space $\tilde{\s}^{(2)}_1 = 
\langle X_{-1},G_{-\half}^{1,2},X_0,N_0,G_{\half}^{1,2},X_1\rangle 
\subset \tilde{\s}^{(2)}$ 
(actually $\tilde{\s}_1^{(2)}\subset \tilde{\s}^{(2)}$ is made of symmetries
of the super-Schr\"odinger model). On the root diagram
figure~\ref{Abb2}, the generators of $\tilde{\s}^{(2)}_1$ are all on the 
$f_1$-axis, hence (as one sees easily) $\tilde{\s}^{(2)}_1$ is  a Lie algebra. 
The following proposition gives several equivalent characterizations of 
$\tilde{\s}^{(2)}_1$. We omit the easy proof.

\noindent {\bf Proposition 3.7}
{\it 
\begin{enumerate}
\item
The embedding $\tilde{\s}^{(2)}\subset{\cal P}^{(2|2)}_{\le 2}$ 
of eq.~(\ref{gl:3:s2Pois}) in Proposition~3.4 maps $\tilde{\s}_1^{(2)}$
onto ${\cal P}_{(2)}^{(2|2)}$. Hence, by Proposition~3.5, 
$\tilde{\s}^{(2)}_1\cong\osp(2|2).$
\item
The Poisson bracket on ${\cal P}_{\le 2}^{(2|2)}$ (see Proposition~3.4) 
is of degree -1 with respect
to the graduation $\widetilde{\deg}$ of ${\cal P}^{(2|2)}$ 
defined by $\widetilde{\deg}(q)=\widetilde{\deg}(p)=
\widetilde{\deg}(\theta^i)=\half$: in other words, 
$\widetilde{\deg}\ \{f,g\}=\wit{\deg} f+\wit{\deg} g-1$ for 
$f,g\in{\cal P}^{(2|2)}$.
Hence the set $\{X\in{\cal P}^{(2|2)}\ |\ 
\widetilde{\deg}(X)=1\}\cong \tilde{\s}_1^{(2)}$ is a Lie subalgebra
of ${\cal P}^{(2|2)}$.
\item
The Poisson bracket on ${\cal P}_{(2)}^{(4|2)}$ 
(see Proposition~3.5) is of degree -1 with respect to the
graduation $\deg$ of ${\cal P}^{(4|2)}$ defined by 
$\deg(q_1)=\deg(p_1)=\deg(\theta^1)=\deg(\theta^2)=\half,
\deg(p_2)=1,\deg(q_2)=0$. Hence the set 
$\{X\in{\cal P}_{(2)}^{(4|2)}\ |\ \deg(X)=1\}
\cong \tilde{\s}_1^{(2)}\oplus\R D$
is a Lie subalgebra of ${\cal P}^{(4|2)}$.
\end{enumerate}  }

Note that points 1 and 2 use the first correspondence 
(see eq.~(\ref{gl:3:s2Pois}) in Proposition 3.4) while point 3 uses the 
second correspondence (see eqs.~(\ref{gl:3:osp1},\ref{gl:3:osp2}) in 
Proposition 3.6).

%%%%%%%%%%%%%%%%%%%%%%%%%%%%%%%%%%%%%%%%%%%%%%%%%%%%%%%%%%%%%%%%%%%%%%%%%%%%%%%%
\section{Extended Schr\"odinger and super-Schr\"odinger transformations}
%%%%%%%%%%%%%%%%%%%%%%%%%%%%%%%%%%%%%%%%%%%%%%%%%%%%%%%%%%%%%%%%%%%%%%%%%%%%%%%%

We shall be looking in this section for infinite-dimensional extensions 
of  various Lie algebras of Schr\"odinger type  
($\sch_1,\tilde{\s}^{(1)},\tilde{\s}^{(2)},\tilde{\s}^{(2)}_1\cong\osp(2|2)$) 
that we introduced until now, hoping that these
infinite-dimensional Lie algebras or super-algebras might  play for 
anisotropic systems a role analogous to that of  the Virasoro
algebra in conformal field theory \cite{Bela84}. Note that the 
Lie superalgebra $\s^{(2)}\cong \osp(2|4)$ was
purposely not included in this list, nor could $\conf_3$ be 
included: it seems that there is a `no-go theorem'
preventing this kind of embedding of Schr\"odinger-type 
algebras into infinite-dimensional Virasoro-like
algebras to extend to an embedding of the whole 
conformal-type Lie algebra (see \cite{RogUnt}).

In the preceding section, we saw that all Schr\"odinger or 
super-Schr\"odinger or conformal or `super-conformal'
Lie symmetry algebras could be embedded in different ways into 
some Poisson algebra or super-algebra ${\cal P}^{(n|N)}$. 

We shall extend the Schr\"odinger-type  Lie algebras by embedding them in a
totally different  way into some of the following `twisted'
Poisson algebras, where, roughly speaking, one is allowed to 
consider the square-root of the coordinate $p$. 

\noindent {\bf Definition 7.}  
{\it The {\em twisted Poisson algebra  $\widetilde{\cal P}^{(2|N)}$} 
is the associative algebra of super-functions
\BEQ
f(p,q;\theta):= f(p,q;\theta^1,\ldots,\theta^N)
=\sum_{i\in \half \Z} \sum_{j\in\Z} \sum_{k=1}^N \sum_{i_1<\ldots<i_k}
c_{i,j,i_1,\ldots,i_k} p^i q^j \theta^{i_1}\ldots \theta^{i_k}
\EEQ\
with  usual multiplication and Poisson bracket defined by
\BEQ \label{gl:4:Pois}
\{ f,g\} := {\partial f\over\partial q}{\partial g\over \partial p}
- {\partial f\over\partial p}{\partial g\over \partial q}
-(-1)^{\del(f)}\sum_{i=1}^N \partial_{\theta^i}f\partial_{\theta^i}g
\EEQ    
with the graduation $\del: \wit{\cal P}^{(2|N)}\to \mathbb{N}$ defined as a 
natural extension of Definition~3  (see subsection~3.5) on the monomials by
\BEQ \label{gl:4:delta}
\del(f(p,q)\theta^{i_1}\ldots\theta^{i_k}) :=k.
\EEQ
}
The Poisson bracket may be defined more loosely by setting 
$\{q,p\}=1,\ \{\theta^i,\theta^j\}=\del^{i,j}$
and applying the Leibniz identity.

\noindent {\bf Definition 8.}
{\it We denote by $\gra: \wit{\cal P}^{(2|N)}\to\{0,\half,1,\ldots\}$ 
the graduation (called {\em grade}) on the associative
algebra $\wit{\cal P}^{(2|N)}$ defined by
\BEQ
\gra(q^n p^m \theta^{i_1}\ldots\theta^{i_k}):=m+k/2
\EEQ
on monomials.}

This graduation may be defined  more simply by setting
$\gra(q)=0, \ \gra(p)=1,\ \gra(\theta^i)=\half$.
Note that it is closely related but clearly different from the 
graduations $\deg,\ \widetilde{\deg}$ defined on
untwisted Poisson algebras in Proposition~3.7.

\noindent {\bf Definition 9.}
{\it We denote by $\wit{\cal P}^{(2|N)}_{\le \kappa}, \ \ \kappa\in\half\Z$ 
(resp. $\wit{\cal P}^{(2|N)}_{(\kappa)}$)   the vector subspace
of $\wit{\cal P}^{(2|N)}$ consisting of all elements of grade $\le \kappa$  
(resp. of grade equal to $\kappa$).}

Since the Poisson  bracket is of grade -1 (as was the case for $\deg$ 
and $\widetilde{\deg}$)
it is clear that $\wit{\cal P}^{(2|N)}_{\le \kappa}$ (resp.
$\wit{\cal P}^{(2|N)}_{(\kappa)}$)  is a Lie algebra if and only if
$\kappa\le 1$ (resp. $\kappa=1$).

It is also easy to check, by the same considerations, 
that  $\widetilde{\cal P}_{\le \kappa}^{(2|N)}$  $(\kappa\le\half)$
is a (proper) Lie ideal of $\widetilde{\cal P}_{\le 1}^{(2|N)}$, 
so one may consider the resulting quotient algebra. 
In the following, we shall restrict to the case $\kappa=-\half$ and 
define the {\it Schr\"odinger-Neveu-Schwarz algebra} $\sns^{(N)}$ by
\BEQ
\sns^{(N)}:= \widetilde{\cal P}_{\le 1}^{(2|N)}/
\widetilde{\cal P}_{\le -1/2}^{(2|N)}
\EEQ 
The choice for the name is by reference to the case $N=1$ (see below).

\subsection{Elementary examples}
%%++++++++++++++++++++++++++++++++++++++++++++++++++++++++++++++++++++++++++++++
\begin{table}
\caption{\small Conformal dimensions $\cdim$ of the generators of the 
three supersymmetric
extensions $\sv$, $\mathfrak{sns}^{(1)}$ and $\mathfrak{sns}^{(2)}$ of the
one-dimensional Schr\"odinger algebra $\mathfrak{sch}_1$. 
\label{tab1}}
\begin{center}
\begin{tabular}{|l|c|ll|} \hline
 & $\cdim$ & pair & impair \\ \hline
$\sv$ & 2     & $X$ & \\
      & $3/2$ & $Y$ & \\
      & 1     & $M$ & \\ \hline
$\mathfrak{sns}^{(1)}$ & 2     & $X$ &           \\
                       & $3/2$ & $Y$ & $G$       \\
		       & 1     & $M$ & $\bar{Y}$ \\
		       & $1/2$ &     & $\bar{M}$ \\ \hline
$\mathfrak{sns}^{(2)}$ & 2     & $X$      &                          \\
                       & $3/2$ & $Y$      & $G^1$, $G^2$             \\
		       & 1     & $M$, $N$ & $\bar{Y}^1$, $\bar{Y}^2$ \\
		       & $1/2$ & $P$      & $\bar{M}^1$, $\bar{M}^2$ \\
		       & 0     & $Q$      &                          \\ \hline
\end{tabular} \end{center}
\end{table}
%%++++++++++++++++++++++++++++++++++++++++++++++++++++++++++++++++++++++++++++++

%%++++++++++++++++++++++++++++++++++++++++++++++++++++++++++++++++++++++++++++++
\begin{table}
\caption{\small Grades $\gra$ of the generators of the three supersymmetric
extensions $\sv$, $\mathfrak{sns}^{(1)}$ and $\mathfrak{sns}^{(2)}$ of the
one-dimensional Schr\"odinger algebra $\mathfrak{sch}_1$. 
\label{tab2}}
\begin{center}
\begin{tabular}{|l|c|ll|} \hline
 & $\gra$ & pair & impair \\ \hline
$\sv$ & 1     & $X$ & \\
      & $1/2$ & $Y$ & \\
      & 0     & $M$ & \\ \hline
$\mathfrak{sns}^{(1)}$ & 1     & $X$ & $G$       \\
                       & $1/2$ & $Y$ & $\bar{Y}$ \\
		       & 0     & $M$ & $\bar{M}$ \\ \hline
$\mathfrak{sns}^{(2)}$ &  1    & $X$, $N$ & $G^1$, $G^2$             \\
                       & $1/2$ & $Y$, $P$ & $\bar{Y}^1$, $\bar{Y}^2$ \\
		       & 0     & $M$, $Q$ & $\bar{M}^1$, $\bar{M}^2$ \\ \hline
\end{tabular} \end{center}
\end{table}
%%++++++++++++++++++++++++++++++++++++++++++++++++++++++++++++++++++++++++++++++

Let us study in this subsection the simplest examples $N=0,1.$
\begin{itemize}
\item \underline{$N=0$.}

The Lie algebra $\sns^{(0)}$ is generated by (images in the quotient
 $\widetilde{\cal P}_{\le 1}^{(2|0)}/
\widetilde{\cal P}_{\le -1/2}^{(2|0)}$ ) 
of the fields $X,Y,M$ defined by
\BEQ
X_{\phi}=\phi(q) p \;\; , \;\; 
Y_{\phi}=\phi(q) p^{\half} \;\; , \;\;
M_{\phi}=\phi(q).
\EEQ
By computing the commutators in
the quotient, we see that $\sns^{(0)}=\sv$ is the 
Schr\"odinger-Virasoro algebra eq.~(\ref{gl:1:sv}), with mode expansion 
$X_n=q^{n+1}p$, $Y_m=q^{m+\half}p^{\half}$,
$M_n=q^n$ (where $n\in\Z$, $m\in\Z+\half$). Each of these three fields 
$A=X,Y$ or $M$ has a mode expansion of the form
$A_n=q^{n+\eps}p^{\eps}$. We may rewrite this as 
$A_{\lambda-\eps}=q^{\lambda}p^{\eps}$ with $\lambda\in\mathbb{Z}+\eps$ and
see that  the shift $\eps$  in the indices 
of the generators (with respect to the power of $q$) is equal to 
the opposite of the power of $p$. This will also hold true for 
any value of $N$.

It is important to understand that successive `commutators' 
$\{Y_{\phi},M_{\psi}\}, \{Y_{\phi_1},\{Y_{\phi_2},M_{\psi}\}\},\ldots$ in 
$\widetilde{\cal P}_{\le 1}^{(2|0)}$ are generally non-zero and yield ultimately
the whole algebra $\widetilde{\cal P}_{\le -\half}^{(2|0)}$. 
This is due to the fact that derivatives of
$p^{\half}$ give $p$ to power $-\half,-{3\over 2},\ldots$, unlike derivatives 
of {\it integer} positive powers of
$p$, which cancel after a finite time and give only polynomials in $p$.

The algebraic structure of $\sv$ is as follows, see (\ref{gl:1:sv}). 
It is the semi-direct product $\sv = \Vir\ltimes\h^{(0)}:=\langle
X_n\rangle_{n\in\Z}\ltimes\langle Y_m,M_p\rangle_{m\in\half+\Z,p\in\Z}$ of a 
centreless Virasoro algebra and
of  a {\it two-step nilpotent} (that is to say, whose brackets are central)
Lie algebra generated by the $Y_m$ and $M_n$, 
extending the  Heisenberg algebra $\h_1$.
The inclusion $\sch_1\cong \slin(2,\R)\ltimes\h_1\subset \Vir\ltimes\h^{(0)}$ 
(see the introduction) respects the semi-direct product structure.
If one considers the generators 
$X_n$, $Y_m$ and $M_n$ as the components of associated conserved currents $X,Y$ 
and $M$, then $X$ is a Virasoro
field, while $Y,M$ are primary with respect to $X$, with conformal dimensions  
$\frac{3}{2}$, respectively $1$.

Note also that the conformal dimension of the $\eps$-shifted field $A^{\eps}$ 
$(A^{\eps}=X,Y,M)$ with mode expansion $A^{\eps}_n=q^{n+\eps}p^{\eps}$
$(\eps=0,\half,1)$ is equal to $1+\eps$. This fact is also a general one 
(see subsection 4.2 below).

For later use, we collect in tables~\ref{tab1} and \ref{tab2} 
the conformal dimensions and grades of the generators
of $\sns^{(N)}$, with $N=0,1,2$.

\item \underline{$N=1$.}

The Lie  algebra $\sns^{(1)}$ is generated by (images in the quotient) of the 
{\em even} functions 
$X_{\phi}=\phi(q) p,Y_{\phi}=\phi(q) p^{\half},M_{\phi}=\phi(q)$, and of the 
{\em odd} functions $G_{\phi}=\phi(q) \theta^1 p^{\half},
\bar{Y}_{\phi}=\phi(q) \theta^1,\bar{M}_{\phi}=\phi(q)\theta^1  p^{-\half}$. 
We use the same notation as in the case $N=0$ for the
mode expansions $X_n=q^{n+1}p,M_n=q^n,\bar{Y}_n=q^n\theta^1$ 
$(n\in \Z)$, $Y_m=q^{m+\half}p^{\half}, 
G_m=q^{m+\half}p^{\half}\theta^1,
\bar{M}_m=q^{m-\half}p^{-\half}\theta^1$ ($m\in\half+\Z$),
with the same shift in the indices, equal to the opposite of the power in $p$.

We have a semi-direct product structure 
$\mathfrak{sns}^{(1)}=\ns\ltimes\h^{(1)}$, where
\BEQ
\ns:=\langle X,G\rangle
\EEQ 
is isomorphic to the Neveu-Schwarz algebra \cite{Neve71} 
with a vanishing central charge, and 
\BEQ
\h^{(1)}=\langle (Y,\bar{Y}),(M,\bar{M})\rangle.
\EEQ
%The grades of the fields are given by (see table~\ref{tab1}) 
%\BEQ
%\gra(X_{\phi})=\gra(G_{\phi})=1 \;\; , \;\; 
%\gra(Y_{\phi})=\gra(\bar{Y}_{\phi})=\half \;\; , \;\; 
%\gra(M_{\phi})=\gra(\bar{M}_{\phi})=0.
%\EEQ
The commutators of  $G$ with these fields read in mode expansion (where we 
identify the Poisson bracket with an (anti)commutator)
\BEA 
{}[G_n,Y_m] &=& \half (n-m) \bar{Y}_{n+m} \;\; , \;\;
{}[G_n,\bar{Y}_m]=Y_{n+m}
\nonumber \\
{}[G_n,M_m] &=&-\half m \bar{M}_{n+m} \;\; , \;\;
{}[G_n,\bar{M}_m]=M_{n+m}.
\EEA
The Lie algebra $\h^{(1)}$ is two-step nilpotent, which is 
obvious from the definition of the quotient: the only
non-trivial brackets are between elements $Y_{\phi}$ and 
$\bar{Y}_{\phi}$ of grade $\half$ and give elements $M_{\phi}$ or 
$\bar{M}_{\phi}$ of grade 0. Explicitly, we have:
\BEQ
{}[Y_n,Y_m]=\half (n-m) M_{n+m} \;\; , \;\;  
{}[\bar{Y}_n,\bar{Y}_m]= M_{n+m} \;\; , \;\;  
{}[Y_n,\bar{Y}_m]=-\half m \bar{M}_{n+m}.
\EEQ
The fields $(X,G),$ $(Y,\bar{Y})$ and $(M,\bar{M})$ can be seen as 
supersymmetric doublets of conformal fields with conformal
dimensions $(2,{3\over 2}),\ ({3\over 2},1),\  (1,{1\over 2})$, 
see also table~\ref{tab1}. 
Once again, the conformal dimension of any of those fields
is equal to the power of $p$ plus one. The grades of the fields are given by,
see table~\ref{tab2} 
\BEQ
\gra(X_{\phi})=\gra(G_{\phi})=1 \;\; , \;\; 
\gra(Y_{\phi})=\gra(\bar{Y}_{\phi})=\half \;\; , \;\; 
\gra(M_{\phi})=\gra(\bar{M}_{\phi})=0.
\EEQ
\end{itemize}

\subsection{General case}
We shall actually mainly be interested in the case $N=2$, but the algebra 
$\sns^{(2)}$ is quite large and one needs new insight to
study it properly. So let us consider first the main features of the 
general case.

By considering the grading $\gra$, 
one sees immediately that $\sns^{(N)}$ has a semi-direct product structure 
\BEQ 
\sns^{(N)} = \g^{(N)} \ltimes \h^{(N)}
\EEQ 
where the Lie algebra $\g^{(N)}$ contains the elements of grade one and the
nilpotent algebra $\h^{(N)}$  contains the elements of grade $\half$ or $0$.
The algebra $\g^{(N)}$ has been studied by Leites and Shchepochkina 
\cite{Leit03} as one of the `stringy' superalgebras, namely, 
the superalgebra $\kk(1|N)$ of 
supercontact vector fields on the supercircle $S^{(1|N)}$. 
Let us just mention that  $\g^{(N)}$ shows up
as a geometric object, namely, as the superalgebra of vector fields 
preserving the (kernel of the) 1-form 
$\D q+\sum_{i=1}^N \theta^i \D\theta^i.$ 
Recall also that a supercontact vector field $X$ can be obtained from 
its {\it generating function}
$f=f(q,\theta^1,\ldots,\theta^N)$ by putting
\BEQ
X_f=-(1-\half E)(f)\partial_q-\half\partial_q f E-(-1)^{\del(f)} 
\sum_{i=1}^N \partial_{\theta^i}f \partial_{\theta^i},
\EEQ
where $E:=\sum_{i=1}^N \theta^i \partial_{\theta^i}$ is the Euler operator 
for odd coordinates, and $\del$ is the eigenvalue of $E$ for homogeneous 
superfunctions as defined in (\ref{gl:4:delta}). Then one has 
\BEQ
[X_f,X_g]=X_{\{f,g\}_{\kk(1|N)}}
\EEQ
where $[\ ,\ ]$ is the usual Lie bracket of vector fields, and 
the contact bracket $\{\ ,\ \}_{\kk(1|N)}$ is given by
\BEQ \label{gl:4:poissk}
\{f,g\}_{\kk(1|N)} :=-(1-\half E)(f)\partial_q g+\partial_q f (1-\half E)(g)
-(-1)^{\del(f)} \sum_{i=1}^N \partial_{\theta^i}f \partial_{\theta^i}g.
\EEQ

\noindent {\bf Proposition 4.1} {\it 
The Lie algebras  $\g^{(N)}$ and $\kk(1|N)$ are isomorphic.}
 
\noindent {\bf Proof:} Let $f=f(q,\theta)$ and $g=g(q,\theta)$ be two 
$E$-homogeneous superfunctions. Then 
\BEQ \label{gl:4:apptilde}
\tilde{f}(q,p,\theta)=f(q,\theta) \cdot
p^{1-\del(f)/2},\quad  \tilde{g}(q,p,\theta)=g(q,\theta) \cdot
p^{1-\del(g)/2}
\EEQ
belong to the subalgebra of elements of grade one in $\wit{\cal P}^{(2|N)}$.
Formula (\ref{gl:4:Pois}) for the Lie bracket of 
$\widetilde{\cal P}^{(2|N)}$ entails
\BD
\{\tilde{f},\tilde{g}\} (q,p,\theta) = 
\left[ (1-{\del(g)\over 2}) (\partial_q f) g
- (1-{\del(f)\over 2}) f (\partial_q g) \right]  
p^{1-{\del(f)+\del(g)\over 2}}
-(-1)^{\del(f)}\left[ 
\sum_{i=1}^N \partial_{\theta^i}f\partial_{\theta^i}g\right] 
p^{2-{\del(f)+\del(g)\over 2}}
\ED
while formula (\ref{gl:4:poissk}) for the contact bracket yields
\BEA
\widetilde{\{f,g\}}_{\kk(1|N)}(q,p,\theta) &=& 
-\left[ (1-\half\del(f))f (\partial_q g)-(1-\half\del(g)) (\partial_q f) g
\right] p^{1-\half(\del(f)+\del(g))} 
\nonumber \\
& & - (-1)^{\del(f)} \left[  \sum_{i=1}^N 
\partial_{\theta^i}f\partial_{\theta^i}g\right] 
p^{1-\half(\del(f)+\del(g)-2)}.
\nonumber 
\EEA
Hence
\BD
\{\tilde{f},\tilde{g}\}_{\wit{\cal P}^{(2|N)}}=\widetilde{\{f,g\}}_{\kk(1|N)}.
\ED
So the assignment $f\to \tilde{f}$ according to (\ref{gl:4:apptilde}) 
defines indeed a Lie algebra isomorphism from $\kk(1|N)$ onto $\g^{(N)}$. 
\eop

The application $f\to\tilde{f}$ just constructed may be extended in the 
following natural way.

\noindent {\bf Proposition 4.2}
{\it
Assign to any superfunction $f(q,\theta)$ on $S^{(1|N)}$ the 
following superfunctions in the Poisson superalgebra $\wit{\cal P}^{(2|N)}$:
\BEQ
f^{(\alpha)}(q,p,\theta):=f(q,\theta)
\cdot p^{\alpha-\del(f)/2},\quad \alpha\in\half \Z
\EEQ
so that, in particular, $f^{(1)}=\tilde{f}$ as defined in (\ref{gl:4:apptilde}).
Then $f\to f^{(\alpha)}$ defines a {\rm linear} isomorphism 
from the algebra of superfunctions on $S^{(1|N)}$
into the vector space of superfunctions in $\wit{\cal P}^{(2|N)}$ 
with grade $\alpha$, and the Lie bracket
(4.2) on the Poisson algebra may be written in terms of the 
superfunctions on $S^{(1|N)}$ in the following
way: let $f,g$ be two $E$-homogeneous functions on $S^{(1|N)}$,  
\BEQ \label{gl:4:poissN}
\{f^{(\alpha)},g^{(\beta)}\}_{\wit{\cal P}^{(2|N)}}=
\left( -(\alpha-\half E)(f)\partial_q g
-\partial_q f (\beta-\half E)(g)-(-1)^{\del(f)} \sum_{i=1}^N 
\partial_{\theta^i}f\partial_{\theta^i} 
g\right)^{(\alpha+\beta-1)}.
\EEQ }

\noindent {\bf Proof.} Similar to the proof of proposition 4.1. \eop

Coming back to $\sns^{(N)}$, we restrict to the values $\alpha=1,\half,0$.
Put $f_n(q)=q^{n+1}$ $(n\in\Z)$ and $g_m(q)=q^{m+\alpha-|I|/2} \theta^I$, 
where $I=\{i_1<\cdots<i_k\}\subset\{1,\ldots,N\}$ and 
$\theta^I:=\theta^{i_1}\wedge\cdots\wedge \theta^{i_k},$
and $m\in\Z-\alpha+|I|/2$. Then
\BEA
\{f_n^{(1)},g_m^{(\alpha)}\}_{\wit{\cal P}^{(2|N)}} &=& 
\{q^{n+1}p,q^{m+\alpha-|I|/2} \theta^I p^{\alpha-|I|/2} \}_{{\cal P}^{(2|N)}}
\nonumber \\
&=& \left[ -(m+\alpha-|I|/2)+(\alpha-\half |I|)(n+1)\right] 
q^{n+m+\alpha-|I|/2}\theta^I p^{\alpha-|I|/2} 
\nonumber\\
&=& ((\alpha-|I|/2)n-m)g_{n+m}^{(\alpha)},
\EEA
so the $\tilde{f}_n = f^{(1)}_n$ may be considered as the components 
of a centreless Virasoro field $X$, while the $g_m^{(\alpha)}$
are the components of a {\em primary} field $Z_{\alpha}^I$, with conformal 
dimension $1+\alpha-|I|/2$, in the sense of \cite{Bela84}. 

Note also that, as in the cases $N=0,1$ studied in subsection 4.1, 
the conformal dimension $\cdim$ of each field is
equal to the power of $p$ plus one, and the shift in the indices 
(with respect to the power of $q$) is equal to the opposite of the 
power of $p$.

%This formula (\ref{gl:4:poissN}) yields  an operator-product formula which 
%is an extension of a formula due to Kac and Cheng
%({\tt see [?],[Roger]}). Consider the formal field in ${\cal P}^{(N)}$
%\BEQ
%\Theta_I^{(\alpha)}(z)=\sum_{n\in\Z} (t^n \theta^I p^{\alpha-|I|/2}) z^{-n-1},
%\EEQ  
%with $I=\{i_1,\cdots,i_k\}\subset\{1,\ldots,N\}$ and  
%$\theta^I=\theta^{i_1}\wedge\cdots\wedge \theta^{i_k}.$
%If $I,J$ are two families of indices, we denote by $I\bullet J$ the 
%concatenation of $I$ and $J$. 
%Note that the fields $\Theta_I^{(1)}$ can be seen as a
% set of formal generators of
%& $\kk(1|N)$ by the 
%&isomorphism $\g^{(N)}\cong \kk(1|N)$, with (almost) the 
% same notations as in {\tt [Roger]}
%. Then one has the following OPE's in ${\cal Q}^{(2|N)}$:

%&\noindent {\bf Proposition 4.3.} {\it The following results holds true.}
%\begin{eqnarray}
%\Theta_I^{(\alpha)}(z) \cdot \Theta_J^{(\beta)}(w) & \sim & 
%(|I|+|J|-2(\alpha+\beta))
% {\Theta_{I\bullet J}^{(\alpha+
%\beta-1)}(w)\over (z-w)^2} \nonumber \\
%& + & (|I|-2\alpha) 
%{\partial_w \Theta_{I\bullet J}^{(\alpha+\beta-1)}(w)\over z-w} \nonumber \\
%& + & (-1)^{|I|} \sum_{i=1}^N { (\partial\Theta_I/\partial \theta_i \ \wedge\ 
%\partial\Theta_J/\partial \theta_i)^{(\alpha+\beta-1)}(w)\over z-w}.
%\end{eqnarray}
%{\tt K\&C correspond \`a $\alpha=\beta=1$.} 

%The OPE's in the quotient ${\cal Q}^{(N)}_0$ are plainly obtained from 
%{\tt (3.3??)} by setting to zero in the above expansion
%all fields $\Theta_I^{(\alpha)}$ with  $\alpha<0$ along with their derivatives.

\subsection{Study of the case $N=2$.}

As follows from the preceding subsection, the superalgebra $\sns^{(2)}$ is 
generated by the fields $Z_{\alpha}^{I}$ where 
$I=\emptyset,\ \{1\},\ \{2\}$ or $\{1,2\}$ and $\alpha=0,\half$ or $1$. Set
\BEQ
X=Z_1^{\emptyset} \;\; , \;\; 
G^{1,2}={1\over \sqrt{2}} \left( Z_1^{\{1\}}\pm \II Z_1^{\{2\}} \right) 
\;\; , \;\; 
N=\II \sqrt{2} Z_1^{\{1,2\}}
\EEQ
for generators of grade one,
\BEQ
Y=\sqrt{2} Z_{\half}^{\emptyset} \;\; , \;\; 
\bar{Y}^{1,2}=\pm Z_{\half}^{\{1\}} +  \II Z_{\half}^{\{2\}} \;\; , \;\;
P=2\II Z_{\half}^{\{1,2\}}
\EEQ
for generators of grade $\half$, and
\BEQ
M=Z_0^{\emptyset} \;\; , \;\; 
\bar{M}^{1,2}={1\over 2\sqrt{2}}\left( \mp Z_0^{\{1\}} + \II Z_0^{\{2\}}\right) 
\;\; , \;\; 
Q={\II\over \sqrt{2}} Z_0^{\{1,2\}}
\EEQ
for generators of grade $0$. Their conformal dimensions are listed in 
table~\ref{tab1}. 

Then the  superalgebra $\sns^{(2)}$ is isomorphic to 
$\kk(1|2)\ltimes \h^{(2)}$, with
\BEA
\kk(1|2) &\cong& \langle X,G^{1,2},N\rangle 
\nonumber \\
\h^{(2)} &\cong& \langle Y,\bar{Y}^{1,2},P\rangle\oplus\langle
     M,\bar{M}^{1,2},Q\rangle
\label{gl:4:Q22}
\EEA
The fields in the first line of eq.~(\ref{gl:4:Q22}) are of grade 1, while the
three first fields in the second line have grade $\half$ and the three 
other grade $0$.
    
Put $\theta=(\theta^1+\II\theta^2)/\sqrt{2},
\bar{\theta}=(\theta^1-\II\theta^2)/\sqrt{2}$, so that 
$\{\theta,\theta\}=\{\bar{\theta},\bar{\theta}\}=0,
\{\theta,\bar{\theta}\}=1$ and 
$\theta \bar{\theta}=-\II\sqrt{2} \theta^1 \theta^2$
(this change of basis is motivated by a need of coherence with 
section 3, see Proposition 4.3 below): 
then these generators are given by the images in 
$\sns^{(2)}=\wit{\cal P}_{\le 1}^{(2|2)} /
\wit{\cal P}_{\le -\half}^{(2|2)}$ of 
\BEA
& & X_{\phi}=\phi(q)p \;\; \hspace{0.55truecm}, \;\;
G_{\phi}^1=\phi(q)\theta p^{\half} \;\; \hspace{0.5truecm}, \;\;
G_{\phi}^2=-\phi(q)\bar{\theta} p^{\half} \;\; \hspace{0.65truecm}, \;\; 
N_{\phi}=-\theta\bar{\theta}\phi(q)
\nonumber \\ 
& & 
\bar{Y}^1_{\phi}=\sqrt{2}\phi(q)\theta \;\; , \;\; 
\bar{Y}^2_{\phi}=-\sqrt{2}\phi(q)\bar{\theta} \;\; \hspace{0.2truecm}, \;\;
Y_{\phi}=\sqrt{2}\phi(q) p^{\half} \;\; \hspace{0.6truecm}, \;\; 
P_{\phi}=-\sqrt{2}\theta\bar{\theta}\phi(q)p^{-\half}
\label{gl:4:suP} \\
& & M_{\phi}=\phi(q) \;\; \hspace{0.65truecm}, \;\; 
\bar{M}^1_{\phi}=\half \phi(q)\theta p^{-\half} \;\; , \;\; 
\bar{M}^2_{\phi}=-\half\phi(q)\bar{\theta} p^{-\half} \;\; , \;\; 
Q_{\phi}=-\half\theta\bar{\theta}\phi(q) p^{-1}. \nonumber
\EEA
Commutators in the Lie superalgebra $\kk(1|2)\cong\langle X,G^{1,2},N\rangle$
are given as follows:
\BEQ
\{ X_{\phi},X_{\psi}\}=X_{\phi' \psi-\phi \psi'} \;\; , \;\; 
\{X_{\phi},G_{\psi}^{1,2}\}=G^{1,2}_{\half \phi'\psi-\phi\psi'} \;\; , \;\; 
\{X_{\phi},N_{\psi}\}=N_{-\phi\psi'}
\EEQ
(in other words, $G^{1,2}_{\phi}$ have conformal dimension ${3\over 2}$, 
and $N_{\phi}$ conformal dimension 1);
\BEA
\{G_{\phi}^i,G_{\psi}^i\}&=&0,\ i=1,2 \;\; ; \;\;
\{N_{\phi},N_{\psi}\}=0 \nonumber\\
\{G_{\phi}^1,G_{\psi}^2\}&=&-X_{\phi\psi}-N_{\half \phi'\psi-\phi\psi'}\\
\{G_{\phi}^{1,2},N_{\psi}\}&=&\mp G_{\phi\psi}^{1,2} \nonumber.
\EEA
Note that $\sns^{(2)}$ is generated (as a Lie algebra) by the fields 
$X, G^{1,2}$ and $Y$ since one has the formula
\BEQ
N_{\half \phi'\psi-\phi\psi'}=\{G_{\phi}^1,G_{\psi}^2\}-X_{\phi\psi}
\EEQ
for the missing generators of grade 1;
\BEQ
\bar{Y}^{1,2}_{\half(\phi'\psi-\phi\psi')}=\{G_{\phi}^{1,2},Y_{\psi}\}
\;\; , \;\; 
-\half P_{\phi\psi'}=\{G_{\phi}^1,\bar{Y}_{\psi}^2\}+Y_{\phi\psi}
\EEQ
for the missing generators of grade $\half$; and
\BEQ
M_{\half(\phi'\psi-\phi\psi')}=\{Y_{\phi},Y_{\psi}\} \;\; , \;\; 
\bar{M}^{1,2}_{\phi\psi'}=\{G_{\phi}^{1,2},
M_{\psi}\}=\half\{Y_{\phi},\bar{Y}_{\psi}^{1,2}\} \;\; , \;\; 
2Q_{(\phi'\psi+\phi\psi')}=\{Y_{\phi},P_{\psi}\}
\EEQ
for the generators of grade 0.

\noindent {\bf Proposition 4.3.} {\it
\begin{enumerate}
\item  
The  subspace ${\cal R}:= \langle M_{\phi}-Q_{\phi'},\bar{M}_{\phi}^1\rangle$, 
(with $\phi'(t)=\D\phi(t)/\D t$)  is an ideal of $\sns^{(2)}$ strictly included
in the ideal of elements of grade zero.
\item
The quotient Lie  algebra $\sns^{(2)}/{\cal R}$ has a realization in terms 
of differential operators of first order that extends
the representation of $\tilde{\s}^{(2)}$ given e.g. in appendix B : 
the formulas read (in decreasing order of conformal dimensions)
\begin{eqnarray}
-X_{\phi} &\to & \phi(t) \partial_t+\half \phi'(t) 
(r\partial_r+\theta^1\partial_{\theta^1})+\frac{x}{2} \phi'(t)
 +{1\over 4}{\cal M} \phi''(t) r^2+{1\over 4} \phi''(t)
	  r\theta^1\partial_{\theta^2} \quad \nonumber\\
-Y_{\phi} &\to & \phi(t)\partial_r+{\cal M} \phi'(t) r+\half\phi'(t) 
\theta^1\partial_{\theta^2}
\quad \nonumber\\
-M_{\phi} &\to & {\cal M}\phi(t) \nonumber\\
-N_{\phi} &\to&  \phi(t)\left(\theta^1\partial_{\theta^1}+
\theta^2\partial_{\theta^2}-x\right)-{{\cal M}\over 4}
\phi'(t) r^2+ {1\over 4}  \phi'(t) r\theta^1\partial_{\theta^2} \quad 
\nonumber \\
-P_{\phi} &\to& \phi(t) \left(\theta^1\partial_{\theta^2}-2{\cal M}r\right)
\nonumber \\
-Q_{\phi} &\to& {\cal M}\int \phi \quad 
\label{gl:4:supair}
\end{eqnarray}
for the even generators, and 
\begin{eqnarray}
-G_{\phi}^2 &\to & \phi(t)\left(\theta^1\partial_t+\theta^2\partial_r\right)
+\phi'(t) \left(\half \theta^1 r\partial_r+x\theta^1+{\cal 
M}r\theta^2-\half\theta^1\theta^2\partial_{\theta^2} \right) 
+ {{\cal M}\over 2} \phi''(t) r^2\theta^1   
\nonumber\\
-G_{\phi}^1 &\to & \phi(t)\partial_{\theta^1}
+\half \phi'(t) r\partial_{\theta^2} \quad
\nonumber \\
-\bar{Y}_{\phi}^1 &\to & \phi(t)\partial_{\theta^2} 
\nonumber\\
-\bar{Y}_{\phi}^2 &\to&  \phi(t) \left(\theta^1 \partial_r+2 {\cal M} 
\theta^2\right)+ 2{\cal M} \phi'(t) r\theta^1
\quad \nonumber\\
-\bar{M}_{\phi}^1 & \to & 0 \nonumber\\
-\bar{M}_{\phi}^2 &\to&  {\cal M}\phi(t)\theta^1 
\label{gl:4:suimpair}
\end{eqnarray}
for the odd generators. Their conformal dimensions are listed in 
table~\ref{tab1} and their grades in table~\ref{tab2}.
\end{enumerate} }

\noindent {\bf Proof.}
\begin{enumerate}
\item
Since $\sns^{(2)}$ is generated as a Lie algebra by the fields $X,G^{1,2}$ 
and $Y$, one only needs to check
that $[X_{\phi},{\cal R}]\subset {\cal R},\ [G_{\phi}^{1,2},{\cal R}]\subset 
{\cal R}$ and $[Y_{\phi},{\cal R}]\subset {\cal R}$ for any $\phi$. 
Then straightforward computations show that
\BEA
{}[X_{\phi},M_{\psi}-Q_{\psi'}] &=& -M_{\phi\psi'} + Q_{(\phi\psi')'} 
\nonumber \\
{}[G_{\phi}^{i},M_{\psi}-Q_{\psi'}] &=& -2\del_{i,1} \bar{M}_{\phi\psi'}^1
\nonumber \\ 
{}[G_{\phi}^i,\bar{M}^1_{\psi}]&=&
-\half \del_{i,2} (M_{\phi\psi}- Q_{(\phi\psi)'})
\nonumber
\EEA
while $[Y_{\phi},M_{\psi}- Q_{\psi'}]=[Y_{\phi},\bar{M}_{\psi}^1]=0$ by the 
definition of $\sns^{(2)}$ as a quotient.
\item 
This is a matter of straightforward but tedious calculations. 
\end{enumerate}
\eop

\noindent {\bf Remarks.} 
\begin{enumerate}
\item
Each of the above generators is homogeneous  with respect to an $\R^3$-valued 
graduation for
which $t,r,\theta^1$ are independent measure units and $[{\cal M}]\equiv 
[t/r^2]$, $[\theta^2]\equiv [\theta^1 r/t]$.
\item
One may read (up to an overall translation) the conformal dimensions of the 
fields by putting $[t]=-1,$ 
$[r]=[\theta^1]=-\half,$ $ [\theta^2]=[{\cal M}]=0$.
\item
Consider the two distinct embeddings $\tilde{\s}^{(2)}\subset\sns^{(2)}$ and 
$\tilde{\s}^{(2)}\subset {\cal P}^{(2|2)}$ with respective graduations $\gra$ 
(see definition~8) and $\widetilde{\deg}$ (defined in Proposition 3.7). 
Then both graduations coincide on $\tilde{\s}^{(2)}$. 
In particular, the Lie subalgebra 
$\tilde{\s}^{(2)}_1\cong\osp(2|2)\subset\tilde{\s}^{(2)}$ may be 
defined either as the set of elements $X$ of $\tilde{\s}^{(2)}$ with 
$\widetilde{\deg} X =1$ or else as the set of elements with
$\gra X=1$, depending on whether
one looks at $\tilde{\s}^{(2)}$ as sitting inside ${\cal P}^{(2|2)}$ or 
inside $\tilde{\cal P}^{(2|2)}$.
\item When we reconsider the four operators (\ref{gl:3:Sops}) for the
supersymmetric equations of motion
\BD
{\cal S} = 2 M_0 X_{-1} - Y_{-\half}^2 \;\; , \;\;
{\cal S}' = 2 M_0 G_{-\half}^1 - Y_{-\half}\bar{Y}_0^1 \;\; , \;\;
\bar{\cal S}' = Y_{-\half} G_{-\half}^1 - X_{-1} \bar{Y}_0^1 \;\; , \;\;
{\cal S}'' = G_{-1/2}^1 \bar{Y}_0^1
\ED
and use the Poisson algebra representation (\ref{gl:4:suP}) of $\sns^{(2)}$, 
then ${\cal S}= {\cal S}'=\bar{\cal S}' ={\cal S}''=0$. The consequences of
this observation remain to be explored. 
\item One may check that the algebra (\ref{gl:1:ext}) cannot be obtained by 
any of the Poisson quotient 
constructions introduced at the beginning of this section. 
\end{enumerate}

%%%%%%%%%%%%%%%%%%%%%%%%%%%%%%%%%%%%%%%%%%%%%%%%%%%%%%%%%%%%%%%%%%%%%%%%%%%%%%%%
\section{Two-point functions}
%%%%%%%%%%%%%%%%%%%%%%%%%%%%%%%%%%%%%%%%%%%%%%%%%%%%%%%%%%%%%%%%%%%%%%%%%%%%%%%%

We shall compute in this section the two-point functions 
$\langle \Phi_1 \Phi_2\rangle$ that are covariant
under some of the Lie subalgebras of $\osp(2|4)$ introduced previously. 
Consider the two superfields
\BEA
\Phi_1 &=& \Phi_1(t_1,r_1,\theta_1,\bar{\theta}_1)=f_1(t_1,r_1)+
\phi_1(t_1,r_1)\theta_1+\bar{\phi}_1(t_1,r_1)\bar{\theta}_1+g_1(t_1,r_1)
\theta_1\bar{\theta}_1
\nonumber \\
\Phi_2 &=&\Phi_2(t_2,r_2,\theta_2,\bar{\theta}_2)=f_2(t_2,r_2)+
\phi_2(t_2,r_2)\theta_2+\bar{\phi}_2(t_2,r_2)\bar{\theta}_2+g_2(t_2,r_2)
\theta_2\bar{\theta}_2
\label{gl:5:sufeld}
\EEA
with respective masses and scaling dimensions $({\cal M}_1,x_1)$ and 
$({\cal M}_2,x_2)$. With respect to equation (\ref{gl:3:superc}), 
we performed a change of notation. 
The Grassmann variable previously denoted by $\theta^1$ is now called 
$\theta$ and the Grassmann variable $\theta^2$ is now called $\bar{\theta}$. 
The {\it lower} indices
of the Grassmann variables now refer to the first and second superfield,
respectively. The two-point function is 
\BEQ
{\cal C}(t_1,r_1,\theta_1,\bar{\theta}_1;t_2,r_2,
\theta_2,\bar{\theta}_2):=\langle \Phi_1(t_1,r_1,\theta_1,\bar{\theta}_1)
\Phi_2(t_2,r_2,\theta_2,\bar{\theta}_2)\rangle.
\EEQ
Since we shall often have invariance under translations in either space-time or 
in superspace, we shall use the following abbreviations
\BEQ \label{gl:5:inva}
t : = t_1 - t_2 \;\; , \;\;
r := r_1 - r_2 \;\; , \;\;
\theta : = \theta_1 - \theta_2 \;\; , \;\;
\bar{\theta} := \bar{\theta}_1 - \bar{\theta}_2
\EEQ
The generators needed for the following calculations are collected in
appendix~B. 

\noindent
{\bf Proposition 5.1.} {\it The $\osp(2|4)$-covariant two-point function is, 
where the constraints $x:= x_1=x_2$ and 
${\cal M} := {\cal M}_1 = - {\cal M}_2$ hold true and $c_{2}$ is 
a normalization constant}
\BEQ \label{gl:5:osp24}
\mathcal{C} = 
c_2 \delta_{x,\half}\, t^{-\half} 
\exp\left(-\frac{\cal M}{2}\frac{r^2}{t}\right)
\left( \bar{\theta} -{r\over 2t}\theta \right)
\EEQ

In striking contrast with the usual `relativistic' $N=2$ superconformal theory, 
see e.g. \cite{Dola02,Nagi05b,Park00}, we find that covariance under a 
finite-dimensional Lie algebra is enough to fix the scaling dimension of 
the quasiprimary fields. We have already pointed out that this surprising 
result can be traced back to our non-relativistic identification of the 
dilatation generator $X_0$ as $-\half E_{11} - \half x$ in proposition 3.1. 
 
It is quite illuminating to see how the result (\ref{gl:5:osp24}) is 
modified when one considers
two-point functions that are only covariant under a subalgebra $\g$ of 
$\osp(2|4)$. We shall consider the following four cases and refer to 
figure~\ref{Abb3} for an illustration how these algebras are embedded into
$\osp(2|4)$. 
\begin{enumerate}
\item $\g=\tilde{\s}^{(1)}$, which describes invariance under an $N=1$ 
superextension of the Schr\"odinger algebra;
\item $\g=\tilde{\s}^{(2)}$, which describes invariance under an $N=2$ 
superextension of the Schr\"odinger algebra;
\item $\g=\osp(2|2)$, where, as compared to the previous case 
$\g=\tilde{\s}^{(2)}$, invariance under spatial
translations is left out, which opens prospects for a future application to
non-relativistic supersymmetric systems with a boundary;
\item $\g=\se(3|2)$, for which  time-inversions are left out.
\end{enumerate}
{}From the cases 2 and 4 together, the proof of the proposition 5.1 will be
obvious. 

\noindent {\bf Proposition 5.2.} {\it The non-vanishing two-point function,
$\tilde{\s}^{(1)}$-covariant under the representation 
\newline \typeout{*** hier ist ein Zeilenvorschub ***}
(\ref{gl:3:se32},\ref{gl:3:N0},\ref{gl:3:X1},\ref{gl:3:Gh}), 
of the superfields $\Phi_{1,2}$ of the form (\ref{gl:5:sufeld}) is given by}
\BEQ \label{gl:5:2P}
{\cal C} = \delta_{x_1,x_2} \delta_{{\cal M}_1+{\cal M}_2,0} \left[ 
c_1 {\cal C}_1 + c_2 {\cal C}_2 \right]
\EEQ
{\it where $c_{1,2}$ are constants and}
\BEA
{\cal C}_1(t,r;\theta_1,\bar{\theta}_1,\theta_2,\bar{\theta}_2) &=&
t^{-x_1}\exp\left(-\frac{{\cal M}_1}{2}\frac{r^2}{t}\right) 
\left\{ 1 + {1\over t} \left(-x_1+\frac{{\cal M}_1}{2} {r^2\over t}\right) 
\theta_1\theta_2 +2{\cal M}_1 \bar{\theta}_1 \bar{\theta}_2 \right.
\nonumber \\
& & \left. 
- {\cal M}_1\ {r\over t}\ \left(\theta_1\bar{\theta_2}
-  \theta_2\bar{\theta_1}\right) 
-{1\over t}\ {\cal M}_1 (2x_1-1)\theta_1\theta_2\bar{\theta_1}\bar{\theta_2}
\right\}
\label{gl:5:2Pdetail} \\
{\cal C}_2(t,r;\theta_1,\bar{\theta}_1,\theta_2,\bar{\theta}_2) &=&
t^{-x_1}e^{-{\cal M}_1r^2/2t} \left\{
  -{r\over 2t} (\theta_1-\theta_2)+(\bar{\theta}_1-\bar{\theta_2})
+{1\over t} \left(\half-x_1\right) 
(\bar{\theta_1}\theta_1\theta_2-\bar{\theta_2}\theta_1
\theta_2) \right\}.
\nonumber
\EEA

The proof is given in appendix~A. 

For ordinary quasiprimary superfields with fixed masses,
the two-point function ${\cal C} =c_1 {\cal C}_1+c_2 {\cal C}_2$ reads
as follows, where we 
suppress the obvious arguments $(t,r)$ and also the constraints 
$x:=x_1=x_2$ and ${\cal M}:={\cal M}_1=-{\cal M}_2$\ : 
\BEA 
\langle f_1 f_2\rangle &=& c_1\  t^{-x} e^{-{{\cal M}r^2\over 2t}}
\nonumber \\
\langle \phi_1\phi_2\rangle &=& c_1\ 
\left(-x+\frac{{\cal M}}{2} {r^2\over t}\right)
\ t^{-x-1} e^{-{{\cal M}r^2\over 2t}}
\nonumber \\
\langle \bar{\phi}_1\bar{\phi}_2\rangle&=& 2{\cal M} c_1\  t^{-x} 
e^{-{{\cal M}r^2\over 2t}}
\label{gl:5:s1kov} \\
\langle \phi_1\bar{\phi}_2 \rangle &=& 
\langle \bar{\phi}_1\phi_2\rangle \:=\: -c_1\,
{\cal M}r\, t^{-x-1} e^{-{{\cal M}r^2\over 2t}}
\nonumber \\
\langle g_1 g_2\rangle &=&-c_1\ {\cal M} (2x-1) 
t^{-x-1} e^{-{{\cal M}r^2\over 2t}}
\nonumber \\
\langle f_1 \phi_2\rangle &=& -\langle \phi_1 f_2\rangle 
\:=\: c_2 \ {r\over 2}\  t^{-x-1} e^{-{{\cal M}r^2\over 2t}}
\nonumber \\
\langle f_1 \bar{\phi}_2\rangle &=& -\langle \bar{\phi}_1 f_2\rangle
\:=\: -c_2 \ t^{-x} e^{-{{\cal M}r^2\over 2t}}
\nonumber \\
\langle \bar{\phi}_1 g_2\rangle &=& \langle g_1 \phi_2\rangle 
\:=\: -c_2\ \left(\half-x\right) t^{-x-1} e^{-{{\cal M}r^2\over 2t}}
\nonumber \\
\langle \phi_1\bar{\phi}_1\rangle &=& \langle \phi_2\bar{\phi}_2\rangle
= \langle \phi_1 g_2\rangle\:=\: \langle g_1\bar{\phi}_2\rangle\:=\: 0.
\nonumber 
\EEA

\noindent {\bf Corollary.} {\it Any $\tilde{\s}^{(2)}$-covariant two-point
function has the following form, where $x=x_1=x_2$ and 
${\cal M}={\cal M}_1=-{\cal M}_2$ 
and $c_{1,2}$ are normalization constants} 
\BEQ \label{gl:5:st2}
\mathcal{C} = c_1\delta_{x,0}\,\delta_{{\cal M},0} 
+ c_2 \delta_{x,\half}\, t^{-\half} 
\exp\left(-\frac{\cal M}{2}\frac{r^2}{t}\right)
\left( \bar{\theta} -{r\over 2t}\theta \right)
\EEQ

For the proof see appendix~A. We emphasize that 
covariance under $\tilde{\s}^{(2)}$
is already enough to fix $x$ to be either $0$ or $\half$. The contrast with 
$\tilde{\s}^{(1)}$ comes from the fact that $\tilde{\s}^{(1)}$ 
does not contain 
the generator $N_0$, while $\tilde{\s}^{(2)}$ does. The only non-vanishing
two-point functions of the superfield components are
\BEA
\langle f_1 f_2 \rangle &=& \delta_{x,0}\delta_{{\cal M},0}\, c_1
\nonumber \\ 
\langle f_1 \phi_2\rangle &=& -\langle \phi_1 f_2\rangle 
\:=\: \delta_{x,\half}\, c_2\,  {r\over 2}\  t^{-3/2} 
\exp\left(-\frac{\cal M}{2}\frac{r^2}{t}\right) 
\label{gl:5:corol} \\
\langle f_1 \bar{\phi}_2\rangle &=&-\langle \bar{\phi}_1 f_2\rangle
\:=\: -\delta_{x,\half}\, c_2\, t^{-1/2} 
\exp\left(-\frac{\cal M}{2}\frac{r^2}{t}\right).
\nonumber
\EEA

We now study the case of covariance under the Lie algebra
\BEQ
\g = \left\langle X_{-1,0,1}, G_{\pm\half}^{1}, G_{\pm\half}^2, N_0
\right\rangle \cong \osp(2|2),
\EEQ
see appendix~B for the explicit formulas. 
We point out that neither space-translations nor phase-shifts are included
in $\mathfrak{osp}(2|2)$, so that the two-point functions will in 
general depend
on both space coordinates $r_{1,2}$, and there will in general be no constraint
on the masses ${\cal M}_{1,2}$. On the other hand, time-translations and
odd translations are included, so that $\cal C$ will only depend on
$t=t_1-t_2$ and $\theta=\theta_1-\theta_2$. 
{}From a physical point of view, the absence of the requirement of 
spatial translation-invariance means that the results might be used to
describe the kinetics of a supersymmetric model close to a boundary surface, 
especially for semi-infinite systems \cite{Henk94,Plei04,Baum05b}. 

\noindent {\bf Proposition 5.3.} {\it There exist non-vanishing, 
$\osp(2|2)$-covariant two-point functions, of
quasiprimary superfields $\Phi_i$ of the form (\ref{gl:5:sufeld}) with 
scaling dimensions $x_i$, $i=1,2$, only  in the three cases 
$x_1+x_2=0,1$ or $2$. Then the two-point functions $\cal C$ 
are given as follows.

\noindent (i) if $x_1+x_2=0$, then necessarily $x_1=x_2=0$,  
${\cal M}_1={\cal M}_2=0$ and
\BEQ
{\cal C} = a_0
\EEQ
where $a_0$ is a constant. 

\noindent (ii) if $x_1+x_2=1$, then 
\BEQ
{\cal C} = t^{-1/2} \left\{ \left( \bar{\theta}_1 - \frac{r_1}{2{t}}
(\theta_1-\theta_2)\right) h_1 +  \sqrt{\frac{{\cal M}_1}{{\cal M}_2}\,}
\left( \bar{\theta}_2 - \frac{r_2}{2{t}}
(\theta_1-\theta_2)\right) h_2\right\}
\EEQ
where
\BEA
h_1 &=& \left(\frac{r_1 r_2}{t}\right)^{1/2} 
\left( \frac{r_1}{r_2}\right)^{-(x_1-x_2)/2}
\left( \alpha J_{\mu}\left(\frac{{\cal M}r_1 r_2}{t}\right) +
\beta J_{-\mu}\left(\frac{{\cal M}r_1 r_2}{t}\right) \right)
\exp\left[ -\frac{{\cal M}_1}{2}\frac{r_1^2}{t} 
+ \frac{{\cal M}_2}{2}\frac{r_2^2}{t}\right] 
\nonumber \\
h_2 &=& \left(\frac{r_1 r_2}{t}\right)^{1/2} 
\left( \frac{r_1}{r_2}\right)^{-(x_1-x_2)/2}
\left( \alpha J_{\mu+1}\left(\frac{{\cal M}r_1 r_2}{t}\right) -
\beta J_{-\mu-1}\left(\frac{{\cal M}r_1 r_2}{t}\right) \right)
\exp\left[ -\frac{{\cal M}_1}{2}\frac{r_1^2}{t} 
+ \frac{{\cal M}_2}{2}\frac{r_2^2}{t}\right] 
\nonumber \\
 & & 
\label{gl:5:ff}
\EEA
and
\BEQ \label{gl:5:abbr}
{\cal M}^2 = {\cal M}_1 {\cal M}_2 \;\; , \;\;
1+2\mu = x_2-x_1
\EEQ
while $J_{\mu}(x)$ are Bessel functions and $\alpha,\beta$ are arbitrary 
constants.

\noindent (iii) if $x_1+x_2=2$, then
\BEQ
{\cal C} = \left( \theta_1-\theta_2\right) \left( 
\bar{\theta}_1 - \frac{r_1}{r_2}\bar{\theta}_2\right) B 
+\bar{\theta}_1\bar{\theta}_2 D
\EEQ
where
\BEA
B &=& t^{-3/2} \left(\frac{r_1^2}{t}\right)^{-(x_1-1/2)}
h\left(\frac{r_1 r_2}{t}\right) \exp\left[ -\frac{{\cal M}_1}{2}\frac{r_1^2}{t} 
+ \frac{{\cal M}_2}{2}\frac{r_2^2}{t}\right] 
\nonumber \\
D &=& \frac{2t}{r_2} B
\label{gl:5:BD}
\EEA
and where $h$ is an arbitrary function.
}

Again, the proof in given in appendix~A.
A few comments are in order. 
\begin{enumerate}
\item In applications, one usually considers either (i) response functions 
which in a standard field-theoretical setting may be written as a correlator
$\langle \phi\wit{\phi}\rangle$ of an order-parameter field $\phi$ with a
`mass' ${\cal M}_{\phi}\geq 0$ and a
conjugate response field $\wit{\phi}$ whose `mass' is non-positive
${\cal M}_{\wit{\phi}}\leq 0$ \cite{Henk02,Pico04} or else (ii) for
purely imaginary masses ${\cal M}=\II m$, correlators
$\langle \phi \phi^*\rangle$ of a field and its complex conjugate \cite{Henk94}.
\item The two supercharges essentially fix the admissible
values of the sum of the scaling dimensions $x_1+x_2$, which is a consequence
of covariance under the supersymmetry generator $N_0$. 
\item For systems which are covariant under the scalar 
Schr\"odinger generators $X_{\pm 1,0}$ only, 
it is known that for scalar quasiprimary fields $\phi_{1,2}$ \cite{Henk94}
\BEQ \label{gl:5:schs}
\langle \phi_1(t_1,r_1)\phi_2(t_2,r_2)\rangle = \delta_{x_1,x_2}
t^{-x_1} h\left(\frac{r_1 r_2}{t_1-t_2}\right)
\exp\left( -\frac{{\cal M}_1}{2}\frac{r_1^2}{t_1-t_2}
           +\frac{{\cal M}_2}{2}\frac{r_2^2}{t_1-t_2}\right)
\EEQ
where $h$ is an arbitrary function. At first sight, there appears 
some similarity of (\ref{gl:5:schs}) with the result (\ref{gl:5:BD}) 
obtained for 
$x_1+x_2=2$ in that a scaling function of a single variable remains arbitrary, 
but already for $x_1+x_2=1$ the very form of the scaling 
function (\ref{gl:5:ff}) is completely distinct. The main difference of 
eq.~(\ref{gl:5:schs}) with our Proposition 5.3 is that here 
we have a condition
on the {\em sum} of the scaling dimensions, whereas (\ref{gl:5:schs}) rather
fixes the {\em difference} $x_1-x_2=0$. 
\end{enumerate}

\noindent {\bf Proposition 5.4} {\it The non-vanishing two-point function which
is covariant under $\mathfrak{se}(3|2)$ is, where $x=x_1+x_2$ and $c_2$ and 
$d_0$ are normalization constants}
\BEA
{\cal C} &=& c_2\, \delta_{x,1} \delta_{{\cal M}_1+{\cal M}_2,0}\, t^{-1/2} 
\exp\left(-\frac{{\cal M}_1}{2}\frac{r^2}{t}\right) 
\left( \bar{\theta}-\frac{r}{2t}\theta\right)
\nonumber \\
& & +d_0\, \delta_{{\cal M}_1+{\cal M}_2,0}\, {\cal M}_1^{(x-1)/2} t^{-(x+1)/2}
\exp\left(-\frac{{\cal M}_1}{2}\frac{r^2}{t}\right) \theta \bar{\theta}
\label{gl:5:se32}
\EEA 

This makes it clear that one needs both $N=2$ supercharges and the
time-inversions $t\mapsto -1/t$  in order to obtain a finite list of
possibilities for the scaling dimension $x=x_1+x_2$. 

%%%%%%%%%%%%%%%%%%%%%%%%%%%%%%%%%%%%%%%%%%%%%%%%%%%%%%%%%%%%%%%%%%%%%%%%%%%%%%%%
\section{Conclusions}
%%%%%%%%%%%%%%%%%%%%%%%%%%%%%%%%%%%%%%%%%%%%%%%%%%%%%%%%%%%%%%%%%%%%%%%%%%%%%%%%

Motivated by certain formal analogies between $2D$ conformal invariance and
Schr\"odinger-invariance, we have attempted to study some mathematical
aspects of Lie superalgebras which contain the Schr\"odinger algebra
$\sch_1$ as a subalgebra. Our discussion has been largely based on the
free non-relativistic particle, either directly through the free
Schr\"odinger (or diffusion) equation, or else as a two-component spinor
which solves the Dirac-L\'evy-Leblond equations. In both cases, it is useful to
consider the (non-relativistic) mass parameter $\cal M$ as an 
additional variable which allows to extend the dynamical symmetry algebra from 
the Schr\"odinger algebra to a full conformal algebra 
$(\conf_3)_{\C}\supset\sch_1$. 

Including these building blocks into a superfield formalism, we have shown
that the solution $f$ of the Schr\"odinger equation, the spinor 
$\left(\vekz{\psi}{\phi}\right)$ and an auxiliary field $g$ form a
supermultiplet such that the equations of motion are supersymmetric invariant,
with  $N=2$ supercharges.
Depending on whether the mass is considered as a constant or as an 
additional variable, we have defined two free-particle models, 
see table~\ref{tab0}. Furthermore, taking the scale-invariance and even the 
invariance under time inversions $t\mapsto -1/t$ into account, we have
shown that the supersymmetries of these models can be extended to the
superalgebra $\tilde{\s}^{(2)}\cong \osp(2|2)\ltimes\sh(2|2)$ for a fixed mass 
$\cal M$ and further to $\s^{(2)}\cong \osp(2|4)$
when $\cal M$ is considered as a variable. These results take on a particularly
transparent form when translated into a Poisson-algebra language. 
In this context, we have seen that several distinct gradings of the 
superalgebras provided useful insight. 

Motivated by the known extension of $\sch_1$ to the infinite-dimensional
Schr\"odinger-Virasoro algebra $\sv$, and by the extension of the Virasoro
algebra by the Neveu-Schwarz algebra, we then looked for similar 
extensions of the Lie superalgebras of Schr\"odinger type found so far. 
By introducing 
{\em twisted} Poisson algebras, we defined
the {\em Schr\"odinger-Neveu-Schwarz algebras} $\sns^{(N)}$ with $N$ 
supercharges and  derived explicit formulas for the generators, both in a 
Poisson geometry setup, see eq.~(\ref{gl:4:suP}), and as linear differential 
super-operators, see eqs.~(\ref{gl:4:supair},\ref{gl:4:suimpair}), and 
obtained in particular
an explicit embedding of $\tilde{\s}^{(2)}$ into $\sns^{(2)}$. 

Finally, we derived  explicit predictions 
(see section~5) for the two-point functions of quasiprimary superfields of 
models satisfying some or all of  the non-relativistic $N=2$ supersymmetries 
of either free-particle model. Remarkably,
the presence of the supersymmetric generator $N_0$ essentially fixes the 
sum of the scaling dimensions $x_1+x_2$ of the two quasiprimary superfields,
rather than their difference as commonly seen in relativistic superconformal
theories. In particular, non-zero  $\osp(2|4)$-covariant two-point functions 
arise only for a scaling dimension equal to $\half$, and are completely 
determined (up to normalization).
This surprising result appears to be peculiar to non-relativistic systems. 
Physically, this result means that only
the simple random walk (or rather its supersymmetric extension) 
has a non-vanishing two-point function which is covariant under the $N=2$ 
super-Schr\"odinger-invariance with time-inversions included, as 
constructed in this paper. 

We have left open many important questions, of which we merely mention two. 
First, it remains to be seen what the possible central extensions of the
superalgebras $\sns^{(N)}$ are; second, are there richer physical models
than the free particle which realize these non-relativistic supersymmetries~?

\newpage 

%%%%%%%%%%%%%%%%%%%%%%%%%%%%%%%%%%%%%%%%%%%%%%%%%%%%%%%%%%%%%%%%%%%%%%%%%%%%%%%%
\appsection{A}{Supersymmetric two-point functions}
%%%%%%%%%%%%%%%%%%%%%%%%%%%%%%%%%%%%%%%%%%%%%%%%%%%%%%%%%%%%%%%%%%%%%%%%%%%%%%%%
\subsection{$\tilde{\s}^{(1)}$-covariant two-point functions}

We prove the formulas (\ref{gl:5:2P}) and (\ref{gl:5:2Pdetail}) of 
proposition 5.2 for the two independent $\tilde{\s}^{(1)}$-covariant 
two-point functions.

Let $\Phi_1$ and $\Phi_2$ two superfields with respective masses and
dimensions $({\cal M}_1,x_1)$, $({\cal M}_2,x_2)$ as in Section 5, and
let ${\cal C}(t_1,r_1,\theta_1,\bar{\theta_1};t_2,r_2,\theta_2,\bar{\theta}_2)
:=\langle \Phi_1(t_1,r_1,\theta_1)\Phi_2(t_2,r_2,\theta_2)\rangle$
be the associated two-point function. One assumes that ${\cal C}$ is covariant 
under the Lie symmetry representation (see e.g. appendix~B for a list of the
generators) of the `chiral' superalgebra $\tilde{\s}^{(1)}$
generated by $X_{-1},X_0,X_1,Y_{\pm\half},
G_{-\half}:=G^1_{-\half}+G^2_{-\half},
\bar{Y}_0:=\bar{Y}_0^1+\bar{Y}_0^2$ and $M_0$.

Because of the invariance under time- and space-translations 
$X_{-1}$, $Y_{-\half}$ and
under the mass generator $M_0$, the two-point function $\cal C$ depends on 
time and space only through the coordinates
$t:=t_1-t_2$ and $r:=r_1-r_2$, and one can assume that $\Phi_1$ and
$\Phi_2$ have opposite masses. We set ${\cal M}={\cal M}_1=-{\cal M}_2.$

Covariance under $X_0$ of the two-point function $C$ gives
\begin{equation}
\left( t\partial_t+\half r\partial_r+\half(\theta_1\partial_1+\theta_2
\partial_2)+\half(x_1+x_2)\right){\cal C}
(t,r,\theta_1,\bar{\theta}_1,\theta_2,\bar{\theta}_2)=0.
\end{equation}
Covariance under $Y_{\half}$ gives
\begin{equation}
\left( t\partial_r+{\cal M}r+\half (\theta_1 \bar{\partial}_1+\theta_2
\bar{\partial}_2) \right) {\cal C}
(t,r,\theta_1,\bar{\theta}_1,\theta_2,\bar{\theta}_2)=0.
\end{equation}
Covariance under $X_1$ entails
\begin{equation}
\left( t^2\partial_t+tr\partial_r+t\theta_1\partial_1+tx_1+\half {\cal M}
r^2+\half r\theta_1 \bar{\partial}_1 \right)
{\cal C}(t,r,\theta_1,\bar{\theta}_1,\theta_2,\bar{\theta}_2)=0.
\end{equation}
Covariance under $G_{-\half}$ yields
\begin{equation} \label{gl:A:4}
\left( \partial_1+\partial_2+\theta \partial_t+\bar{\theta} \partial_r \right)
{\cal C}(t,r,\theta_1,\bar{\theta}_1,\theta_2,\bar{\theta}_2)=0.
\end{equation}
Finally, covariance under $\bar{Y}_0$ yields
\begin{equation}
\left( \bar{\partial}_1+\bar{\partial}_2+(\theta_1-\theta_2)\partial_r
+2{\cal M}(\bar{\theta}_1-\bar{\theta}_2) \right)
{\cal C}(t,r,\theta_1,\bar{\theta}_1,\theta_2,\bar{\theta}_2)=0.
\end{equation}
In general, the two-point function may be written as
\begin{eqnarray}
{\cal C}(t,r,\theta_1,\bar{\theta}_1,\theta_2,\bar{\theta}_2)&=&
A(t,r)+B_i(t,r)\theta_i+\bar{B}_{i}(t,r)\bar{\theta}_i \nonumber \\
& & +C_{12}\theta_1\theta_2+C_{\bar{1}\bar{2}}\bar{\theta}_1\bar{\theta}_2
+C_{1\bar{1}}\theta_1\bar{\theta}_1+C_{2\bar{2}} \theta_2\bar{\theta}_2
+C_{1\bar{2}}\theta_1\bar{\theta}_2+C_{2\bar{1}}\theta_2\bar{\theta}_1
\nonumber \\
& & +D_1(t,r) \bar{\theta}_1\theta_1\theta_2+D_2(t,r)\bar{\theta_2}\theta_1
\theta_2+\bar{D}_1(t,r) \theta_1\bar{\theta}_1\bar{\theta}_2 + \bar{D}_2(t,r)
\theta_2\bar{\theta}_1\bar{\theta}_2 \nonumber \\
& & +E(t,r) \theta_1\theta_2\bar{\theta}_1\bar{\theta}_2.
\end{eqnarray}
where $i$ is summed over $i=1,2$.

Covariance under $G_{-\half}$ eq.~(\ref{gl:A:4}) gives the following 
system of linearly independent equations:
\newpage \typeout{ *** hier ein Seitenvorschub im Anhang A *** }
\begin{eqnarray}
B_1 &=&-B_2 %\quad {\mathrm{(coefficient\ of\ 1)}}
\\
C_{12} &=&\partial_t A %\quad {\mathrm{(coefficient\ of\ }} \theta_i)
\\
C_{1\bar{1}}+C_{2\bar{1}}\:=:
-\partial_r A,\ C_{1\bar{2}}+C_{2\bar{2}}&=&\partial_rA 
%\quad {\mathrm{(coefficient\ of\ }} \bar{\theta}_i)
\\
\partial_r (\bar{B}_1+\bar{B}_2)+(\bar{D}_1+\bar{D}_2)&=&0 
%\quad{\mathrm{(coefficient\ of\ }} \bar{\theta}_1\bar{\theta}_2)
\\
\partial_t \bar{B}_1-\partial_r B_1-D_1&=& 0 \label{gl:A:A11} \\
\partial_t \bar{B}_2+\partial_r
B_1-D_2&=&0 
%\quad {\mathrm{(coefficients\ of\ }} \theta_i\bar{\theta}_j)
\\
\partial_t C_{\bar{1}\bar{2}}-\partial_r C_{1\bar{1}}-\partial_r
C_{1\bar{2}}-E&=&0 
%\quad{\mathrm{(coefficients\ of\ }} \theta\bar{\theta}^2,\bar{\theta}\theta^2)
\\
\partial_t (\bar{D}_1+\bar{D}_2)+\partial_r (D_1+D_2)&=&0 
%\quad{\mathrm{(coefficient\ of\ }} \theta^2\bar{\theta}^2).
\end{eqnarray}
%where we used the short-hand $\theta\bar{\theta}^2
%=\theta_1\bar{\theta_1}\bar{\theta}_2$ or
%$\theta_2\bar{\theta_1}\bar{\theta}_2$ and so on. 

Covariance under $\bar{Y}_0$ gives the following system of linearly
independent equations:
\begin{eqnarray}
\bar{B}_1&=&-\bar{B}_2 %\quad {\mathrm{(coefficient\ of\ 1)}}
\\
\partial_r A \:=\: C_{1\bar{1}}+C_{1\bar{2}}, -\partial_rA
&=& C_{2\bar{1}}+C_{2\bar{2}}
%\quad {\mathrm{(coefficient\ of\ }} \theta_i)
\\
2{\cal M}A &=& C_{\bar{1}\bar{2}}
%\quad {\mathrm{(coefficient\ of\ }} \bar{\theta}_i)
\\
\partial_r (B_1+B_2)+(D_1+D_2)&=&0 
%\quad{\mathrm{(coefficient\ of\ }} \theta_1\theta_2
\\
\partial_r \bar{B}_1-2{\cal M}  B_1+\bar{D}_1&=&0 \label{gl:A:A19} \\
- \partial_r \bar{B}_2+2{\cal M} B_2-\bar{D}_2&=&0 
%\quad {\mathrm{(coefficients\ of\ }} \theta_i\bar{\theta}_j)
\\
\partial_r C_{1\bar{1}}+\partial_r C_{2\bar{1}}+2{\cal M}C_{12}-E&=&0 
%\quad {\mathrm{(coefficients\ of}}\theta\bar{\theta}^2,\bar{\theta}\theta^2)
\\
\partial_r (\bar{D}_1+\bar{D}_2)+2{\cal M} (D_1+D_2)&=&0 
%\quad{\mathrm{(coefficient\ of\ }} \theta^2\bar{\theta}^2).
\end{eqnarray}
Combining these relations, we can express all the coefficients
of ${\cal C}$ in terms of $A,B_1,\bar{B}_1,D_1,\bar{D}_1$ and $\Gamma:=
C_{1\bar{1}}=C_{2\bar{2}}$ through the obvious relations $B_1+B_2=
\bar{B}_1+\bar{B}_2=D_1+D_2=\bar{D}_1+\bar{D}_2=0$ and the 
(less obvious) relations
\BEQ 
C_{2\bar{1}}=-\partial_r A-\Gamma,\ C_{1\bar{2}}=\partial_r A-\Gamma,\
C_{12}=\partial_t A,\ C_{\bar{1}\bar{2}}=2{\cal M}A.
\EEQ
There remain only three supplementary equations : 
(\ref{gl:A:A11}), (\ref{gl:A:A19}) and
\BEQ \label{gl:A:A24}
(2{\cal M}\partial_t-\partial_r^2)A-E=0.
\EEQ
Recall that the only solution (up to scalar multiplication) of the
equations 
\BEQ
(t\partial_r+{\cal M}r)F(t,r)=(t\partial_t+\half r\partial_r
+\lambda)F(t,r)=0
\EEQ
is $F(t,r)={\cal F}_{\lambda}(t,r):=t^{-\lambda}e^{-{\cal M}r^2/2t}$, 
which one might call a `Schr\"odinger quasiprimary function'. 
Looking now at the consequences of $X_0$- 
and $Y_{-\half}$-covariance, one understands easily that the coefficients in 
${\cal C}$ of the polynomials
in $\theta_i,\bar{\theta}_i$ that depend on $\theta_1,\theta_2$ only through
$\theta_1\bar{\theta}_1$ and $\theta_2\bar{\theta}_2$ are Schr\"odinger
quasiprimary functions, namely:
\BEA
A\:=\:a{\cal F}_{\half(x_1+x_2)} \;\; , \;\;
\bar{B}_1 &=& \bar{b} {\cal F}_{\half(x_1+x_2)} 
\nonumber \\
\bar{D}_1\:=\:\bar{d} {\cal F}_{\half(x_1+x_2)+\half} \;\; , \;\;
\Gamma&=&\gamma{\cal F}_{\half(x_1+x_2)+\half}
\\
E &=&e{\cal F}_{\half(x_1+x_2)+1}
\nonumber 
\EEA
with yet undetermined constants $a,\bar{b},\bar{d},\gamma,e$. This, together
with the previous relations, allows one to express all the coefficients
of $\cal C$ in terms of these constants, since all other coefficients
are derived directly from $A,\bar{B}_1,\bar{D}_1,\Gamma$ and $E$. 
Equation~(\ref{gl:A:A24}) gives
\BEQ 
e=-a{\cal M} (x_1+x_2-1).
\EEQ 
Finally, it remains to check covariance under $X_1$, which gives
constraints on the scaling dimensions of the Schr\"odinger quasiprimary
coefficients, namely:  $x_1=x_2$ unless $a=\bar{b}=0$; $\gamma=\bar{d}=0$ 
(otherwise we would have simultaneously $x_1-x_2=1$ and $x_1-x_2=-1$). 
In order to get a non-zero solution, we have to put 
in the constraint $x_1=x_2=:x,$ and find 
${\cal C}=a{\cal C}_1+\bar{b}{\cal C}_2$.

One then checks that all supplementary relations coming from $Y_{-\half}$,
$X_0$- and $X_1$-covariance are already satisfied. \eop

\subsection{$\tilde{\s}^{(2)}$-covariant two-point functions}

Starting from an $\tilde{\s}^{(1)}$-covariant two-point function ${\cal C}=
a {\cal C}_1+\bar{b}{\cal C}_2$, all there is to do is to postulate invariance
of $\cal C$ under the vector field $G_{-\half}^1=-\partial_1-\partial_2$. 
We find that either $\bar{b}=0$ and then
also $x=0$ and ${\cal M}=0$, or else $a=0$ and furthermore $x=\half$ which
establishes (\ref{gl:5:st2}).  \eop

\subsection{$\mathfrak{osp}(2|2)$-covariant two-point functions}

Here, we prove proposition 5.3. From the definition of
$\mathfrak{osp}(2|2)=\langle X_{\pm 1,0},G_{\pm\half}^{1,2},N_0\rangle$ 
we see that time-translations $X_{-1}=-\partial_{t_1}-\partial_{t_2}$ and odd
translations $G_{-\half}^1=-\partial_{\theta_1}-\partial_{\theta_2}$ 
are included, hence $\cal C$ will only depend on
$t=t_1-t_2$ and $\theta=\theta_1-\theta_2$. From the explicit 
differential-operator representation 
(\ref{gl:3:se32},\ref{gl:3:N0},\ref{gl:3:X1},\ref{gl:3:Gh})
we obtain the following covariance conditions
for ${\cal C}={\cal C}(t,r_1,r_2;\theta,\bar{\theta}_1,\bar{\theta}_2)$
\BEA
-X_0 {\cal C} &=& \left[ t\partial_t 
+\half\left(r_1\partial_{r_1}+r_2\partial_{r_2}+\theta\partial_{\theta}\right)
+\frac{x_1+x_2}{2}\right] {\cal C} = 0
\nonumber \\
-X_1 {\cal C} &=& \left[ t^2\partial_t 
+t\theta\partial_{\theta}+t r_1 \partial_{r_1} + x_1 t 
+\half\left( {\cal M}_1 r_1^2 +{\cal M}_2 r_2^2\right)
+\frac{r_1}{2} \theta\partial_{\bar{\theta}_1} \right]{\cal C}=0
\nonumber \\
-G_{\half}^1 {\cal C} &=& \left[ t\partial_{\theta} +
\half\left(r_1 \partial_{\bar{\theta}_1}+r_2 \partial_{\bar{\theta}_2}
\right)\right] {\cal C} = 0
\nonumber \\
-G_{-\half}^2 {\cal C} &=& \left[ \theta\partial_t +\partial_{r_1}
\bar{\theta}_1 +\partial_{r_2}\bar{\theta}_2\right] {\cal C} = 0 
\nonumber \\
-G_{\half}^2 {\cal C} &=& \left[ t\theta\partial_{t} 
+t\partial_{r_1}\bar{\theta}_1 + \half\left( 
\left( r_1\partial_{r_1}+2x_1\right)\theta +2\left(
{\cal M}_1 r_1\bar{\theta}_1+{\cal M}_2 r_2\bar{\theta}_2\right)
-\theta\bar{\theta}_1\partial_{\bar{\theta}_1}\right)\right]{\cal C} = 0
\nonumber \\
-N_0 {\cal C} &=& \left[ \theta\partial_{\theta} +
\bar{\theta}_1\partial_{\bar{\theta}_1}+\bar{\theta}_2\partial_{\bar{\theta}_2}
-x_1-x_2 \right] {\cal C} = 0 
\EEA
The solutions of this system of equations can be written in the form
\BEQ \label{A3:Gzerlegung}
{\cal C} = A + \theta A_0 + \bar{\theta}_1 A_1 + \bar{\theta}_2 A_2 +
\theta \bar{\theta}_1 B_1 + \theta \bar{\theta}_2 B_2 +
\theta \bar{\theta}_1\bar{\theta}_2 C +\bar{\theta}_1 \bar{\theta}_2 D
\EEQ
where the functions $A=A(t,r_1,r_2), \ldots$ depend on the variables 
$t,r_1,r_2$ and are to be determined. In what follows, the arguments 
of these functions will usually be suppressed. 

First, we consider the condition $G_{\half}^1{\cal C}=0$ which together with
(\ref{A3:Gzerlegung}) leads to the following equations
\newpage\typeout{ *** Seitenvorschub im Anhang A vor Gl. (A30) ***}
\BEA
t A_0 +\frac{r_1}{2} A_1 + \frac{r_2}{2} A_2 &=& 0 
\nonumber \\
t B_1 -\half r_2 D &=& 0 
\nonumber \\
t B_2 + \half r_1 D &=& 0 
\nonumber \\
C &=& 0
\label{A3:Gh1}
\EEA
Next, we use the condition $N_0{\cal C}=0$, which together with
(\ref{A3:Gzerlegung}) leads to
\BEA
(x_1+x_2) A &=& 0 \nonumber \\
(1 -x_1 -x_2) A_i &=& 0 \nonumber \\
(2 - x_1 - x_2) B_j &=& 0 \nonumber \\
(2 - x_1 -x_2) D &=& 0 
\EEA
for $i=0,1,2$ and $j=1,2$. Therefore, we have to distinguish the three cases
$x_1+x_2=0,1,2$, respectively. 

We begin with the case (i) $x_1+x_2=0$. Then $A_i=B_j=C=D=0$. From 
$G_{-\half}^{2}{\cal C}=\left[\theta\partial_t +\bar{\theta}_1\partial_{r_1}
+\bar{\theta}_2\partial_{r_2}\right]A=0$ it follows that
$A=a_0$ is a constant. Furthermore, the covariance $X_1{\cal C}=0$ implies
$x_1=x_2=0$ and ${\cal M}_1={\cal M}_2=0$. 

Next, we consider the case (ii) $x_1+x_2=1$. Then $A=B_i=C=D=0$ and it remains
to find $A_{1,2}$, whereas $A_0$ is given by the first of eqs.~(\ref{A3:Gh1}). 
{}From the condition $G_{-\half}^2{\cal C}=0$, we have
\BEQ \label{A3:glA33}
\partial_t A_1 = \partial_{r_1} A_0 \;\; , \;\;
\partial_t A_2 = \partial_{r_2} A_0 \;\; , \;\;
\partial_{r_1} A_2 = \partial_{r_2} A_1
\EEQ
{}From the condition $G_{\half}^2{\cal C}=0$, we find
\BEA
\left( t\partial_t +\half r_1\partial_{r_1}+\left(x_1-\half\right)\right) A_1
&=& \left( t\partial_{r_1}+{\cal M}_1 r_1\right) A_0 
\nonumber \\
\left( t\partial_t +\half r_1\partial_{r_1}+x_1\right) A_2
&=& {\cal M}_2 r_2 A_0 
\nonumber \\
\left( t\partial_{r_1}+{\cal M}_1 r_1\right) A_2 &=& {\cal M}_2 r_2 A_1
\label{A3:glA34}
\EEA 
Dilatation-covariance $X_0 {\cal C}=0$ gives
\BEA
\left( t\partial_t +\half\left(r_1\partial_{r_1}+r_2\partial_{r_2}\right)
+\half(1+x_1+x_2) \right) A_0 &=& 0 
\nonumber \\
\left( t\partial_t +\half\left(r_1\partial_{r_1}+r_2\partial_{r_2}\right)
+\half(x_1+x_2) \right) A_{1,2} &=& 0
\label{A3:glA35}
\EEA
and finally, covariance under the special transformations $X_1{\cal C}=0$
leads to
\BEA
\left( t\left(r_1\partial_{r_1}-r_2\partial_{r_2}\right) +t(1+x_1-x_2)
+\left( {\cal M}_1 r_1^2 + {\cal M}_2 r_2^2 \right) \right) A_0 
+ r_1 A_1 &=& 0 
\nonumber \\
\left( t\left(r_1\partial_{r_1}-r_2\partial_{r_2}\right) +t(x_1-x_2)
+\left( {\cal M}_1 r_1^2 + {\cal M}_2 r_2^2 \right) \right) A_{1,2}
&=& 0 
\label{A3:glA36}
\EEA 
To solve 
eqs.~(\ref{A3:Gh1},\ref{A3:glA33},\ref{A3:glA34},\ref{A3:glA35},\ref{A3:glA36}),
we use that $x_1+x_2=1$ and have the scaling ansatz
\BEQ
A_{1,2} = t^{-1/2} {\cal A}_{1,2}(u_1,u_2) \;\; , \;\;
A_0 = t^{-1} {\cal A}_0 (u_1,u_2)
\EEQ
where $u_i=r_i/\sqrt{t\,}$, $i=1,2$. Then (\ref{A3:Gh1}) becomes
${\cal A}_0+\half u_1 {\cal A}_1 + \half u_2 {\cal A}_2=0$. On the other hand,
(\ref{A3:glA36}) gives
\BEQ
{\cal A}_{1,2}(u_1,u_2) = h_{1,2}(u_1 u_2) u_1^{x_2-x_1} 
\exp\left[ -\frac{{\cal M}_1}{2} u_1^2 + \frac{{\cal M}_2}{2} u_2^2 \right]
\EEQ
It is now easily seen that the remaining equations all reduce to the 
following system of equations for the two functions $h_{1,2}(v)$
\BEA
\frac{\D h_1(v)}{\D v} &=& - {\cal M}_1 h_2(v) \nonumber \\
\frac{\D h_2(v)}{\D v} &=& {\cal M}_2 h_1(v) + (x_1-x_2) \frac{1}{v} h_2(v)
\EEA
The general solution of these equations is found with standard techniques
\BEA
h_1(v) &=& \alpha' \left(\frac{{\cal M}v}{2}\right)^{-\mu} 
J_{\mu}({\cal M}v) + \beta' \left(\frac{{\cal M}}{2v}\right)^{\mu} 
J_{-\mu}({\cal M}v) 
\nonumber \\
h_2(v) &=& \sqrt{\frac{{\cal M}_2}{{\cal M}_1}\,} \left[ 
\alpha' \left(\frac{{\cal M}v}{2}\right)^{-\mu} 
J_{\mu+1}({\cal M}v) - \beta' \left(\frac{{\cal M}}{2v}\right)^{\mu} 
J_{-\mu-1}({\cal M}v) \right]
\EEA
where we used eq.~(\ref{gl:5:abbr}), $J_{\mu}$ is a Bessel function and 
$\alpha',\beta'$ are arbitrary constants.  Combination 
with (\ref{A3:Gzerlegung}) establishes the second part of the assertion. 

Finally, we consider the third case (iii) $x_1+x_2=2$. Then $A=A_j=C=0$ and 
we still have to find $B_{1,2}$ and $D$. Going through the covariance
conditions, we obtain the following system of equations
\BEQ
D \:=\: \frac{2t}{r_2} B_1 \;\; , \;\; D = -\frac{2t}{r_1} B_2
\EEQ
and
\BEA
\partial_{r_1} B_2 - \partial_{r_2} B_1 &=& \partial_t D 
\nonumber \\
\left( t\partial_t+\half\left( r_1\partial_{r_1}+r_2\partial_{r_2}\right)
+1\right) D &=& 0 
\nonumber \\
\left( t\partial_t+\half\left( r_1\partial_{r_1}+r_2\partial_{r_2}\right)
+\frac{3}{2}\right) B_{1,2} &=& 0 
\nonumber \\
\left( t^2\partial_t +tr_1\partial_{r_1} +t(x_1+1) +\half{\cal M}_1 r_1^2
+\half{\cal M}_2 r_2^2 \right) B_1 &=& 0 
\nonumber \\
\left( t^2\partial_t +tr_1\partial_{r_1} +t x_1 +\half{\cal M}_1 r_1^2
+\half{\cal M}_2 r_2^2 \right) B_2 &=& 0 
\nonumber \\
\left( t^2\partial_t +tr_1\partial_{r_1} +t x_1 +\half{\cal M}_1 r_1^2
+\half{\cal M}_2 r_2^2 \right) D &=& 0 
\nonumber \\
\left( t\partial_t +\half r_1\partial_{r_1} +x_1-\half\right) D -
\left(t\partial_{r_1}+{\cal M}_1 r_1\right) B_2 +{\cal M}_2 r_2 B_1 &=& 0
\label{A3:glA42}
\EEA
We see that $B_2 = -(r_1/r_2) B_1$ and it further follows that 
eqs.~(\ref{A3:glA42}) can be reduced to the system
\BEA
\left( t\partial_t+\half\left( r_1\partial_{r_1}+r_2\partial_{r_2}\right)
+\frac{3}{2}\right) B_{1} &=& 0 
\nonumber \\
\left( t^2\partial_t +tr_1\partial_{r_1} +t(x_1+1) +\half{\cal M}_1 r_1^2
+\half{\cal M}_2 r_2^2 \right) B_1 &=& 0 
\EEA
with the general solution
\BEQ
B_1 = t^{-3/2} f\left(\frac{r_1 r_2}{t}\right) 
\left(\frac{r_1^2}{t}\right)^{\half-x_1}
\exp\left[-\frac{{\cal M}_1}{2}\frac{r_1^2}{t}
+\frac{{\cal M}_2}{2}\frac{r_2^2}{t}\right]
\EEQ
where $f=f(v)$ is an arbitrary function. We have hence found the 
function $B=B_1$. Combining this with (\ref{A3:Gzerlegung})
then yields the last part of the assertion. \eop

\subsection{$\mathfrak{se}(3|2)$-covariant two-point functions}

In order to prove proposition 5.4, we first observe that because of the 
covariance under the generators $X_{-1}, Y_{-\half}, M_0, G_{-\half}^1$ and
$\bar{Y}_0^1$, we have
\BEQ
{\cal C} = \delta({\cal M}_1+{\cal M}_2) G(t,r,{\cal M}_1,\theta,\bar{\theta})
\EEQ
where the notation of eq.~(\ref{gl:5:inva}) was used. The remaining
six conditions become
\BEA
\left[ t\partial_t + \half r\partial_r +\half\theta\partial_{\theta} +
\frac{x_1+x_2}{2} \right] G &=& 0 
\label{gl:A:se1} \\ 
\left[ t\partial_r + {\cal M}_1 r + \half\theta\partial_{\bar{\theta}} \right]G
&=& 0
\label{gl:A:se2} \\
\left[ -\partial_{{\cal M}_1}\partial_r +r\partial_t 
+\bar{\theta}\partial_{\theta} \right] G &=& 0 
\label{gl:A:se3} \\
\left[ t\partial_t + r\partial_r +\half\theta\partial_{\theta}
+\half\bar{\theta}\partial_{\bar{\theta}} -{\cal M}_1\partial_{{\cal M}_1}
+(x_1+x_2-1) \right] G &=& 0 
\label{gl:A:se4} \\
\left[ \theta \partial_t + \bar{\theta}\partial_r \right] G &=& 0 
\label{gl:A:se5} \\
\left[ \theta\partial_r + 2{\cal M}_1 \bar{\theta} \right] G &=& 0 
\label{gl:A:se6}
\EEA
These are readily solved through the expansion
\BEQ
G = A + \theta B + \bar{\theta} C + \theta\bar{\theta} D
\EEQ
where $A=A(t,r,{\cal M}_1)$ and so on. Now, from eq.~(\ref{gl:A:se5}) we have
$\partial_t A = \partial_r A = \partial_t C - \partial_r B =0$. Similarly,
{}from eq.~(\ref{gl:A:se6}) we find $2{\cal M}_1 A=0$ and 
$\partial_r C= 2{\cal M}_1 B$. 

First, we consider the coefficient $A$. 
{}From (\ref{gl:A:se1}) it follows that $x_1+x_2=0$ and from (\ref{gl:A:se4}) 
it can be seen that $({\cal M}_1\partial_{{\cal M}_1}+1)A=0$, 
hence $A=a_0/{\cal M}_1$. Because of $2{\cal M}_1 A=0$ as derived above it 
follows $a_0=0$. 

Next, we find $C$ from eqs.~(\ref{gl:A:se1}) and (\ref{gl:A:se2}) which give
$(t\partial_t +\half r\partial_t +\frac{x}{2})C=0$ and
$(t\partial_r +{\cal M}_1 r)C=0$ with the result
$C=c({\cal M}_1) t^{-x/2} \exp\left(-{\cal M}_1 r^2/(2t)\right)$. From the
above relation $\partial_r C= 2{\cal M}_1 B$ it follows that
$B=-r/(2t) C$ and the relation $\partial_t C - \partial_r B =0$ derived before 
then implies $x=x_1+x_2=1$. Hence the terms parametrized jointly by $B$ and $C$ 
reads $(\bar{\theta}-\theta r/t) c({\cal M}_1) t^{-1/2}
e^{-{\cal M}_1 r^2/(2t)}$. Its covariance
under $V_-$ and $D$ eqs.~(\ref{gl:A:se3},\ref{gl:A:se4}) leads to 
$c({\cal M}_1)=c_2=\mbox{\rm cste.}$.

Finally, it remains to find $D$, which completely decouples from the 
other coefficients. From 
eqs. (\ref{gl:A:se1},\ref{gl:A:se2},\ref{gl:A:se3},\ref{gl:A:se4}) we have,
with $x=x_1+x_2$
\BEA
\left[ t\partial_t +\half r \partial_r +\half(x+1) \right] D &=& 0 
\nonumber \\
\left[ t\partial_r + {\cal M}_1 r \right] D &=& 0 
\nonumber \\
\left[ - \partial_{{\cal M}_1} \partial_r + r\partial_t \right] D &=& 0
\\
\left[t\partial_t +r \partial_r -{\cal M}_1\partial_{{\cal M}_1}+x \right] D
&=& 0 
\nonumber
\EEA
whose general solution is
\BEQ
D = d_0 {\cal M}_1^{(x-1)/2} t^{-(x+1)/2} \exp\left(-\frac{{\cal M}_1}{2}
\frac{r^2}{t} \right)
\EEQ
which proves the assertion. \eop

\subsection{$\osp(2|4)$-covariant two-point functions}

In order to prove the proposition 5.1 it is enough to observe that
$\osp(2|4)$ includes both $\tilde{\s}^{(2)}$ and $\se(3|2)$, hence
$\osp(2|4)$-covariant two-point functions must be also covariant under these
subalgebras. The assertion follows immediately by comparing
eqs.~(\ref{gl:5:st2}) and (\ref{gl:5:se32}). \eop 

%%%%%%%%%%%%%%%%%%%%%%%%%%%%%%%%%%%%%%%%%%%%%%%%%%%%%%%%%%%%%%%%%%%%%%%%%%%%%%%%
\appsection{B}{ }
%%%%%%%%%%%%%%%%%%%%%%%%%%%%%%%%%%%%%%%%%%%%%%%%%%%%%%%%%%%%%%%%%%%%%%%%%%%%%%%%

In order to help the reader find his way through the numerous 
Lie superalgebra defined all along the article,
we recall here briefly their definitions and collect the formulas 
for the realization of $\osp(2|4)$
as Lie symmetries of the $(3|2)$-supersymmetric model. 

The super-Euclidean Lie algebra of $\R^{3|2}$ is
\BEQ
\se(3|2) = \left\langle X_{-1,0}, Y_{\pm\half}, M_0, D, V_{-}, 
G_{-\half}^{1,2},
\bar{Y}_0^{1,2} \right\rangle
\EEQ
whose commutator relations are given at the end of section 3.1 
(see the root diagram on figure~\ref{Abb3}c). From this,
the super-Galilean Lie algebra $\mathfrak{sgal}\subset\se(3|2)$ is obtained 
by fixing the mass
\BEQ
\mathfrak{sgal} = \left\langle X_{-1,0}, Y_{\pm\half}, M_0, G_{-\half}^{1,2},
\bar{Y}_{0}^{1,2} \right\rangle
\EEQ
The super-Schr\"odinger algebras with $N=1$ or $N=2$ supercharges are called
$\tilde{\s}^{(1)}$ and $\tilde{\s}^{(2)}$ and read
\BEQ
\tilde{\s}^{(1)} = \left\langle
X_{\pm 1,0}, Y_{\pm\half}, M_0, G_{-\half}^1+G_{-\half}^2, 
G_{\half}^1+G_{\half}^2, \bar{Y}_0^1+\bar{Y}_0^2 \right\rangle
\EEQ
and
\BEQ
\tilde{\s}^{(2)} = \left\langle
X_{\pm 1,0}, Y_{\pm\half}, M_0, G_{\pm\half}^{1,2}, \bar{Y}_0^{1,2}, N_0
\right\rangle \cong \osp(2|2) \ltimes \sh(2|2)
\EEQ
The commutators of $\tilde{\s}^{(2)}$ are coherent with the root diagram of  
figure~\ref{Abb3}a and those of
$\tilde{\s}^{(1)}\subset\tilde{\s}^{(2)}$ follow immediately. 
Finally, all these Lie
superalgebras can be embedded into the Lie superalgebra $\s^{(2)}$
\BEQ
\s^{(2)} = \left\langle
X_{\pm 1,0}, Y_{\pm\half}, M_0, D, N_0, G_{\pm\half}^{1,2}, \bar{Y}_0^{1,2}, 
V_{\pm}, W, \bar{Z}_0^{1,2} 
\right\rangle \cong \osp(2|4),
\EEQ
see figure~\ref{Abb2} for the root diagram. This is
the largest dynamical symmetry algebra of the $(3|2)$-supersymmetric model with
equations of motion (\ref{gl:SSzeta}). To make the connection with 
the infinite-dimensional
Lie superalgebras introduced in section 4, let us mention that the Lie algebra 
\BEQ
\tilde{\s}^{(2)}_1= \left\langle X_{\pm 1,0},N_0, G_{\pm\half}^{1,2} 
\right\rangle \cong \osp(2|2)
\EEQ
is the subalgebra of $\tilde{\s}^{(2)}$ made up of all 
grade-one elements, with the identification
of $\tilde{\s}^{(2)}$ as a subalgebra of $\sns^{(2)}/{\cal R}$ given in 
Proposition~4.3.

Let us finally give explicit formulas for
the realization of $\osp(2|4)$ as Lie symmetries of the 
$(3|2)$-supersymmetric model, using the notation
of section 3 and 4. In formulas (B7) through (B21), 
the indices $n$ range through $-1,0,1$ while
$m=\pm\half$. Note that these formulas are compatible with 
those of Proposition 4.3 if one
substitutes $\theta^1$ for $\theta$, $\theta^2$ for 
$\bar{\theta}$, and $2{\cal M}$ for $\partial_{\zeta}$.

\newpage\typeout{ *** Seitenvorschub im Anhang B ***}
\BEA
X_n &=& -t^{n+1}\partial_t - \frac{n+1}{2} t^n \left( r\partial_r +
\theta^1\partial_{\theta^1}\right) - \frac{(n+1)x}{2} t^n 
-\frac{n(n+1)}{8}t^{n-1}r^2\partial_{\zeta} \nonumber \\ && 
-\frac{n(n+1)}{4} t^{n-1} r\theta^1\partial_{\theta^2} 
\\
Y_m &=& -t^{m+1/2}\partial_r 
-\half\left(m+\half\right)t^{m-1/2}r\partial_{\zeta}
-\half\left(m+\half\right)t^{m-1/2}\theta^1\partial_{\theta^2}
\\
M_0 &=& - \half \partial_{\zeta} \\
D &=& -t\partial_t -\zeta\partial_{\zeta}-r\partial_r
-\half\left(\theta^1\partial_{\theta^1}+\theta^2\partial_{\theta^2}\right) 
-x 
\\
N_0 &=& -\theta^1\partial_{\theta^1}-\theta^2\partial_{\theta^2} +x \\
G_m^{1} &=& -t^{m+1/2}\partial_{\theta^1}-\half\left(m+\half\right)
t^{m-1/2}r\partial_{\theta^2} 
\\
G_m^{2} &=& -t^{m+1/2} \left( \theta^1\partial_t +\theta^2\partial_r\right)
-\left(m+\half\right)t^{m-1/2}\left( \half\theta^1r\partial_r
+\half r\theta^2\partial_{\zeta}-\half\theta^1\theta^2\partial_{\theta^2}
+x\theta^1 \right) 
\nonumber \\
& & -\frac{1}{4}
\left(m^2-\frac{1}{4}\right)t^{m-3/2}r^2\theta^1\partial_{\zeta}
\\
\bar{Y}_0^1 &=& -\partial_{\theta^2} \\
\bar{Y}_0^2 &=& -\theta^1\partial_r - \theta^2\partial_{\zeta} \\
V_{-} &=& -\half r\partial_t-\zeta\partial_r-\half\theta^2\partial_{\theta^1}
\\
V_+ &=& 
       -2tr\partial_t-2\zeta r\partial_{\zeta}-(r^2+4\zeta t)\partial_r
       -r(\theta^1\partial_{\theta^1}+\theta^2\partial_{\theta^2})
       -2t\theta^2\partial_{\theta^1}-2\zeta\theta^1\partial_{\theta^2}-2xr
\\ 
W &=& -2\zeta^2\partial_{\zeta}
      -2\zeta(r\partial_r+\theta^2\partial_{\theta^2})
      -{r^2\over 2}\partial_t-r \theta^2\partial_{\theta^1}-2x\zeta
\\
\bar{Z}_0^{1} &=& 
                  -\half\left(\zeta\partial_{\theta^2}
		   + \half r\partial_{\theta^1}\right)
\\
\bar{Z}_0^2 &=& 
-\half\left(\zeta (\theta^2\partial_{\zeta}+\theta^1\partial_r)
+\half \theta^2 r\partial_r+\half r\theta^1 \partial_t
+\half \theta^1\theta^2\partial_{\theta^1} + x\theta^2\right).
\EEA

%%%%%%%%%%%%%%%%%%%%%%%%%%%%%%%%%%%%%%%%%%%%%%%%%%%%%%%%%%%%%%%%%%%%%%%%%%%%%%%%

\end{document}